\documentclass[journal,12pt, onecolumn,draftclsnofoot]{IEEEtran}
\usepackage{cite}
\usepackage{pdfpages}

\usepackage[implicit=false]{hyperref}

\usepackage{graphicx}
\usepackage{caption}
\usepackage{subcaption}
\usepackage{amsthm,amsmath}
\usepackage{amsfonts}
\usepackage[font=footnotesize,labelfont=bf]{caption}

\theoremstyle{definition} 

\newcommand\T{\rule{0pt}{2.6ex}}       
\usepackage{tabularx}
\usepackage{multirow}

\newcommand{\cN}{\mathcal{N}}
\newcommand{\C}{\mathrm{C}}
\newcommand{\D}{\mathrm{D}}
\newcommand{\vs}{\mathbf{s}}

\newcommand{\fracc}[2]{\, \displaystyle \frac{ #1}{ #2}}

\newcommand{\morabba}[1]{\,\begin{flushright}
 \Rectsteel \\
\end{flushright}}

\newcommand{\al}[1]{\,\begin{align}
                   #1 
                   \end{align}
}
\newcommand{\all}[2]{\,\begin{align}
                   #1 
                    \label{#2}
                   \end{align}
}

\newcommand{\vast}{\bBigg@{4}}
\newcommand{\Vast}{\bBigg@{5}}

\title{  
\textbf{
Conjoining uncooperative societies 
\\~~
facilitates evolution of cooperation  \\ [12mm]
}
 }

\author{
\Large
 Babak Fotouhi$^{1,2}$,  Naghmeh Momeni$^{1,3}$, \\ Benjamin Allen$^{1,4,5}$,    Martin A. Nowak$^{1,6,7}   $ \\ [12mm]
    \footnotesize
    $^1 $ Program for Evolutionary Dynamics, Harvard University, Cambridge, MA, USA\\[2mm]
 $^2 $ Institute for Quantitative Social Sciences, Harvard University, Cambridge, MA, USA\\[2mm]
 $ ^3$ Massachusetts Institute of Technology (MIT) - Sloan School of Management, Cambridge, MA, USA\\[2mm]
 $ ^4$ Department of Mathematics,  Emmanuel College, Boston, MA, USA\\[2mm]
 $ ^5$ Center for Mathematical Sciences and Applications, Harvard University, Cambridge, MA, USA\\[2mm]
 $ ^6$ Department of Mathematics, Harvard University, Cambridge, MA, USA\\[2mm]
 $ ^7$ Department of Organismic and Evolutionary Biology, Harvard University, Cambridge, MA, USA
}

\begin{document}
 \maketitle

\clearpage
 \tableofcontents
 
 \clearpage
\begin{abstract}
\textbf{Social structure affects the emergence and maintenance of cooperation. 
Here we study    the evolutionary dynamics of cooperation in fragmented societies, and show that conjoining segregated  cooperation-inhibiting  groups, if done properly,   rescues the fate of collective cooperation. 
  We highlight the essential role of       inter-group ties,  that  sew the patches of the social network together and facilitate   cooperation. 
  We point out several examples of this phenomenon in actual  settings.
   We explore random and non-random graphs, as well as empirical networks.
    In many cases we find a marked reduction of the critical benefit-to-cost ratio needed for sustaining cooperation. 
   Our finding gives hope that the increasing worldwide connectivity, if managed properly, can promote global cooperation.
   }
   \end{abstract}
   
    \renewcommand{\thesection}{\arabic{section}}
   
   \section{Introduction}
 A core problem in evolutionary game theory  is that of cooperation. 
Cooperation involves individuals  paying a cost to benefit others, and is a ubiquitous  feature of the  social life~\cite{nowak2006five,simpson2015beyond}. 
The structure of social networks  affect  pathways of information, exchange,   and other interpersonal mechanisms  which undergird cooperation~\cite{simpson2015beyond}.
      Thus a natural question in the mathematical study of evolutionary dynamics of cooperation is how network structure influences collective cooperative outcomes~\cite{hauert2004spatial,lieberman2005evolutionary,ohtsuki2006simple,szabo2007evolutionary,allen2017evolutionary}.

Here we look at the evolution of cooperation from a new perspective. 
We ask the question of how the interconnection between segregated  \emph{groups} can promote cooperation. 
Similar to individuals forming groups towards collective individually-implausible accomplishments,  sometimes groups come together to form larger composite structures.
Examples abound throughout  history, 
from  trade and intermarriage relations between tribes and communities in antiquity, to the waves of globalization which increasingly connect  local   entities for  economic, cultural,  and technological exchange.
Another example is  project management at different levels in corporations and organizations, which involves   the cooperative division of labor  between  sparsely-interconnected distinctly-specialized units.

  We use the framework of evolutionary graph theory~\cite{lieberman2005evolutionary,ohtsuki2006simple,allen2017evolutionary} to  study settings where groups that are individually undesirable for cooperation can be conjoined to build larger cooperation-promoting structures.
  We first study the conjoining of cohesive communities (clique-like structurally-homogeneous groups) under  different connection schemes. 
 We then focus on extremely-heterogeneous structures. 
 We  study  stars and their various interconnection schemes, as well as rich clubs, and introduce ensuing topologies that are \emph{super-promoters} of cooperation.
  Then we focus on  bipartite graphs. 
In addition to these ideal graph families, we consider several random graph models. 
Finally, we consider empirical social networks and investigate the role of community structure on the evolution of cooperation. 
The findings are consistent across topologies: sparse interconnections of cooperation-inhibiting graphs leads to composite structures that are better for the evolution of cooperation.

Under the framework of mathematical graph theory,  social structure is described  by a graph, in which nodes represent individuals and links represent interactions and/or relations.
 In the simplest setting, individuals are conventionally envisaged with  two possible strategies pertaining to a $2\times2$ context-specific payoff matrix  which  characterizes their interaction. 
  The outcomes of these interactions (`games') determine the   `fitness' values of the individuals: those who accrue more benefits are endowed with higher fitness, which  governs their  influence over the peers' choices of strategy. 
The most stringent form of cooperation is found in the Prisoner's Dilemma (PD) game, in which individuals  are either cooperators (paying a cost $c$ and bestowing benefit $b>c$  upon the interaction partner) or defectors (who seek to benefit without paying a cost).
 The analysis throughout this paper uses the so-called `donation game' version of PD (as shall be discussed, generalization to arbitrary symmetric 2-player games is straightforward). 
  In this game, mutual cooperation has payoff $b-c$, unilateral cooperation has payoff $-c$ for the cooperator and $b$ for the defector, and mutual defection has payoff 0. 
 The ratio  $  {b}/{c}$  characterizes the trade-off players face. 
 Throughout this paper, without loss of generality, we set $c=1$. 
 This is simply equivalent to a change of scale in payoffs, and helps brevity of notation. 
The strategies of the agents change according to death-birth (dB) updating: a random individual is chosen to update; it adopts one of the the neighbors' strategies  proportional to payoff. The small `d' indicates that death is random, while the large `B' indicates that birth is under selection.
The probability that the chosen node copies the strategy of neighbor $y$ is proportional to ${1+ \delta \pi_y}$, where $\delta$  denotes  the selection strength and  $\pi_y$ is the average payoff that node $y$ gleans playing with its own neighbors. 
We consider the limit of weak selection.
   To see if natural selection favors or hinders collective cooperation, we must calculate the probability that a single cooperator emerging at a random place in the network takes over the population. 
Natural selection favors cooperation if this fixation probability   exceeds that of the fixation probability of  a defector.  
Otherwise, natural selection inhibits cooperation.

Before we proceed, we point out a central feature of network models of cooperation, such as ours. In these models, social influence spreads beyond immediate neighbors. In conventional models of social contagion, such as simple contagion models, which often describe information diffusion, and complex contagion models, which often describe spread of behaviors~\cite{centola2010spread,centola2007complex}, the ego's activation probability depends on the states of the alters. 
 An activated alter exerts the same influence on the ego regardless of the states of the neighbors of that alter. 
 For example, in the simple-contagion model of information  diffusion, the ego needs to have heard the news from only one alter to have become informed,  and is agnostic to  how many  neighbors of that alter have heard the news. 
  Or in threshold models of complex contagion,  the ego is activated once a certain number or fraction of alters are   activated, regardless of the ego's second neighbors. 
  In contrast,  due to the strategic nature of cooperative dynamics, in our model, the radius of influence is two~\cite{nowak1992evolutionary}.  
 Ego is influenced directly by the strategies of the alters (from whom ego copies its strategy), and also indirectly by those of the neighbors of   each alter (who contribute to the payoff of that alter). 
 Our model, with a setting similar to  the  previous theoretical~\cite{hauert2004spatial,lieberman2005evolutionary,ohtsuki2006simple,allen2017evolutionary} and experimental~\cite{jordan2013contagion,rand2014static} studies of human cooperation,  thereby adds a strategic element to pure imitation dynamics. 
  Our model shares one similarity  with  simple contagion processes:  having one alter who has adopted each of the strategies makes the ego's adoption probability for that strategy nonzero.

A recently-discovered formulation gives the exact condition under which natural selection favors cooperation on a given network~\cite{allen2017evolutionary}. 
 The solution utilizes the mathematical equivalence of the problem to that of coalescing random walks on the graph, and the solution is in terms of the remeeting times of random walkers initiated at each node. 
 In the Methods section, we provide a brief overview of the framework.
 For a given network, the framework produces a quantity (which we denote by $b^*$) that determines the fate of cooperation. 
 For any network, we have $|b^*|>1$. 
If $b^*$ is positive, then $b^*$ is  the critical benefit-to-cost ratio. That is, natural selection favors the fixation of cooperation over that of defection on the given network if the benefit-to-cost-ratio is greater than $b^*$. 
 The closer $b^*$ is to unity, the better the network is for promoting cooperation. 
Conversely, if $b^*$ is negative, natural selection inhibits cooperation for any benefit-to-cost ratio. In these cases, the network promotes `spite'  instead of cooperation.  That is,  individuals are willing to pay a cost to reduce the payoff of   others.
The closer the value of $b^*$ is to $-1$, the more strongly the network promotes spite.  
 The convention of the literature has   hitherto been using $b^*$ to characterize the conduciveness of networks for cooperation~\cite{lieberman2005evolutionary,ohtsuki2006simple,nowak2006five,rand2014static,allen2017evolutionary}. 
 In  the SI, we remark   that $1/b^*$ can also be used, we discuss   the  advantages of each measure, and we find that some of the numerical results are visually better presentable using $1/b^*$ instead of $b^*$. 
 For consistency with the previous literature, we  use $b^*$ to present the   results in the main text.

We consider distinct settings in which structures that are known to  inhibit cooperation can be connected under various schemes to create larger structures that  promote cooperation. 
In the main text, for brevity, we only provide  the simplified version of the results in the large-$n$ limit (that is, the leading term), and present the full expressions  in  the corresponding Supplementary material. 
We use the   terminology of asymptotic analysis throughout. We say   $b^*$ \emph{grows as} $a n^b$, denoted $b^* \sim an^b$, if $\lim_{n \to \infty} b^*/(an^b)=1$. 
Equivalently, we call $an^b$  the  \emph{leading term} of $b^*$.

 \section{Cohesive communities}
Suppose there is a complete graph (clique) of $n$ nodes.
For a clique, selection does not favor cooperation, regardless of $b$. 
Namely, the value of $b^*$ that the method gives is negative. 
This means that  cliques promote spite.

\begin{figure}
\centering
\includegraphics[width= \columnwidth]{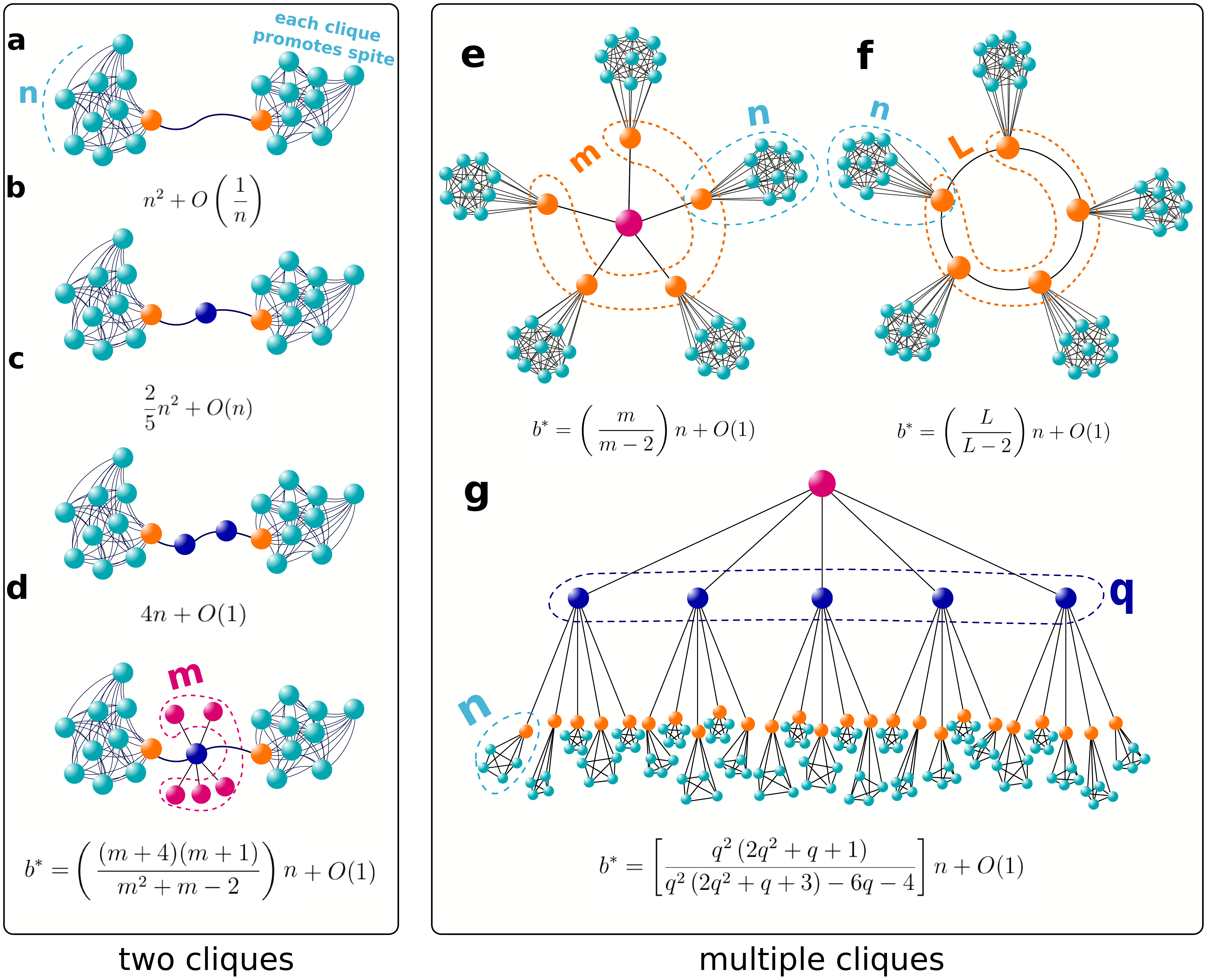}
\caption  
{
\footnotesize{
 \textbf{From spite to cooperation by conjoining cliques.}
  Cohesive communities (cliques) hinder the flourishing of cooperation. Each clique    promotes spiteful behavior. Conjoining cliques to build larger networks facilitates cooperation. 
  This figure illustrates several topologies of conjoining  two (a-d) or multiple (e-g) cliques to build composite cooperation-promoting structures. 
  If we connect two cliques, either directly (a) or via an intermediary node (b), then the composite structure is a promoter of cooperation: the critical benefit-to-cost ratio, $b^*$, grows with the square of the clique size, $n^2$. This is a steep increase of $b^*$ with network size, thus although cooperation is in principle possible, it might be impractical for actual settings. 
  (c) Having two intermediary nodes leads to further improvement: the critical benefit-to-cost ratio now grows linearly with $n$. 
 This is a much slower increase of $b^*$ with network size, as compared to the previous case. Thus, this is a more desirable interconnection scheme for actual scenarios.
   (d) The broker node who bridges two cliques can also be  connected to leaf nodes. In this case, too, $b^*$  grows linearly with $n$.
   The following conjoining schemes for multiple cliques   produce composite structures that promote cooperation with a critical benefit-to-cost ratio, $b^*$, that grows linearly with   the size of individual cliques. 
   (e) A broker node connects multiple cliques. (f) A ring of cliques which  represents the `caveman graph'. (g) Hierarchical organization of cliques.   
}
} \label{2cliques0and1} 
\end{figure}

In real-world networks, communities can join together to form larger structures that are better for cooperation. 
Suppose there are two cliques (which for simplicity we consider to be of the same size, and the analytical steps for the general case are the same) and we connect a node from the first one to a node in the second one (Fig.~1a).
We call these two nodes `gate nodes', and the rest of the nodes in the two communities `commoners'. 
In organizational  settings, for example, these  gate nodes are called `boundary spanners'. 
They are essential for   intergroup  flow of information and ideas, intergroup coordination and collaboration, and     organizational effectiveness and novelty~\cite{long2013bridges}. 
For two communities of size $n$ with the described interconnection, $b^*$ is positive and finite, but it grows as ${1\times n^2}$.
Thus it is in principle possible that natural selection favors cooperation, but the necessary $b^*$ grows quickly with network size. This might be infeasible for actual settings.
Connecting the gate nodes via an intermediary `broker' node (Fig.~1b) reduces  the leading term to ${(2/5)\times n^2}$, 
which is slightly better, but it still grows quickly with $n$. 
Marked reduction of $b^*$ ensues  if instead of one broker, there are two brokers on the path between the gate nodes (Fig.~1c). 
Each group is connected to a third-party trustee node, or representative, and exchange is done via these two nodes. 
With two broker nodes in the middle,  then the leading term of $b^*$ drops to $4n$,  thus $b^*$ grows considerably slower with network size. 
This interconnection scheme offers a substantial improvement and the two communities which individually promote spite can now be conjoined to form a new composite network which supports cooperation with more plausible values of $b^*$.

Longer chains of intermediary nodes between the two cliques is mathematically possible, but relatively less common in actual settings. 
The possible exceptions are chain-of-command structures  which   resemble this topology: a group of   decision-makers sit at one end (the first clique) and through a chain of intermediary units, the agenda reaches the bottom-most unit (the second clique)  which is  in charge of implementation. 
For chains with more than two intermediary nodes, the analytical results become too lengthy to be presented. But fortunately the employed coalescing random walks framework  enables numerical  extraction of the leading term.
If the chain of intermediaries has length $L$, with $L \ll n$,  then the leading term  of $b^*$ drops further to ${n\times 4/(L-1)}$. 
 The results for intermediate values of $L$, with the possibility of $L>n$,  are presented in   the SI.

There are also alternative  intercommunity connection schemes that offer a marked reduction in $b^*$. 
For example, if  there is one broker node between the gate nodes, and the broker is  connected to $m>1$  peripheral leaf nodes  (Fig.~1d), then  the leading term of $b^*$     is  given by  ${n (m+2)(m+5)/[m(m+3)]   }$, which is linear in $n$.

In actual settings, often there are more than two communities (local social networks, production units, etc.). 
Urbanization has led to a proliferation of diverse subcultures and enhanced interaction and  diffusion between  them as a daily principle of contemporary life~\cite{fischer1975toward,wellman2001persistence}.
In organizational settings,  `network brokers' can bridge  existing  `structural holes' and  connect multiple segregated sectors and facilitate cooperation   among them~\cite{burt2009structural,rosenthal1997social}.  
An simple example of such a setting would be a star of cliques:   $m>2$  communities   connected via a highly-central broker node (Fig.~1e). 
With  this $m$-community structure, with $m>2$ communities,  $b^*$  has the leading term $n \times m/(m-2)$. Linear growth in  community size $n$ indicates  a substantial improvement over a single community or two communities  is attained. 

Another  interconnection scheme of multiple cohesive communities is the so-called  `caveman graph' from the sociological literature~\cite{watts1999networks} (Fig.~1f). 
With  $L>2$   cliques,  situated on a ring,  the leading term of $b^*$ is given by  $n \times L/(L-2)$, which is    linear in clique size.  

Cliques can also be organized hierarchically, such as in modern organizational bureaucracies (Fig.~1g). 
In this case, too, for large cliques, the leading term of $b^*$  grows linearly with clique size, as shown in Fig.~1g.

 \section{Star-like structures}
  A star graph comprises a hub and  $n$ leaf nodes connected to the hub. 
 In this strictly-centralized system, natural selection does not promote cooperation regardless of $b$. 
 Similar to the case of cliques,  stars can be connected  to  promote collective cooperation.
 If we have two stars, one with $n$ leaf nodes  and the other with $\alpha n$ leaf nodes (Fig.~2a), 
 then if we connect the hubs, $b^*$ for large $n$ approaches  a constant ${(8+\alpha+1/\alpha)/4}$.  
 The smallest possible $b^*$ for two stars is $5/2$, which pertains to $\alpha=1$ (identical stars). 
 The independence from network size is a remarkable feature that star structures exhibit. 

 If we connect the hubs via one intermediary broker node, we get ${b^* \sim (10+\alpha+1/\alpha)/4}$. 
 For two identical stars, this simplifies to $b^* \sim 3$. 
We can also connect the hubs via a chain of $L$ intermediary brokers, such as in a chain-of-command structure with a decision-making unit at the top and and an implementation unit at the bottom. For $L \geq 1$,  the leading term  of  $b^*$  is given by ${(8+2L+\alpha+1/\alpha)/(L+3)}$. 
 In all these cases, it is remarkable that  
  for large network size, $b^*$ tends to a constant.  This   independence from network size  evinces the high merit of locally-star-like structures in the promotion of cooperation.

\begin{figure}
\centering
\includegraphics[width=.9 \textwidth]{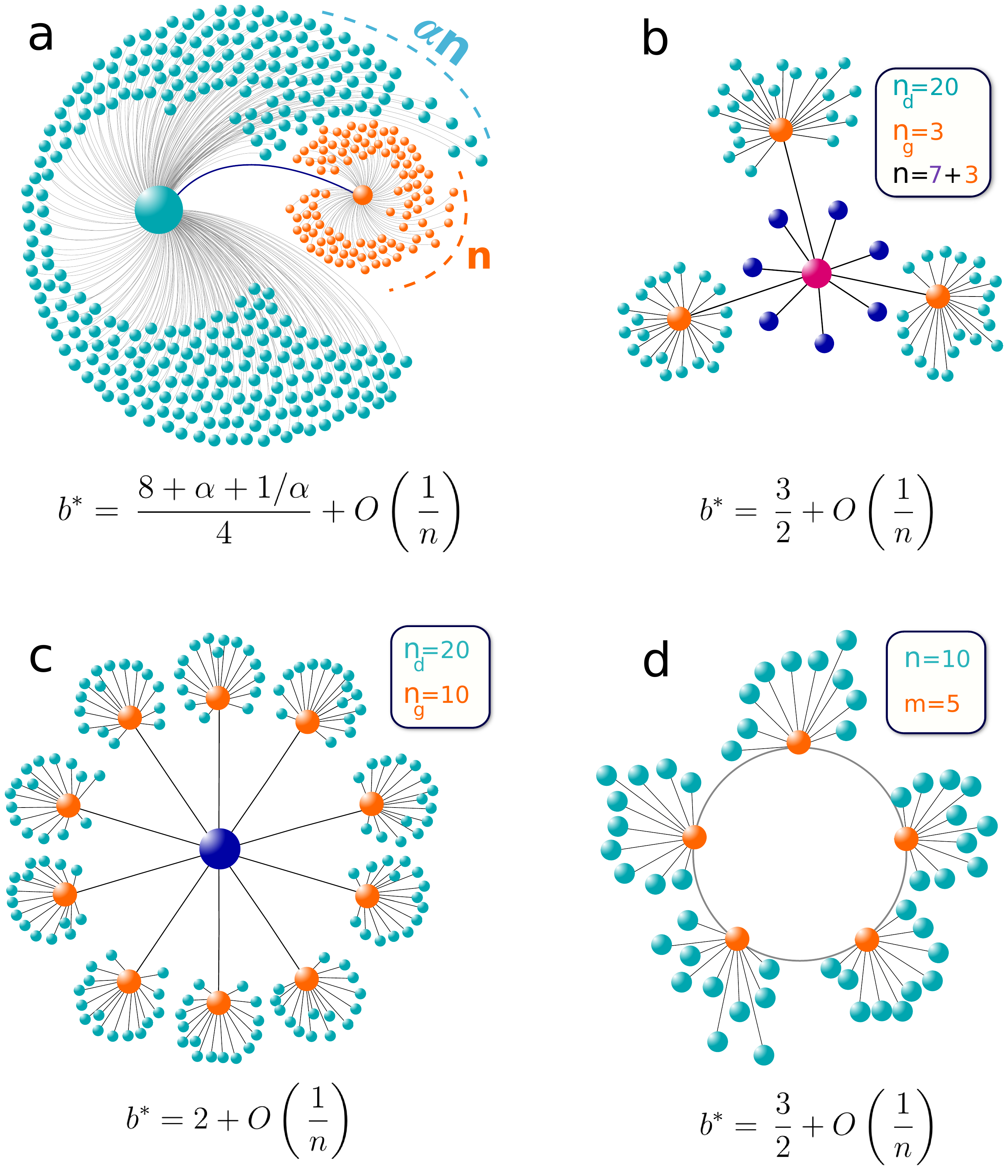}
\caption  
{
\footnotesize{
\textbf{Super-promoters of cooperation.} 
Star graphs represent extreme core-periphery structures where a central node is connected to many leaf nodes. Although a single star hinders cooperation, connecting stars promotes cooperation. All reported critical benefit-to-cost ratios, $b^*$, pertain to the limit of large population size. Exact formulas are  presented  in   the SI, Section~\ref{SI:star}. (a) Two stars, one with $n$ leaf nodes and the other with $\alpha n$ leaf nodes. (b) An imperfect meta-star: a central node has n peripheral nodes, $n_g$ of them are hubs, while $n_d$ of them are leaves. If $n_g \ll n \ll n_d$, then $b^*$ tends to 3/2 and the average degree tends to 2. Thus, the structure is a super-promoter of cooperation, since $b^*$  is less than the average degree. (c) The perfect meta-star is a hierarchical structure with a head node connected to $n$ subsidiary nodes, each of them connected to $n_d$ peripheral nodes. The reported result is for the case $n \ll n_d$, which means most of the population belongs to the bottom layer. (d) A more flat hierarchical structure: there are $m$ head nodes connected on a ring, each with $n$  peripheral nodes. For $m \ll n$, this graph becomes a super-promoter of cooperation, outperforming the strict hierarchy.
}
} \label{stars_combined}
\end{figure}

 In many actual settings,    star-like structures are not directly connected as we envisaged above. 
 Rather, global hubs  are connected to local large-scale hubs, which are in turn connected to local peripheral nodes. 
This leads to a hierarchical organization: the head unit  connects to a number of subsidiary units, each of them connect in turn to subordinate units, and so an. 
To  study  this interconnection scheme, we consider graphs   with megahubs and hubs in a nested manner. 
We consider only two levels, though the calculation can be in principle extended to more.
 Out of the $n$ total leaf nodes  of a star graph, we take $n_g$ of them   and attach $n_d$ nodes to each  (Fig.~2b).
 The total number of nodes will be ${1+n+n_g n_d}$, and the number of links is $n+n_g n_d$. 
 The full expressions for $b^*$ are long (presented in the SI, Section~\ref{SI:imperfect}), but simplifications can be obtained in some interesting limits. 
We consider the case  where the number of leaf nodes  are much larger than the number of hubs. 
This is the case in many actual settings, to the extent that the  marked imbalance between the latter two numbers  constitutes the cornerstone of many egalitarian social  discourses and movements. 
    If we have $n_g\ll n$, 
 then the leading term  of $b^*$ approaches   ${3/2 }$. 
 Whether $n_d$ and $n$ are of the same order of magnitude, or  if we have$ n \ll n_d$, only affects the second leading order terms. 
 This leading behavior of $b^*$  is particularly interesting because the average degree approaches 2 in  these cases, and $b^*$ being less than the average degree is a rare property of graphs.  Hence we can dub these structures `super-promoters' of cooperation. 

 We  can readily generalize these results to the fully-hierarchical structure (where $n_g=n$), that is, a star of stars (Fig.~2c). 
 A mega-hub is connected to $n$ hubs which are each connected to $n_d$ leaf nodes. 
 In the limit of $n \ll n_d$, $b^*$ approaches 2, which indicates that this structure is a strong promoter  of  cooperation. 
 Full results are presented in the SI, Section~\ref{SI:perfect}. 
 
Hierarchies can also be more `flat', which is getting popular in certain management approaches~\cite{benkler2011penguin}. 
The simplest model would be to have the  upper layer of nodes  connect horizontally instead of hierarchically. 
 We consider the simple case where the hubs of $m$ stars, each with $n$ leaf nodes,  are connected on a ring (Fig.~2d).
 For large $n$, the leading term of $b^*$  approaches  ${(3m-1)/(2m-2) }$, which is independent of $n$. 
 This means that for large $m$, the value of $b^*$ approaches $3/2$. 
 The average degree in this limit approaches 2. 
 Hence, a ring of stars is another super-promoter of cooperation. 
 The full results for this setup are presented in the SI, Section~\ref{SI:star_ring}.

 \section{The Rich Club}
  A rich-club network is one comprised of a small dense core of connection-rich high-degree nodes and  a large sparse periphery. 
  These structures are found across social and technological networks. 
  The notion of `oligarchy' in institutions and organizations is usually linked to structures that can be characterized by such a rich-club feature~\cite{ansell2016says}. 
Other examples with this feature  include the social network of company executives and directors (within-company~\cite{burt2010neighbor}, national inter-company~\cite{fracassi2016corporate}, and international inter-company~\cite{heemskerk2016corporate}), 
 the collaboration network  between  academics~\cite{crane1969social},   and the Internet~\cite{zhou2004rich}. 
  
 As a simple example with  this characteristic, we consider  a clique of $n_c$ nodes (where $c$ denotes `core') and $n_p$ peripheral nodes. 
  Each core node is connected to every other core node and every peripheral node. 
  Each peripheral node is connected to every core node but to none of the other peripheral nodes. 
  In the special case of $n_c=1$, this becomes a star graph. 
For a single rich-club network, natural selection does not favor cooperation, regardless of $b$. 
Similar to the case of a single clique, single rich-club networks promote spite.

\begin{figure}
\centering
\includegraphics[width=.8 \columnwidth]{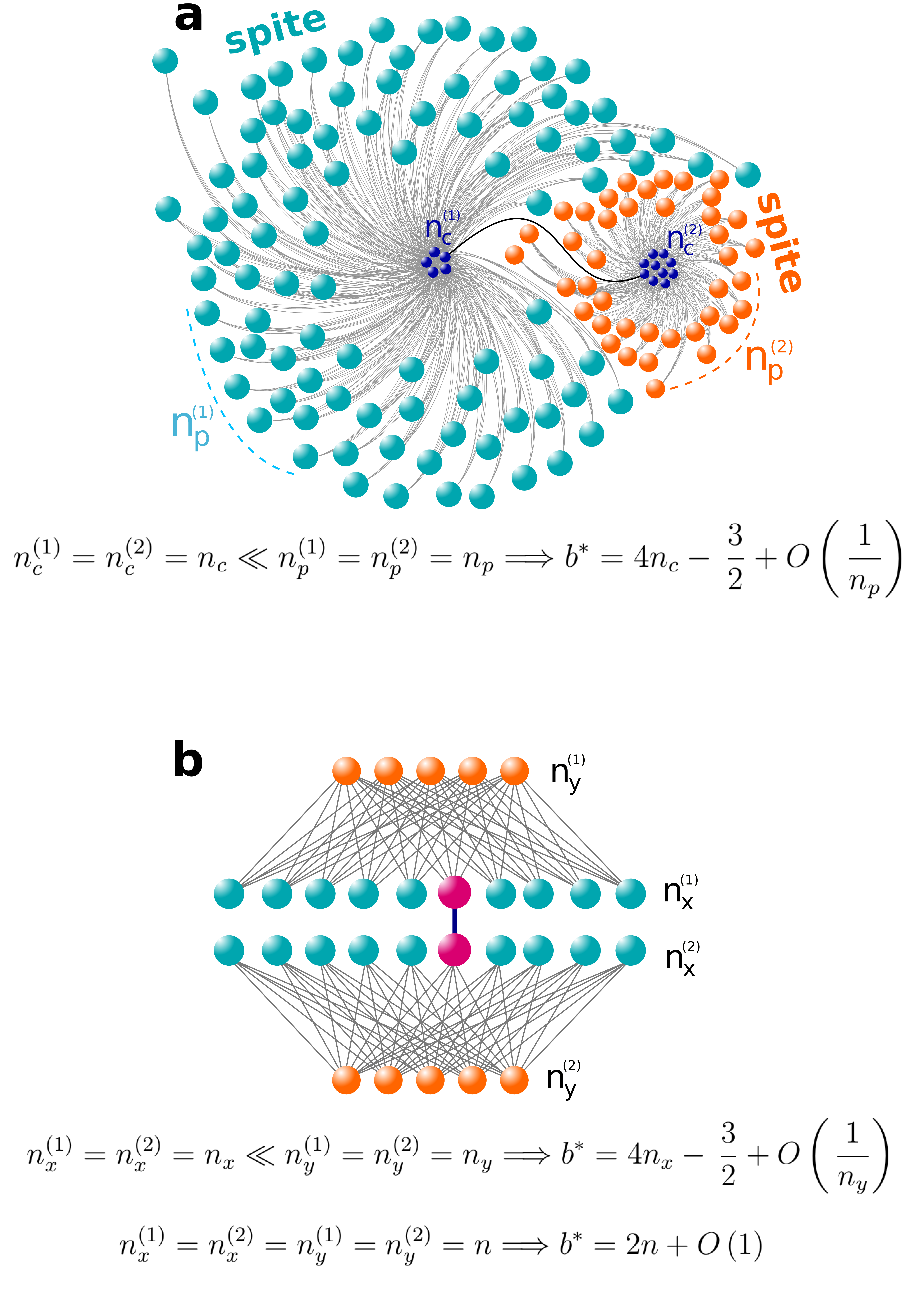}
\caption  
{
\footnotesize{
\textbf{Rich clubs and bipartite graphs.} 
(a) Rich-club graphs comprise a dense core and a large, sparse periphery. A single rich club hinders cooperation, but conjoined rich clubs promotes cooperation. For the simple case of two identical rich clubs, with the periphery size, $n_p$, much larger than the core, $n_c$, the critical benefit-to-cost ratio $b^*$ grows linearly with $n_c$. (b) Complete bipartite graphs comprise two distinct groups of nodes, where links exist only between the two groups, but not within each group. Examples are buyer-seller networks or heterosexual marriage networks. A single bipartite graph hinders cooperation, but connecting them promotes cooperation. For the simple case of two identical graphs, each with two groups of the same size,  $b^*$ grows linearly with group size $n$.
}
} \label{RichClubs_and_Bipartite}
\end{figure}

To improve the situation,  we connect two rich-club networks by connecting a `gate' node in the first core to a gate node in the second (Fig.~3a).
An actual example of conjoining rich clubs via cores is that director networks of different companies often connect, and they do so predominantly via their cores, rather than the peripheries---creating  `interlocked directorates'~\cite{davis1996significance}. 
  In the simple case of two identical rich-club networks with $n_c \ll n_p$ (small core and large periphery), the leading term  of $b^*$  is given by $4n_c -3/2$, which is a linear function of ${n_c}$.
  That is, the leading behavior in the large-$n_p$ limit  only depends on the number of core nodes and is independent of the number of peripheral nodes. 
  In the case of $n_c=1$,  this leading term is  $5/2$, which is consistent with our previous findings for star graphs. 
  For $n_c=2$, the  leading term of $b^*$  is 13/2. 
  The results point out a remarkable   feature of these structures: when the periphery is large, the fate of the collective outcome is determined solely by the core.

 \section{Bipartite structures}
  In a bipartite network, nodes can be  divided into two distinct groups, where  there is no intra-group link. 
 For example, traditional  heterosexual marriage networks   comprised two disjoint sets;  males only connected to females and vice versa.
 Other examples include buyer/seller~\cite{kranton2003theory}, and   employer/employee~\cite{rocha2010information} bipartite networks.
  
  Here we present the results for the simplest case of a bipartite graph which is analytically tractable: we consider a complete bipartite graph.
  A complete bipartite network is one which has two groups, and each node is connected to every node in the other group but no node in its own group. 
  Natural selection does not promote cooperation on a complete bipartite graph, regardless of $b$. 
  If we connect two bipartite networks, however, the situation improves (Fig.~3b). 
  Consider  a bipartite graph  comprising two groups of nodes with sizes $n_x$ and $n_y$, respectively. 
  Suppose we connect two identical such bipartite graphs by connecting a type-$x$ node in the first graph in a type-$x$ node in the second. 
  In the special case of $n_x=n_y=n$, $b^*$ grows linearly with $n$. 
  The leading term of $b^*$  in this case is given by  $2n$. 
  Alternatively, if $n_x \ll n_y$, then  the leading term of $b^*$ only depends on $n_x$, and is given by $4n_x-3/2$.
 Hence, similar to rich clubs, if a large group of nodes are not interconnected within themselves and are all connected only to another small group of nodes, the collective outcome will be determined by that small group.

\begin{figure}
\centering
\includegraphics[width=.85 \columnwidth]{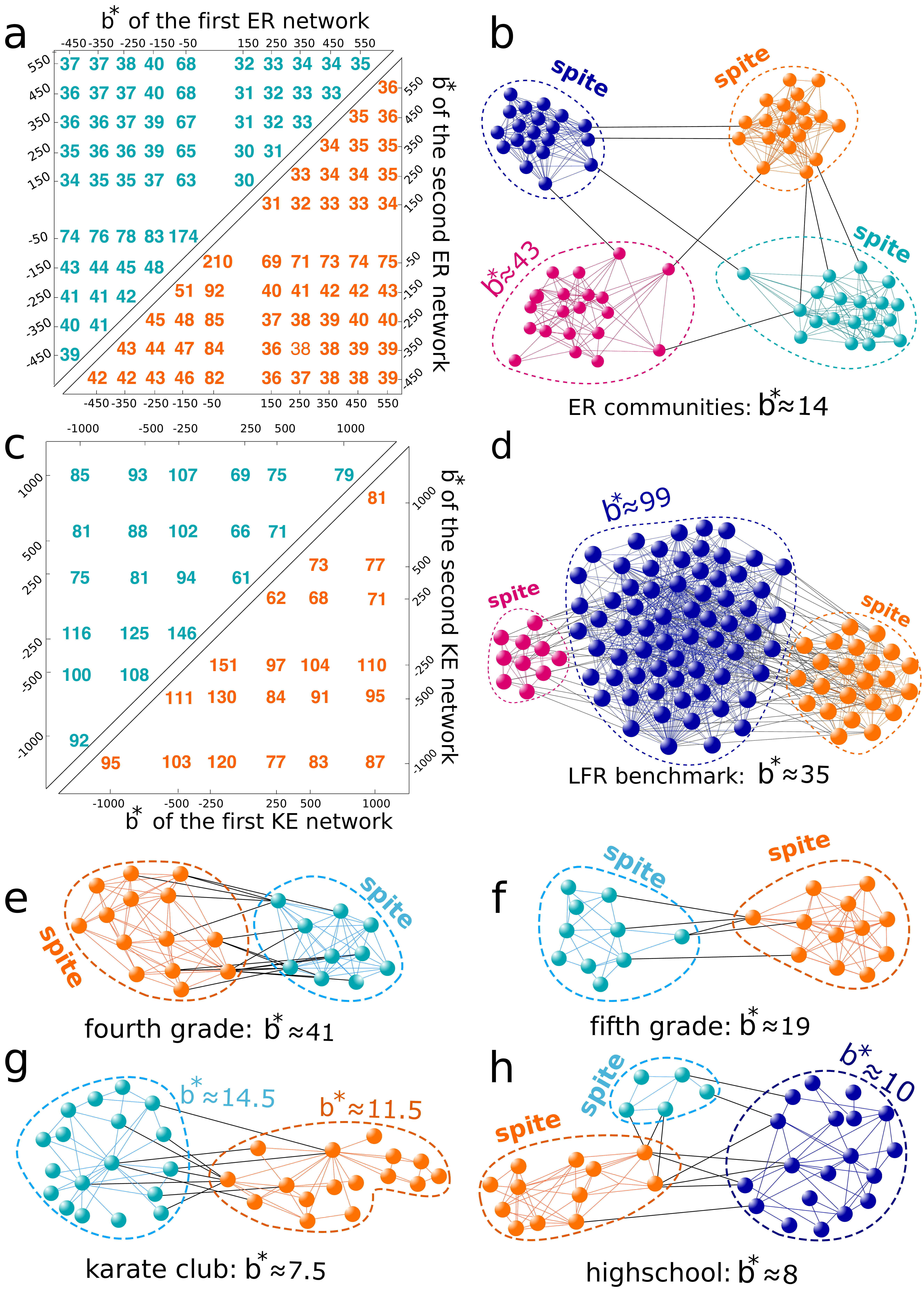}
\caption  
{
\footnotesize{
\textbf{Conjoining random graphs and empirical networks.}
a) Conjoining two   Erd\H{o}s-R\'enyi  random graphs directly (lower triangle) and via one broker node (upper triangle). 
b) Extension to more than two graphs: four ER graphs with link-formation probabilities 0.8 (top left), 0.7 (top right), 0.45 (bottom right), and 0.35 (bottom left). The inter-community link probability is 0.01. 
 c) Conjoining two scale-free networks generated by the model of Klemm and Eguiluz~\cite{klemm2002growing}. 
 d) An example network with community structure generated by the LFR benchmark~\cite{lancichinetti2008benchmark}.
Four  empirical   friendship networks (e-h).
We employed standard community detection algorithms to partition the data set into communities, and calculated  $b^*$ for the whole network and for each community separately. 
In every case, the whole network  is better than individual subnetworks in promoting cooperation. 
}
} \label{ER_conjoin}
\end{figure}

   \section{Random graphs and empirical social networks} 
Since actual social networks typically have more randomness than the ideal structured considered above, we investigate random networks to check if they have qualitatively similar properties. 
Our first test (discussed in the SI, Section~\ref{SI:robustness})  is to add structural noise to the above-considered topologies and verify that the $b^*$ values are indeed robust against  structural deviations. 
For the next check, we investigate how conjoining cooperation-inhibiting random networks can promote cooperation. 
We   generate 10 random  Erd\H{o}s-R\'enyi graphs~\cite{ER1}, with   values of $b^*$ that are undesirable for cooperation:   negative (promoting spite)  or  highly positive (hindering cooperation). 
Network size is fixed at 40.
There are 55 possible network pairs (45 pairs in which the two networks are different and 10 pairs in which they are identical), 
and there are 1600 ways to conjoin  two networks via one gate node in each. 
We calculate the median value of  $b^*$ among all these possible conjoinings for each pair of networks.  
The lower triangle in  Fig.~4a presents the resulting  $b^*$ of the conjoined network against the  $b^*$  of the first and the second network. 
The upper triangle presents the results for the same procedure, except the gate nodes  are connected via one broker node, instead of being directly connected. 
It can be seen that in most cases, a substantial improvement is achieved in both conjoining schemes. 
The most resistant case is the one with $b^*=-50$. 
Note that networks whose $b^*$  is negative are promoters of spite, and the closer to zero the value of $b^*$  is, the more strongly the structure promotes spite. 
The results indicate that if the spite-promotion capacity of either  group is high, conjoining them would be less helpful collectively. 
In Fig.~4b we illustrate that the conjoining mechanism works also  for more than two ER networks. In the example case shown, three of the four ER networks promote spite, and one of them promotes cooperation with $b^* \approx 43$. 
Creating inter-community links between these four groups with probability $0.01$ begets a marked improvement: the overall structure has $b^* \approx14$, which is considerably better than each of the individual groups for promoting cooperation. 
 A generalization of this procedure gives rise to the stochastic block model, which we investigate in  the SI. 
 
%

The same conjoining procedure is applicable to networks with heavy-tailed degree distributions, which emulate actual social networks more realistically than ER networks. 
Here we use the model proposed by Klemm and Eguiluz~\cite{klemm2002growing} to generate scale-free networks with both small-world property and high clustering coefficient, which are both ubiquitous features in social networks. The results are presented in Fig.~4c. 
Conjoining every pair of networks  produces a composite network with positive $b^*$.
In the SI, we present results for four  additional scale-free models. The results are qualitatively similar, and the improvement in $b^*$ via conjoining  ensues  consistently.

To study the effect of community structure on the cooperative outcome, we employ the   Lancichinetti-Fortunato-Radicchi (LFR) benchmark~\cite{lancichinetti2008benchmark} that are used for comparing community-detection algorithms. The procedure generates networks with  community structure in which   the degree distribution within each community and the distribution of community sizes are both heavy-tailed.
Fig.~4d depicts an example case with 100 nodes divided into three communities. The degree distribution is scale-free with exponent 2. The community sizes are 10, 23, and 67. Only the largest community  has a positive  $b^*$, with $b^*\approx 99$. The composite network (with mixing parameter 0.1) has $b^* \approx 35$.
In Supplementary Method 1.8, we provide a systematic investigation for LFR networks and show that, consistent with the above findings, when communities are not conducive to cooperation, sparse interconnections tend  to generate composite networks better than the  individual modules.

We can apply  the same mathematical formalism to real-world social network data. 
We use offline social networks that pertain to friendships, to ascertain that cooperative dynamics would be reasonable. 
We use two children friendship networks of fourth grade and fifth grade students~\cite{parker1993friendship,anderson1999p} (for the third grade, no community structure is detected because the network is dense and most people are friends with most others, so we did not use it). 
The second data set is the well-known friendship network of the members of  a  Karate club~\cite{zachary1977information}, and the third data set we use is Coleman's classic highschool friendship network data set~\cite{coleman1964introduction}. 
The results are presented in Fig.~4, panels e-h (more detailed results are presented in Supplementary Table 1). 
We divided the graphs into two communities using the Girvan-Newman method~\cite{girvan2002community}. 
In cases where using three as the number of communities returned meaningful results, we considered both two and three communities separately. 
For all networks, the algorithm returned single-node communities for more than three communities, so we did not consider those cases. 
In all cases, the collective cooperative merit of the network is markedly better than that of the individual communities. 
This reaffirms the advantageousness of inter-group connection vis-\`a-vis cooperation.

\section{Discussion}
  Each population structure can be quantified according to its intrinsic propensity to promote cooperation  (paying a cost to benefit others) or spite (paying a cost to harm others)~\cite{allen2017evolutionary}.  Here we report the observation that sparsely conjoining cooperation-inhibiting  structures tend  to produce cooperation-promoting structures. 
  We have explored this effect when joining together fully connected cliques, star-like structures (which are dominated by a single individual), rich-clubs, and even random graphs. We have found the phenomenon in examples of real social networks that already consist of conjoined sub-structures.

 In our findings, conjoining two graphs that are already favorable for cooperation always results in a cooperation-promoting composite structure, though sometimes the composite graph might not  promote cooperation as strongly as the two individual graphs did. 
 But we did not find any example in which the composite graph would  inhibit cooperation, that is, either with $b^*$ significantly larger than those of the two initial graphs, or with a negative $b^*$. 
  We investigated  random and non-random graph families considered in this paper,  and  several others.

   An extension to our work would be finding better conjoining schemes for cliques. 
  Here we showed that conjoining cliques in the manners described  above results in composite networks that are considerably better than individual cliques. These conjoining methods yield  $b^*$ values that grow linearly with $n$. For very large networks, this improvement might still not be enough. 
  A valuable extension would be to find structures that, similar to the case of stars and rich clubs, would produce $b^*$ that reaches a constant for large clique size.

  We note that evolutionary graph theory, which we employed in this paper,  is a general approach to study the effect of population structure on natural selection.
It is not limited to any particular game and not restricted to one shot interactions. 
The  results   are generalizable to any matrix game (see Methods). 
Hence the competing strategies could instantiate repeated interactions and conditional behavior~\cite{ohtsuki2007direct}. Extensions of
evolutionary graph theory can be used to study direct reciprocity with crosstalk~\cite{reiter2018crosstalk}, and  indirect reciprocity with optional interactions and private information~\cite{olejarz2015indirect}. 
   On the other hand,  there are   social settings   our model is not applicable to. 
  For example, if  each individual interacts with only a subset of its neighbors, 
  then exclusion and inclusion become essential elements of network power. 
  This is an important  feature in Network Exchange Theory~\cite{willer1999network}. 
  In this case, broker nodes have leverage over others due to the high exclusion/inclusion asymmetry.  
  Our model does not consider the possibilities of exclusion and inclusion, and each player plays with every neighbor. 
  Thus an interesting extension to the present paper would be to study analytically the said effects of exclusion/inclusion in a game-theoretical setting to build on the previous experimental work, particularly Network Exchange Theory. 
%

  Finally, we highlight that our results are qualitatively  consistent with several  simulation studies   in the literature across different contexts: 
 cooperation is promoted by interdependence between networks   in spatial public goods games and the Prisoner's Dilemma on interdependent networks~\cite{wang2013optimal,wang2013interdependent,jiang2013spreading}, 
  even if it is endogenous and inter-population links are only rewarded to high-payoff individuals~\cite{wang2014rewarding}.
  The same is true  if multiple types of interactions  are  considered, resulting in a multiplex network~\cite{battiston2017determinants}. 

  Our    findings suggest  a recipe for how to build societal structures that effectively promote cooperation, 
  and together with the ensemble of previous results in the literature, 
  they  engender  hope regarding the increasing interconnection of the contemporary world.

  \clearpage

\section{Methods}
 
  We follow a recently-discovered  framework for unweighted, undirected graphs without self-loops~\cite{allen2017evolutionary}. 
Let us denote the degree of node $x$ with  $k_x$ and its set of neighbors by $\mathcal{N}_x$. Then, we define $p_x$ as the probability that a random walk of length 2 initiated at node $x$ will terminate at node $x$:
\all{
p_x \stackrel{\text{def}}{=} \fracc{1}{k_x} \sum_{y\in \mathcal{N}_x} \fracc{1}{k_y}
.}{px}
We then solve the following   system of $\binom{N}{2}$ linear equations for symmetric quantities $\tau_{xy}$, which are the meeting times of two random walkers initiated at nodes $x$ and $y$: 
\all{
\tau_{xy}= \tau_{yx} = 
(1-\delta_{xy} )
\Bigg[ 
1+\frac{1}{2 k_x} \sum_{z\in \mathcal{N}_x}   \tau_{zy}
+\frac{1}{2 k_y} \sum_{z\in \mathcal{N}_y}  \tau_{zx}
    \Bigg] 
.}{sys}
Here,  $\delta_{xy}$  equals  unity if $x=y$ and is zero otherwise. 
Using these  quantities, we define $\tau_x$ for each node as the expected remeeting time of two random walkers initiated at node $x$ as follows: 
\all{
\tau_x \stackrel{\text{def}}{=} 1+\fracc{1}{k_x}\sum_{y\in \mathcal{N}_x}   \tau_{yx}
.}{taux}
The necessary condition for cooperation to be favored by natural selection  is that ${b ( \sum_x p_x \tau_x k_x   -2N\overline{k} )}$ is greater than ${ c  (\sum_{x} \tau_x k_x-2N\overline{k})}$. 
 If the coefficient of $b$ in this inequality is nonpositive, cooperation is never favored. If the coefficient is positive, then 
the critical benefit-to-cost  ratio is given by the following relation:
\all{
b^* = \fracc{ \sum_{x} \tau_x k_x-2N\overline{k} }{ \sum_x p_x \tau_x k_x   -2N\overline{k} }
.}{Gamma}

  The calculations for specific graphs discussed in the main text can be simplified utilizing their structural  symmetry.
  For example, for a single community (a complete graph), there is only one variable: the remeeting time between any pair of nodes (because $\tau_{xx}$ values are zero). 
  For two communities connected directly by a link, there are only four distinct values for $\tau_{xy}$: the remeeting time between two commoners, between a commoner and the gate node of the same community, between a commoner and the gate node of the other community, and between the two gate nodes. 
  This reduces Equation~\eqref{sys} to a system of four equations with four unknowns. 
  
The results are generalizable to arbitrary $2\times2$ games~\cite{tarnita2009strategy}. For a game with strategies A and B with corresponding payoff matrix $R,S,T,P$, the condition that  natural selection favors strategy A over B in the limit of weak selection is: ${ (T-S)<(R-P)(b^*+1)/(b^*-1)}$.

For the KE networks used in Figure.~4c, we used the model of Klemm and Eguiluz~\cite{klemm2002growing}. We generated many networks, with the cross-over parameter $\mu$ and the number of initial active nodes $m$ both selected randomly in their valid ranges. We selected 6 networks whose $b^*$ differed from the corresponding values used in Fig.~4 by less than 5\%. 

\textbf{Data Availability}. 
All the network data sets used in this paper are freely and publicly available in The Colorado Index of Complex Networks (ICON) collection: \url{https://icon.colorado.edu}

\textbf{Code  Availability.}
For the LFR benchmark, we used the publicly-available code that the authors of Ref.~\cite{lancichinetti2008benchmark} have provided: 
\url{https://sites.google.com/site/santofortunato/inthepress2}
\\
For the coalescing random walks framework, the code for computing   $b^*$   is 
publicly available in Zenodo  at 
 \url{http://dx.doi.org/10.5281/zenodo.276933}

 \newpage
 
\begin{itemize}
  \item \textbf{Correspondence:} Correspondence and requests for materials
should be addressed to B.F. \\(email: babak\_fotouhi@fas.harvard.edu).
 \item \textbf{Acknowledgments:}
 This work was supported by the James S. McDonnell Foundation (B.F.),  NSF grant 1715315 (B.A.), and the John Templeton Foundation (M.A.N.).
 The funders had no role in study design, data collection and analysis, decision to publish, or preparation of the manuscript. 
 B. F. thanks Steven Rytina for insightful and stimulating  conversations. 
 \item \textbf{Author contributions:} All authors contributed   to  all aspects of the paper.
 \item \textbf{Competing Interests:} The authors declare that they have no
competing   interests.

\end{itemize}


\clearpage

\clearpage
\setcounter{figure}{0} \renewcommand{\thefigure}{SI.\arabic{figure}} 
\setcounter{table}{0} \renewcommand{\thetable}{SI.\arabic{table}} 
 
 \vspace{3cm}
  \section*{\LARGE \textbf{Supplementary Information}}
  \addcontentsline{toc}{section}{Supplementary   Information}
%
%
 \setcounter{secnumdepth}{0}
 \setcounter{secnumdepth}{3}
 \setcounter{section}{0}
 \renewcommand{\thesection}{S\arabic{section}}
  \renewcommand{\thesubsection}{\thesection.\Alph{subsection}}

 \vspace{3cm}
 \section{Steps for the calculation of $b^*$}
 Here we repeat the formulas for convenience of  reference. 
 The return probability of a length-2 random walk staring at node $x$ is: 
\all{
p_x \stackrel{\text{def}}{=} \fracc{1}{k_x} \sum_{y\in \mathcal{N}_x} \fracc{1}{k_y}
.}{px}
The system of recurrence equations for meeting times $\tau_{xy}$ are: 
\all{
\tau_{xy}= \tau_{yx} = 
(1-\delta_{xy} )
\Bigg[ 
1+\frac{1}{2 k_x} \sum_{z\in \mathcal{N}_x}   \tau_{zy}
+\frac{1}{2 k_y} \sum_{z\in \mathcal{N}_y}  \tau_{zx}
    \Bigg] 
.}{sys}
Using the solution of this  system of linear equations, we obtain  the remeeting times  $\tau_x$ as follows: 
\all{
\tau_x \stackrel{\text{def}}{=} 1+\fracc{1}{k_x}\sum_{y\in \mathcal{N}_x}   \tau_{yx}
.}{taux}

Then the critical benefit-to-cost  ratio is given by the following relation:
\all{
b^* = \fracc{ \sum_{x} \tau_x k_x-2N\overline{k} }{ \sum_x p_x \tau_x k_x   -2N\overline{k} }
.}{Gamma}

 Below we consider several different topologies and  calculate the critical cost to benefit ratio for them. 
 Since some cases are nested within others, we could first solve the general cases and then present others as special cases, but we chose to present 
  the cases in increasing complexity for pedagogical purposes. 
 
 \subsection{Using  $b^*$ and $1/b^*$}
The convention of the literature has    been using $b^*$ to characterize the conduciveness of networks for cooperation. 
 We remark here that $1/b^*$ can also be used, and it has two advantages: it is confined to the range $(-1,1)$, and it is monotonic. 
 We find that some of the results are visually better presented using $1/b^*$ instead of $b^*$. 
 With this alternative measure, strong promoters of spite (whose $b^*$ are negative but small in absolute value) will fall close to -1, weak promoters of spite will be close to zero  but on the negative side, weak promoters of cooperation will be close to zero but on the positive side, and strong promoters of cooperation will be close to $+1$. 
 Moreover, as we discuss here and as is shown in~\cite{allen2017evolutionary}, for  complete bipartite graphs $b^*$ is infinite, and with  the new measure, zero is assigned to these graphs. 
In this paper  we use $b^*$ to present the analytical results, to be consistent with the previous literature and for the results to be easily comparable.
It is also more intuitive because we fix the cost to $c=1$ and practically, one would seek the required benefit to inject into a system so that cooperation would flourish. Although $1/b^*$ is mathematically more suitable, $b^*$ is   more readily interpretable. 
We use $b^*$ for the analytical calculations in the main text and in the SI, and for some numerical results we present both. In some numerical cases we find that $1/b^*$  produces better visual comprehension, so we use it instead of $b^*$ for producing those plots.

\section{Star graphs}\label{SI:star}

\subsection{Single star graph}\label{sec:star}
Consider a star graph comprising a hub and $n$ leafs. 
 Due to symmetry, there  are only two remeeting time: $\tau_{\ell \ell}$ between two leafs and $\tau_h \ell$ between a hub and a leaf. 
Let us  write the system of equations~\eqref{sys} for this network: 
 \al{
 \begin{cases}
 \tau_{h \ell} = 1+\frac{1}{2n} \big[ (n-1) \tau_{\ell \ell} \big]  
 \\
 \tau_{\ell \ell} = 1 + \frac{1}{2} \tau_{h\ell} + \frac{1}{2} \tau_{h \ell}.
\end{cases}
}
Solving this, we get
\all{
\begin{cases}
 \tau_{h \ell}= 3 - \frac{4}{n+1} \\
 \tau_{\ell \ell}=4 - \frac{4}{n+1}
\end{cases}
}{sys3}

Inserting this into~\eqref{taux}, we get  ${\tau_h=\tau_{\ell}= {4n}/{(n+1)}}$.  Also note that  ${p_h=1}$ and ${p_\ell=1/n}$. Let us use these values to calculate the denominator of Equation~\eqref{Gamma}.  The sum in the denominator becomes $ {4n}$. Noting that the average degree of the network is $ {2n/(n+1)}$, the second term in the denominator becomes $4n$. we observe that the denominator of~ Equation\eqref{Gamma} becomes zero. This indicates that cooperation is not favored  regardless of $ b/c$.

\clearpage

\subsection{Extended star graph}
\begin{figure}[h]
	\centering
	\includegraphics[width=.4\linewidth]{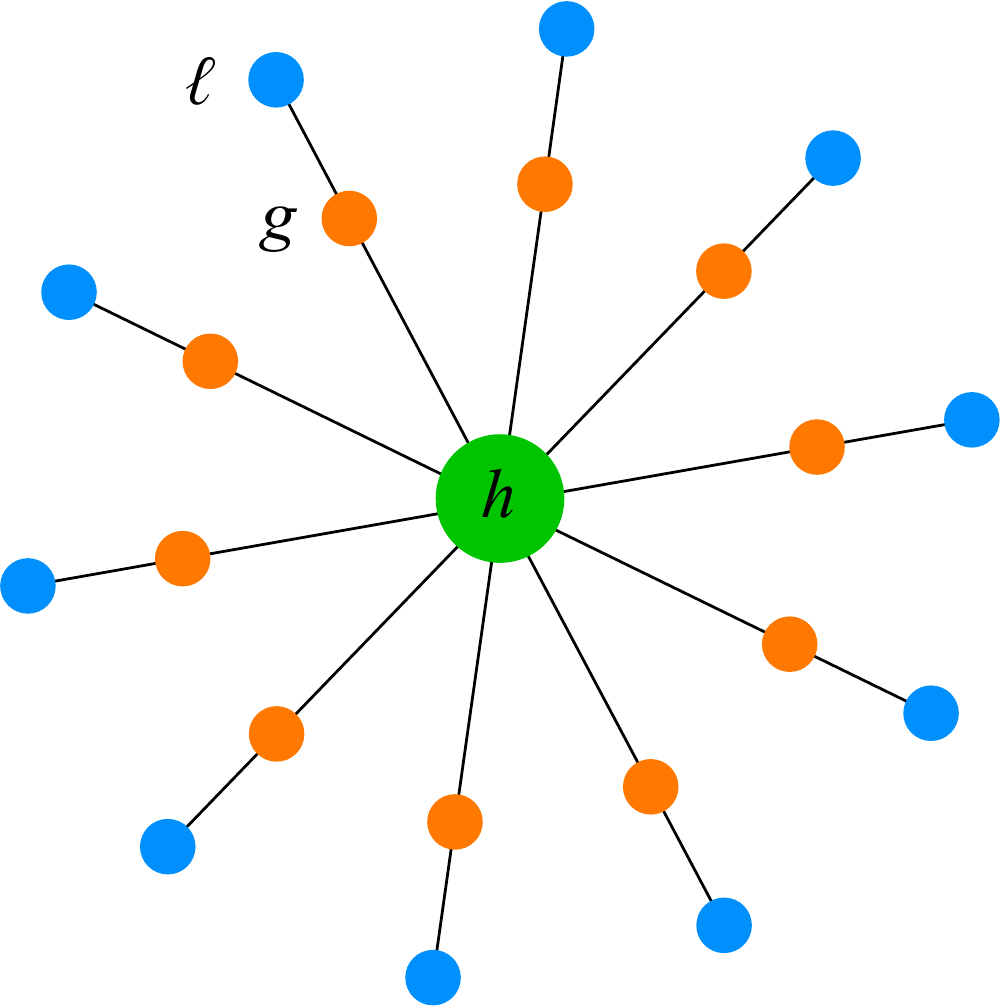}
	\caption{An extended star graph with $n=10$ }
	\label{fig_extended_star}
\end{figure}

Consider an extended star graph, which is made by taking a star graph and then, to each leaf, attaching one new node. 
So the resulting graph has one hub, $n$ nodes with degree 2, and $n$ leafs with degree 1.
An extended star graph with $n=10$ is depicted in Figure~\ref{fig_extended_star}. 
Let us denote each leaf with $\ell$, and the hub by $h$, and degree-2 nodes with $g$ (where $g$ stands for `gate'). 
 There are   distinct values for remeeting times: 
 $\tau_{h g}$, $\tau_{h \ell}$, $\tau_{\ell \ell'}$ (between two leafs), $\tau_{g g'}$ (between two gates), 
 $\tau_{g \ell}$ (between a gate and the leafs attached to it), 
 and $\tau_{g \ell'}$ (between a gate and a leaf attached to another gate). 

  \all{
  \begin{cases}
  \tau_{h g  } = 1 + \frac{1}{2n} \Big[ (n-1) \tau_{g g'} \Big] +  \frac{1}{4} \Big[ \tau_{h \ell} \Big] \\
\tau_{h \ell  } = 1 + \frac{1}{2n} \Big[ (n-1) \tau_{g \ell'} + \tau_{g \ell} \Big] +  \frac{1}{2} \Big[ \tau_{h g} \Big] \\
\tau_{\ell \ell'  } = 1 + \frac{2}{2 } \Big[   \tau_{g \ell'} \Big]    \\
\tau_{g g' } = 1 + \frac{2}{4} \Big[  \tau_{g \ell'} + \tau_{h g} \Big]  \\
\tau_{g \ell } = 1 + \frac{1}{4 } \Big[  \tau_{h \ell} \Big]  \\
\tau_{g \ell' } = 1 + \frac{1}{4} \Big[  \tau_{h \ell}  + \tau_{\ell \ell'} \Big] +  \frac{1}{2} \Big[ \tau_{g g'} \Big] .
  \end{cases}
  }{extended_star_system}
Solving this system and plugging the results into~\eqref{taux}, we get

 \all{
  \begin{cases}
  \tau_{h} =  \fracc{64 n (3n-1)}{12n^2 + 35 n + 1}  \\ \\
\tau_{g } =  \fracc{8 n (17n-1)}{12n^2 + 35 n + 1}  \\ \\
\tau_{\ell   } =  \fracc{16 n (3+5n)}{12n^2 + 35 n + 1}    
  \end{cases}
}{extended_star_taux}

It is also straightforward to see that $p_h=p_\ell=\frac{1}{2}$, and ${p_g=\frac{n+1}{2n}}$. Inserting these values into Equation~\eqref{Gamma}, we get: 
\all{
b^*= 
\fracc{56 n^2 - 39 n -1} {22 n^2 - 20 n -2}
}{extended_Star_Gamma}

Which means that in the large $n$ limit, $b^*$ approaches $28/11 $.

\clearpage
\subsection{Imperfect extended star graph}
We consider the previous setup, but instead of attaching   a new node to every node that is adjacent to hub, we only do so for $n_g$ of them. 
So, the  graph comprises a hub, $n_g$ nodes with degree 2 that are connected to a hub and to a leaf node, and ${n-n_g}$ nodes with degree 1 that are connected directly to the hub. 
An example graph with $n=10$ and $n_g=3$ is depicted in Figure~\ref{fig_imperfect_extended_star}.
 We denote the leafs connected to degree-2 nodes with $d$ and  we denote the leafs directly connected to the hub by $p$. 
 The quantities of interest are 
 $\tau_{p p'}$ (between two $p$ nodes), 
 $\tau_{ ph}$, 
 $\tau_{ pg}$, 
 $\tau_{p d}$, 
 $\tau_{h g}$, 
 $\tau_{h d}$, 
 $\tau_{g g'} $ (between two $g$ nodes), 
 $\tau_{g d}$ (between a $g$ node and its adjacent $d$-node), 
 $\tau_{g d'}$ (between a $g$ node and a $d$-node adjacent to another $g$ node), 
 $\tau_{d d'}$ (between two $d$ nodes attached to the same $g$-node), 
 and $\tau_{ d d''}$ (between two $d$-nodes adjacent to two distinct $g$ nodes). 
 The system of equations to solve for the remeeting times is the following:

 \begin{figure}[h]
	\centering
	\includegraphics[width=.4\linewidth]{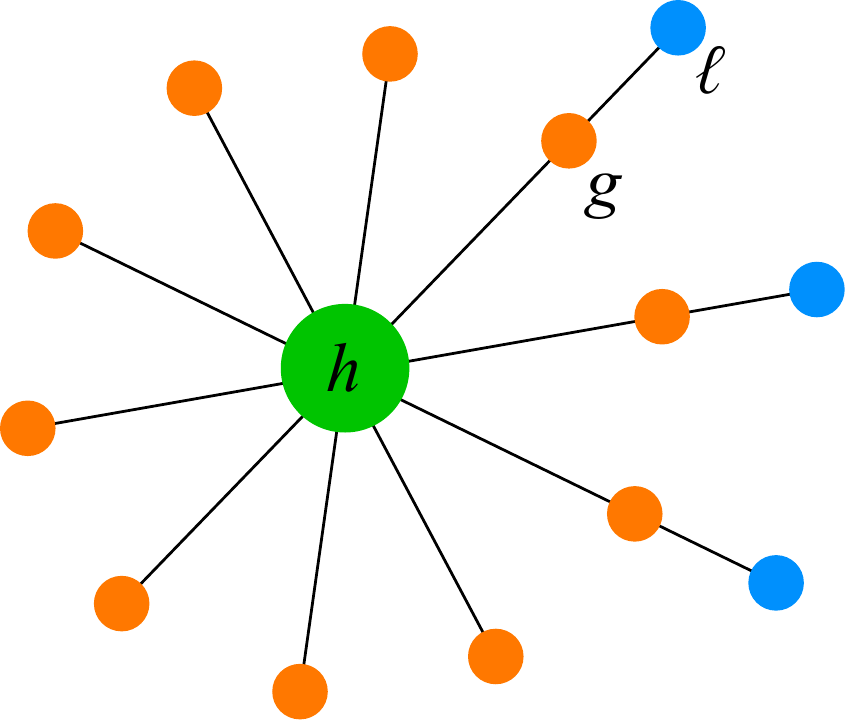}
	\caption{An example imperfect  extended star graph  with $n=10$ and $n_g=3$ }
	\label{fig_imperfect_extended_star}
\end{figure}

\all{
\begin{cases}
\tau_{p p'} =  1 +  \frac{2}{2} \tau_{ph} \\ 
\tau_{ph} + 1+ \frac{1}{2n} \Big[ (n-n_g-1) \tau_{p p'}  + n_g \tau_{pg} \Big] \\
\tau_{p g}= 1 + \frac{1}{2}  \Big[ \tau_{hg} + \frac{1}{4} \Bigg[ \tau_{pd} + \tau_{ph} \Big] \\
\tau_{p d} = 1 + \frac{1}{2} \tau_{hd} + \frac{1}{2} \tau_{pg} \\
\tau_{hg} = 1 + \frac{1}{2n}  \Big[ (n_g-1) \tau_{g g'} + (n-n_g) \tau_{pg} \Big] + \frac{1}{4} \tau_{hd} \\
\tau_{hd}= 1+ \frac{1}{2n} \Big[ (n_g-1) \tau_{g d'} + \tau_{gd} + (n-n_g) \tau_{pd} \Big] + \frac{1}{2} \tau_{hg} \\
\tau_{gg'}= 1 + \frac{1}{2} \Big[ \tau_{hg} + \tau_{g d'} \Big] 
\tau_{g d'} = 1 + \frac{1}{4} \Big[ \tau_{hd} + \tau_{dd'} \Big] + \frac{1}{2} \tau_{gg'} \\ 
\tau_{g d} = 1 + \frac{1}{4} \tau_{hd} \\
\tau_{d d'}= 1 + \tau_{g d'}. 
\end{cases}
}{imperfect_extended_star_system}

\all{
\resizebox{\linewidth}{!}{$
\begin{cases}
\tau_h &=  \fracc{4 \bigg[136 n^4+8 n^3 (35 n_g+29)+n^2 (n_g (152 n_g-139)+7)+4 n n_g \Big(2 n_g (n_g+6)-27 \Big)+n_g^2 (3 n_g-11)\bigg] }{n \bigg[136 n^3+8 n^2 (n_g+46)+n (129 n_g+239)+n_g (7 n_g+18)+7\bigg] } \\ \\
\tau_g &= \fracc{2 \bigg[428 n^3+3 n^2 (97 n_g+111)+n \ (92 n_g^2+90 n_g-26 ) +n_g \Big(5 n_g (n_g+1)-2 \Big)\bigg] }{136 n^3+8 n^2 (n_g+46)+n (129 n_g+239)+n_g (7 n_g+18)+7}\\ \\
\tau_d &= \fracc{4 \bigg[5 (4 n+1) n_g^2+(n (71 n+74)+10) n_g+n \Big(n (148 n+205)+74 \Big)+n_g^3\bigg] }{136 n^3+8 n^2 (n_g+46)+n (129 n_g+239)+n_g (7 n_g+18)+7}\\ \\
\tau_p &= \fracc{4 \bigg[136 n^3+8 n^2 (17 n_g+29)+n \Big(n_g (68 n_g-35)+7 \Big)+(n_g-1) n_g (4 n_g+1)\bigg] }{136 n^3+8 n^2 (n_g+46)+n (129 n_g+239)+n_g (7 n_g+18)+7} 
\end{cases}
$}
}{b}

Using these values, we arrive at
\all{
\resizebox{\linewidth}{!}{$
b^*= 
 \fracc{  136 n^5 + n^4 (712 n_g + 96) + n^3 (438 n_g^2 - 365 n_g - 225) + n^2 (56 n_g^3 + 108 n_g^2 - 325 n_g - 7) + n (2 n_g^4 + 9 n_g^3 - 20 n_g^2 - 7 n_g)}
 {n^4 (356 n_g) + n^3 (185 n_g^2 - 269 n_g) + n^2 (-12 n_g^3 + 141 n_g^2 - 445 n_g) + n (- n_g^4 - 41 n_g^3 + 106 n_g^2 - 28 n_g ) + (11 n_g^3 - 3 n_g^4)  }
$}
}{imperfect_star_of_stars_Gamma}

A more compact way to represent the result is with two matrices for the polynomial coefficients of the numerator and the denominator. 
For the numerator, the $i-j$ element of the following matrix yields the coefficient of $n^{i-1} n_g^{j-1}$ in the numerator:  
\all{
\left[
\begin{array}{ccccc}
 0 & 0 & 0 & 0 & 0 \\
 0 & -7 & -20 & 9 & 2 \\
 -7 & -325 & 108 & 56 & 0 \\
 -225 & -365 & 438 & 0 & 0 \\
 96 & 712 & 0 & 0 & 0 \\
 136 & 0 & 0 & 0 & 0 \\
\end{array} 
\right]
}{numer1}
and for the denominator we have: 
\all{
\left[
\begin{array}{ccccc}
 0 & 0 & 0 & 11 & -3 \\
 0 & -28 & 106 & -41 & -1 \\
 0 & -445 & 141 & -12 & 0 \\
 0 & -269 & 185 & 0 & 0 \\
 0 & 356 & 0 & 0 & 0 \\
\end{array}
\right]
}{denom1}

For $n_g \ll n$, we have: 
\al{
b^*=
\frac{34  }{89 n_g} n + \frac{28539 n_g+8845}{15842 n_g} + O\left(\fracc{1}{n}\right)
\approx  \fracc{0.38}{n_g} n + \left( 1.80+\frac{0.56}{n_g} \right) +  O\left(\fracc{1}{n}\right)
.}

For example, if there is only one gate node, that is, $n_g=1$, we have
\al{
b^* \bigg|_{n_g=1}
= \fracc{34}{89} n + \fracc{18692}{7921} + O\left(\fracc{1}{n}\right)
\approx 
0.38 n + 2.36 + + O\left(\fracc{1}{n}\right)
.}

\clearpage
\subsection{Imperfect star of stars}\label{SI:imperfect}
Consider the previous setup, but instead of attaching one   $d$-node to each  $g$-node, we attach  $n_d$ of them to each $g$-node. 
So,  the number of $g$-nodes is still $n_g$, 
but the number of $d$-nodes is now ${n_g \times n_d}$.

An example graph with $n=10$,  $n_g=3$ , and $n_d=20$ is depicted in Figure~\ref{fig_imperfect_star_of_stars}.
 The quantities of interest are 
 $\tau_{p p'}$ (between two $p$ nodes), 
 $\tau_{ ph}$, 
 $\tau_{ pg}$, 
 $\tau_{p d}$, 
 $\tau_{h g}$, 
 $\tau_{h d}$, 
 $\tau_{g g'} $ (between two $g$ nodes), 
 $\tau_{g d}$ (between a $g$ node and its adjacent $d$-node), 
 $\tau_{g d'}$ (between a $g$ node and a $d$-node adjacent to another $g$ node), 
 $\tau_{d d'}$ (between two $d$ nodes attached to the same $g$-node), 
 and $\tau_{ d d''}$ (between two $d$-nodes adjacent to two distinct $g$ nodes). 
 The system of equations to solve for the remeeting times is the following:

 \begin{figure}[h]
	\centering
	\includegraphics[width=.4\linewidth]{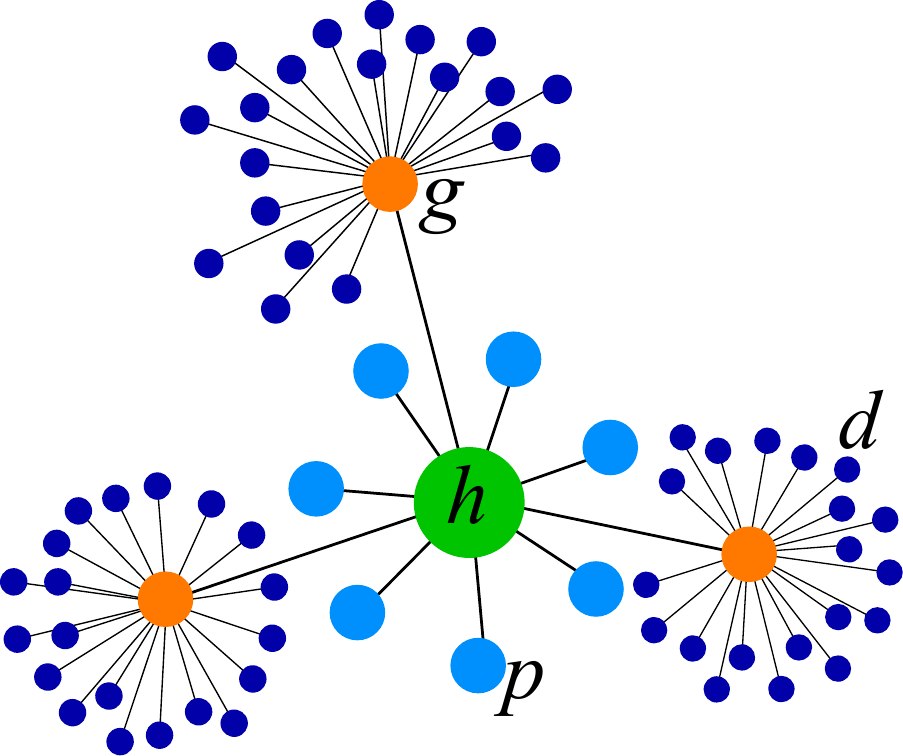}
	\caption{An example of imperfect star of stars,  with $n=10$,  $n_g=3$ , and $n_d=20$ }
	\label{fig_imperfect_star_of_stars}
\end{figure}

\all{
\begin{cases}
\tau_{p p'} =  1 +  \frac{2}{2} \tau_{ph} \\ 
\tau_{ph} + 1+ \frac{1}{2n} \Big[ (n-n_g-1) \tau_{p p'}  + n_g \tau_{pg} \Big] \\
\tau_{p g}= 1 + \frac{1}{2}  \Big[ \tau_{hg} + \frac{1}{2 (n_d+1)} \Bigg[n_d  \tau_{pd} + \tau_{ph} \Big] \\
\tau_{p d} = 1 + \frac{1}{2} \tau_{hd} + \frac{1}{2} \tau_{pg} \\
\tau_{hg} = 1 + \frac{1}{2n}  \Big[ (n_g-1) \tau_{g g'} + (n-n_g) \tau_{pg} \Big] + \frac{1}{2(n_d+1)} n_d \tau_{hd} \\
\tau_{hd}= 1+ \frac{1}{2n} \Big[ (n_g-1) \tau_{g d'} + \tau_{gd} + (n-n_g) \tau_{pd} \Big] + \frac{1}{2} \tau_{hg} \\
\tau_{gg'}= 1 + \frac{1}{n_d+1} \Big[ \tau_{hg} + n_d  \tau_{g d'} \Big] 
\tau_{g d'} = 1 + \frac{1}{2(n_d+1)} \Big[ \tau_{hd} + n_d \tau_{dd''} \Big] + \frac{1}{2} \tau_{gg'} \\ 
\tau_{g d} = 1 + \frac{1}{2(n_d+1)} \Big[ \tau_{hd} + (n_d-1) \tau_{dd'} \Big] \\
\tau_{d d'}= 1 + \tau_{g d}   \\
\tau_{d d''}= 1 + \tau_{g d'}
\end{cases}
}{imperfect_extended_star_system}

The expressions for the solution to this system are long, so we only present the final solution after inserting into Equation~\eqref{Gamma}. 
We have: 
\al{
b^*= \frac{\alpha}{\beta}, 
}
where the numerator is given by: 
\all{
&\alpha=  n^5 \bigg[  16 n_d^3  + 82 n_d^2 + 120 n_d + 54 \bigg]     \nonumber \\ & +
n^4 \bigg[  n_g(112 n_d^4 + 460 n_d^3 + 600 n_d^2 + 252 n_d)  + 16 n_d^4 + 74 n_d^3 + 92 n_d^2 + 22 n_d - 12  \bigg]     \nonumber \\ & +
\resizebox{.95\linewidth}{!}{$
n^3 \bigg[ n_g^2 (48 n_d^5+256 n_d^4+390 n_d^3+182 n_d^2)  +n_g  (-32 n_d^5-138 n_d^4-256 n_d^3-227 n_d^2-77 n_d)  + -16 n_d^4-90 n_d^3-171 n_d^2-135 n_d-38   \bigg]  $}
  \nonumber \\ & +
\resizebox{.95\linewidth}{!}{$
n^2 \bigg[ n_g^3 (22 n_d^5+56 n_d^4+34 n_d^3) + n_g^2 (28 n_d^5+80 n_d^4+80 n_d^3+28 n_d^2) + n_g (-16 n_d^5 - 108 n_d^4 - 238 n_d^3 - 217 n_d^2 - 71 n_d)   -3 n_d^2-7 n_d-4    \bigg]   $}  \nonumber \\ & +
\resizebox{.95\linewidth}{!}{$
n  \bigg[  n_g^4 (2 n_d^5+2 n_d^4) + n_g^3 (4 n_d^5+9 n_d^4+5 n_d^3) + n_g^2 (-2 n_d^5 - 11 n_d^4 - 18 n_d^3 - 9 n_d^2) + n_g (-3 n_d^3 - 7 n_d^2 - 4 n_d)     \bigg]   
 $}
 .}{numer1}

 \all{
 & \beta = 
  n ^4  \Big[  n_g \left(64  n_d ^4+256  n_d ^3+296  n_d ^2+96  n_d  \right)    \Big] \nonumber \\ &
  + n ^3\Big[n_g ^2\left(32  n_d ^5+148  n_d ^4+162  n_d ^3+28  n_d ^2 \right)  + n_g \left(-32  n_d ^5-148  n_d ^4-226  n_d ^3-124  n_d ^2-8  n_d  \right)     \Big]
   \nonumber \\ &+
 \resizebox{.95\linewidth}{!}{$
    n ^2  \Big[ n_g^3 \left(4  n_d ^5-6  n_d ^4-22  n_d ^3 \right)   +  n_g^2 \left(28  n_d ^5+98  n_d ^4+108  n_d ^3+48  n_d ^2 \right)   +  n_g \left(-32  n_d ^5-180  n_d ^4-342  n_d ^3-264  n_d ^2-72  n_d  \right)    \Big]
    $}
     \nonumber \\ &
  \resizebox{.95\linewidth}{!}{$
     + n   \Big[ n_g^4 \left(-2  n_d ^4 \right)   +  n_g^3 \left(-12  n_d ^5-42  n_d ^4-28  n_d ^3 \right)  +  n_g^2 \left(12  n_d ^5+60  n_d ^4+96  n_d ^3+44  n_d ^2 \right)  + n_g \left(-12  n_d ^3-28  n_d ^2-16  n_d  \right)      \Big]
     $}
     \nonumber \\ &
     + 
     n_g ^4 \left(-2  n_d ^5-4  n_d ^4 \right)+ n_g ^3 \left(2  n_d ^5+12  n_d ^4+8  n_d ^3 \right)  
 }{denom1}

 Suppose there are many leafs but very few gates: $n_g \ll n_d=\mu n $, where $\mu$ is of $O(1)$. That is, $n$ and $n_d$ are  of the same order of magnitude and are both much greater than $n_g$. We can use the following expansion: 
 \al{
 b^*=  
 \frac{\mu +\mu ^2 n_g (3 n_g-2)+7 \mu  n_g+1}{2 \mu  n_g (\mu  (n_g-1)+2)}
 + O\left(\frac{n_g}{n} \right)
 }
 Maintaining the previous regime, we can expand this in $n_g$ as follows: 
 \al{
 b^* = 
 \frac{3}{2} + \left(\frac{1}{2} +  \frac{1 }{2 \mu}  \right)  \frac{1}{n_g} + O\left(\frac{1}{n_g^2}\right) + O\left(\frac{n_g}{n} \right).
 }

Since the average degree approaches 2 in the regime considered above, the graph is a significant promoter of cooperation because $b^*$ (which approaches $3/2$) is smaller than the average degree. 

In the  regime $n_g \ll n \ll n_d$, too, we find that $b^*$ is smaller than average degree. In this regime, we have: 
\al{
b^*= \frac{3}{2} + \frac{1}{2(n_g-1)} + O\left( \frac{n_g}{n}\right) + O\left( \frac{n }{n_d}\right) 
 }

\clearpage
\subsection{Star of stars}\label{SI:perfect}
If we take the solution of the last section and set $n_g$ and $n$ equal, the resulting graph is a star of stars,  that is, 
a hub that is connected to $n$ nodes, and each of these $n$ nodes is a hub to a star of $n_d$ nodes. 
The critical benefit to cost ratio simplifies to the following: 
\all{
b^*= 
\fracc{n^2 \Big( 12 n_d^3 + 47 n_d^2 + 44 n_d + 9 \Big) - n \Big( 6 n_d^3 + 31 n_d^2 + 33 n_d + 8\Big) - \big(n_d+1 \big) }
{2 n_d (n-1)  \Big[2 + n (3n_d+8)(n_d+1)  \Big] }
}{star_of_stars_Gamma}

For $n \ll n_d$, we can use the following expansion

\al{
b^*= \left(2+ \frac{1}{n-1} \right)+
\left( \frac{1}{1-n}+\frac{1}{2}\right)\frac{1}{n_d} + O\left(\frac{1}{n_d^2}\right)
.}

So $b^*$ approaches ${2+1/(n-1)}$  as $n_d$ grows. 
The average degree, on the other hand, approaches 2 from above as $n_d$ grows. 
So $b^*$ never becomes smaller than the average degree.

\clearpage

\subsection{Three-layer extended star}

\begin{figure}[h]
	\centering
	\includegraphics[width=.4\linewidth]{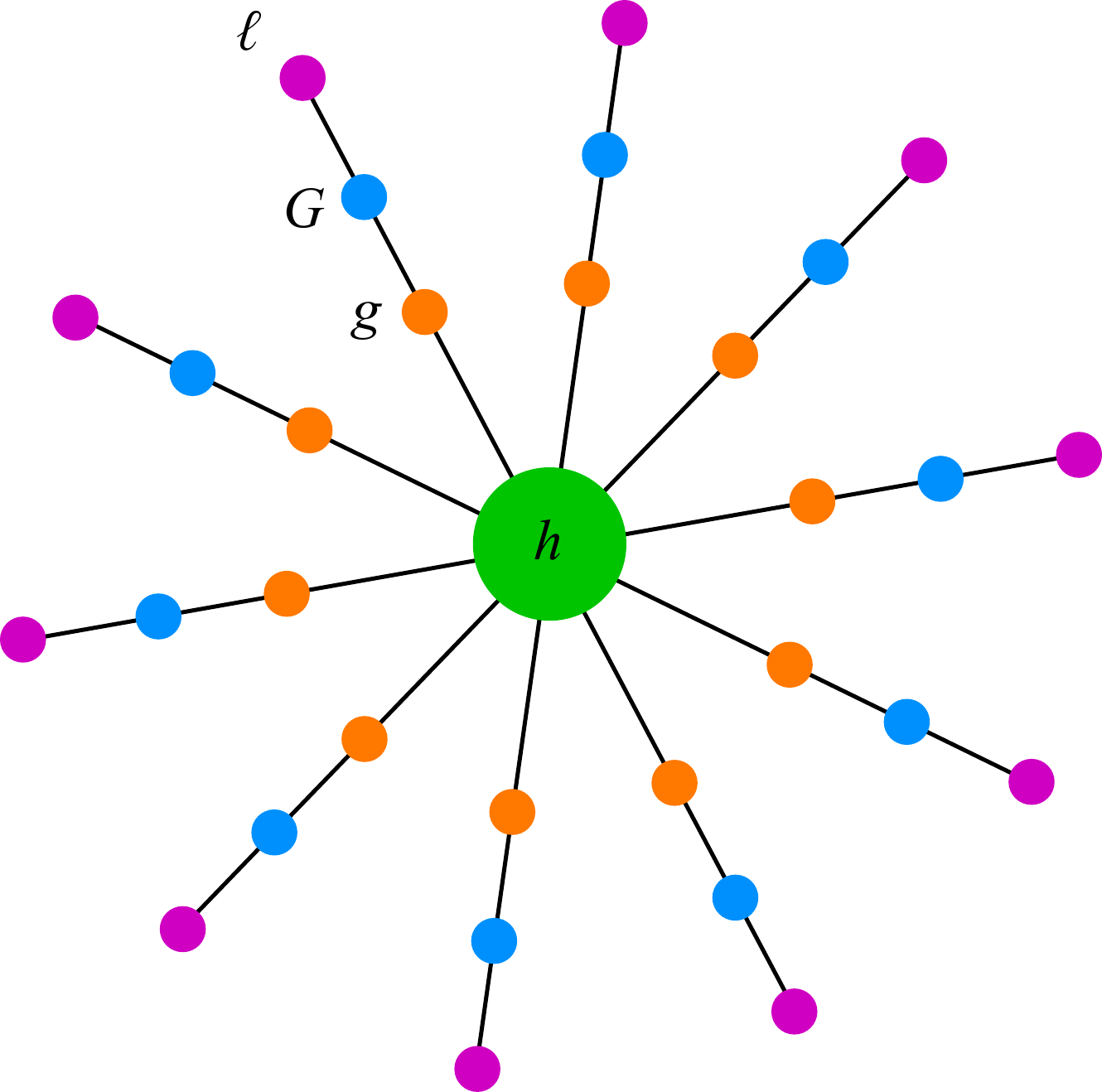}
	\caption{An example three-layer extended star graph  with $n=10$  }
	\label{fig:3layer}
\end{figure}

Consider the extended star, but with three layers instead of two. 
 There is one hub $h$, attached to $n$ nodes $g$ in the first layer, there are $n$ nodes $G$ in the second layer each connected to one $g$ nodes,  and there are $n$ nodes $\ell$ in the third layer each attached to a $G$ node. 
 Figure~\ref{fig:3layer} depicts an example case with $n=10$. 
 The remeeting times of interest are: 
 $\tau_{ \ell \ell' }$ (between two $\ell$ nodes),
 $\tau_{G \ell  }$ (between a $G$ node and its adjacent $\ell$ node),
 $\tau_{ G' \ell }$ (between a $G$ node and an $\ell$ node not adjacent to it),
 $\tau_{ g \ell }$ (between a $g$ node and an $\ell$ node on the same spoke),
 $\tau_{ g' \ell }$ (between a $g$ node and an $ \ell$ node on another spoke),
 $\tau_{ h \ell }$ (between the hub and an $\ell$ node),
 $\tau_{ g G }$ (between a $g$ node and its adjacent $G$ node),
 $\tau_{ g g' }$ (between two distinct $g$ nodes),
 $\tau_{ g h }$ (between the hub and a $g$ node),
 $\tau_{ G G' }$ (between two distinct $G$ nodes),
 $\tau_{ G g' }$ (between a $G$ node and a $g$ node on another spoke), and 
 $\tau_{  G h}$ (between the hub and a $g$ node).

\begin{equation}
\begin{cases}
\tau_{\ell\ell'}=1+\tau_{G'\ell} \\
\tau_{G\ell}= 1+\frac{1}{4}\tau_{g\ell} \\
\tau_{G'\ell}=1+\frac{1}{4}\big[ \tau_{\ell\ell'}+\tau_{g'\ell}\big] +\frac{1}{2} \tau_{GG'}  \\
\tau_{g\ell}=1+\frac{1}{4}\big[ \tau_{h\ell}+\tau_{G\ell}\big] +\frac{1}{2} \tau_{gG} \\
\tau_{g'\ell}=1+\frac{1}{4}\big[ \tau_{G'\ell}+\tau_{h\ell}\big] +\frac{1}{2} \tau_{Gg'}\\
\tau_{h\ell}=1+\frac{1}{2n}\big[ \tau_{g\ell}+(n-1)\tau_{g'\ell}\big] +\frac{1}{2} \tau_{Gh}\\
\tau_{gG}=1+\frac{1}{4}\big[ \tau_{Gh}+\tau_{g\ell}\big]\\
\tau_{gg'}=1+\frac{1}{2}\big[ \tau_{Gg'}+\tau_{gh}\big]\\
\tau_{gh}=1+\frac{1}{2n}\big[ (n-1)\tau_{gg'}\big] +\frac{1}{4} \tau_{Gh}\\
\tau_{GG'}=1+\frac{1}{2}\big[ \tau_{G'\ell}+\tau_{Gg'}\big]\\
\tau_{Gg'}=1+\frac{1}{4}\big[ \tau_{g'\ell}+\tau_{gg'}+\tau_{GG'}+\tau_{gG}\big]\\
\tau_{Gh}=1+\frac{1}{2n}\big[\tau_{gG}+(n-1)\tau_{Gg'}\big] +\frac{1}{4} \big[ \tau_{\ell h}+\tau_{gh}\big]
\end{cases}
\end{equation}

The solution leads us to the following result for the critical benefit-to-cost ratio:

\begin{equation}
b^*=\frac{106940n^3-52194 n^2-4820 n+6}{41723 n^3-18453 n^2-10790n}
\end{equation}

 For large $n$, we can use the following expansion: 
 
 \al{
 b^*=
 \frac{106940}{41723}-\frac{204326442}{1740808729 n} + O\left(\frac{1}{n^2}\right)
 \approx 
 2.56 - \frac{0.12}{n}  + O\left(\frac{1}{n^2}\right)
 }

\clearpage

\subsection{Two stars: hub-to-hub connection}
If we have two star graphs, one with $n_1$ leafs and the other with $n_2$ leafs, we can connect these two graphs in several different ways to restore the faith of cooperation (which is not favored by natural selection in either of the star graphs alone, as shown in Section~\ref{sec:star} above). 

First suppose that the hub of the first star is connected to the hub of the other via a link.  
There are  8  distinct remeeting times to consider: 
$\tau_{\ell_1 \ell_1'} $(between two distinct leafs of the first star), 
$\tau_{\ell_1 h_1}$ (between a leaf and the hub of the first star), 
$\tau_{\ell_1 h_2}$ (between a leaf of the first star and the hub of the second star), 
$\tau_{\ell_1 \ell_2}$ (between a leaf in  one star and a leaf in the other), 
$\tau_{h_1 h_2}$ (between the hubs), 
$\tau_{h_1 \ell_2}$ (between the hub of the first star and a leaf in the second star), 
$\tau_{h_2 \ell_2}$ (between a leaf of the second star and its hub), 
and $\tau_{\ell_2 \ell_2'}$ (between two leafs of the second star). 
The remeeting times satisfy the following system of equations:

\all{
\begin{cases}
 \tau_{\ell_1 \ell_1'} & =1+  \tau_{\ell_1 h_1}                       \\  
 \tau_{\ell_1 h_1} & =1+  \frac{1}{2(n_1+1)} \bigg[ \tau_{\ell_1 h_2} + (n_1-1) \tau_{\ell_1 \ell1'} \bigg]                      \\ 
 \tau_{\ell_1 h_2} & =1+ \frac{1}{2} \tau_{h_1 h_2} + \fracc{1}{2(n_2+1)} \bigg[ n_2 \tau_{\ell_1 \ell_2} + \tau_{\ell_1 h_1} \bigg]                         \\ 
 \tau_{\ell_1 \ell_2}   & =1+    \frac{1}{2}  \tau_{h_1 \ell_2} + \frac{1}{2} \tau_{\ell_1 h_2}                \\  
 \tau_{h_1 h_2}   & =1+    \fracc{1}{2(n_1+1)} \big[ n_1 \tau_{\ell_1 h_2}  \big]     + \fracc{1}{2(n_2+1)} \big[ n_2 \tau_{h_1 \ell_2}   \big]                                 \\ 
 \tau_{h_1 \ell_2}  & =1+       \fracc{1}{2(n_1+1)} \bigg[ n_1 \tau_{\ell_1 \ell_2} + \tau_{h_2 \ell_2} \bigg]     + \frac{1}{2} \tau_{h_1 h_2}                          \\ 
 \tau_{h_2 \ell_2}  & =1+    \frac{1}{2(n_2+1)}   \bigg[  \tau_{h_1 \ell_2} + (n_2-1) \tau_{\ell_2 \ell_2'}      \bigg]             \\ 
  \tau_{\ell_2 \ell_2'} & =1+    \tau_{h_2 \ell_2}                       
  .
  \end{cases}
}{system_doubleStar}

Solving this and inserting into Equation~\eqref{taux} and then plugging the results into Equation~\eqref{Gamma}, we arrive at the solution:
\al{
b^*= \frac{\alpha}{\beta}  
,}
where the numerator is 
\all{
\alpha &= 
   n_1 ^5  \Big(8   n_2 ^3+41   n_2 ^2+60   n_2 +27\Big) \nonumber \\ &
   +  n_1 ^4  \Big(64   n_2 ^4+307   n_2 ^3+551   n_2 ^2+437   n_2 +129 \Big)
    \nonumber \\ &
    +  n_1 ^3  \Big(8   n_2 ^5+307   n_2 ^4+1170   n_2 ^3+1686   n_2 ^2+1042   n_2 +227  \Big)
     \nonumber \\ &
     +  n_1 ^2  \Big(41   n_2 ^5+551   n_2 ^4+ 1686   n_2 ^3+2066   n_2 ^2+1065   n_2 +175 \Big)
      \nonumber \\ &
      +  n_1   \Big(60   n_2 ^5+437   n_2 ^4+1042   n_2 ^3+1065   n_2 ^2+450   n_2 +50  \Big)  \nonumber \\ &
      + \Big(27   n_2 ^5+129   n_2 ^4+227   n_2 ^3+175   n_2 ^2+50   n_2 \Big)
,}{double_star_1_numer}
and the denominator is
\all{
\beta &= 
   n_1^4 \left(32  n_2^4+128  n_2^3+148  n_2^2+48  n_2 \right)
    \nonumber \\ &
    + n_1^3 \left(128  n_2^4+480  n_2^3+544  n_2^2+188  n_2 \right)
     \nonumber \\ &
     + n_1^2 \left(148  n_2^4+544  n_2^3+636  n_2^2+240  n_2 \right)
      \nonumber \\ &
      + n_1 \left(48  n_2^4+188  n_2^3+240  n_2^2+100  n_2  \right) 
.}{double_star_1_denom}

If the two stars are identical, with $n_1=n_2=n$, then the expression simplifies to: 
\all{
b^*= \fracc{(n+1)^2  (10 n^2 + 17 n + 5) }{4 n^4 + 12 n^3 + 11 n^2 + 5 n}
.}{double_star_1_identical}

When $n$ is large, this converges to $10/4$. With first-order correction in the large-$n$ limit, we can write 
\all{
b^*= \frac{5}{2} + \frac{7}{4n} + O\left(\frac{1}{n^2} \right)
.}{2onim}

If the stars are not identical, but both are very large, such that $n_1=n$ and $n_2= \alpha  n$, the in the limit as ${n \rightarrow \infty}$, we have: 
\all{
b^*= \left( 2 + \fracc{1}{4\alpha}+ \fracc{\alpha}{4} \right) + \fracc{(\alpha+1) (9 \alpha^2 + 10 \alpha + 9)}{32 \alpha^2 n } + O\left(\frac{1}{n^2} \right)
}{double_star_1_lambda}

\clearpage

\subsection{Two stars: hub-to-leaf connection}
I
In the previous scenario, if instead of connecting the hubs, we connect the hub of the first star to a leaf in the second star, 
the equations for the remeeting times change. 
 We denote the leaf that is connected to the hub of the first star by $g$. 
There are 12  distinct remeeting times to consider: 
$\tau_{\ell_1 \ell_1'} $,
$\tau_{\ell_1 h_1}$ ,
$\tau_{\ell_1 h_2}$,
$\tau_{\ell_1 \ell_2}$,
$\tau_{h_1 h_2}$,
$\tau_{h_1 \ell_2}$,
$\tau_{h_2 \ell_2}$, 
  $\tau_{\ell_2 \ell_2'}$,
  $\tau_{\ell_1 g}$, 
  $\tau_{h_1 g}$, 
  $\tau_{g h_2}$, 
  and   $\tau_{g \ell_2} $. 
  Without loss or generality,  we  consider the case where the second star  has size $n_2+1$. 
  Equivalently, we can assume the stars have sizes $n_1$ and $n_2$, and the hubs are being connected via one intermediary node. 
  This is merely for aesthetic reasons: with  this change of notation, the analysis will be manifest-symmetric in $n_1 $ and $n_2$. 
The remeeting times satisfy the following system of equations: 

\all{
\begin{cases}
 \tau_{\ell_1 \ell_1'} & =1+  \tau_{\ell_1 h_1}                       \\  
 \tau_{\ell_1 h_1} & =1+  \frac{1}{2(n_1+1)} \bigg[ \tau_{\ell_1g} + (n_1-1) \tau_{\ell_1 \ell1'} \bigg]                      \\ 
 \tau_{\ell_1 h_2} & =1+ \frac{1}{2} \tau_{h_1 h_2} + \fracc{1}{2(n_2+1)} \bigg[ n_2 \tau_{\ell_1 \ell_2} + \tau_{\ell_1 g} \bigg]                         \\ 
 \tau_{\ell_1 \ell_2}   & =1+    \frac{1}{2}  \tau_{h_1 \ell_2} + \frac{1}{2} \tau_{\ell_1 h_2}                \\  
 \tau_{h_1 h_2}   & =1+    \fracc{1}{2(n_1+1)} \big[ n_1 \tau_{\ell_1 h_2} + \tau_{g h_2}   \big]     + \fracc{1}{2(n_2+1)} \big[ n_2 \tau_{h_1 \ell_2}  + \tau_{h_1 g} \big]                                 \\ 
 \tau_{h_1 \ell_2}  & =1+       \fracc{1}{2(n_1+1)} \bigg[ n_1 \tau_{\ell_1 \ell_2} + \tau_{g  \ell_2} \bigg]     + \frac{1}{2} \tau_{h_1 h_2}                          \\ 
 \tau_{h_2 \ell_2}  & =1+    \frac{1}{2(n_2+1)}   \bigg[  \tau_{g \ell_2} + (n_2-1) \tau_{\ell_2 \ell_2'}      \bigg]             \\ 
  \tau_{\ell_2 \ell_2'} & =1+    \tau_{h_2 \ell_2}        \\
   \tau_{\ell_1 g}   & =1+ \frac{1}{2} \tau_{h_1 g} + \frac{1}{4} \bigg[ \tau_{\ell_1 h_1} + \tau_{\ell_1 h_2}    \bigg]                                                           \\
   \tau_{h_1 g}   & =1+  \fracc{1}{2(n_1+1)} n_1 \tau_{\ell_1 g} + \frac{1}{4} \tau_{h_1 h_2}                                                               \\
   \tau_{g h_2}   & =1+   \frac{1}{2(n_2+1)} n_2  \tau_{g \ell_2} + \frac{1}{4} \tau_{h_1 h_2}                                                             \\
  \tau_{g \ell_2}  & =1+   \frac{1}{2} \tau_{g h_2} + \frac{1}{4} \bigg[ \tau_{h_1 \ell_2}  + \tau_{h_2 \ell_2}  \bigg] 
  .
  \end{cases}
}{system_doubleStar_hubtoleaf}

Solving this and using Equation~\eqref{Gamma}, we arrive at

\al{
b^*= \frac{\alpha}{\beta}  
,}
where the numerator is 
\all{
\alpha &= 
  n_1 ^6  \Big( 288  n_2 ^4+1746  n_2 ^3+3738  n_2 ^2+3402  n_2 +1122 \Big) \nonumber \\ &
   n_1 ^5  \Big( 2880 n_2^5 + 19566 n_2^4 + 53836 n_2^3 + 74006 n_2^2 + 50424 n_2 + 13568 \Big) \nonumber \\ &
   +  n_1 ^4  \Big( 288 n_2^6 + 19566 n_2^5 + 113908 n_2^4 + 276152 n_2^3 + 338086 n_2^2 + 
 207330 n_2 + 50766  \Big)
    \nonumber \\ &
    +  n_1 ^3  \Big(1746 n_2^6 + 53836 n_2^5 + 276152 n_2^4 + 610544 n_2^3 + 687478 n_2^2 + 
 389244 n_2 + 88248 \Big)
     \nonumber \\ &
     +  n_1 ^2  \Big(3738 n_2^6 + 74006 n_2^5 + 338086 n_2^4 + 687478 n_2^3 + 716764 n_2^2 + 
 375760 n_2 + 78656  \Big)
      \nonumber \\ &
      +  n_1   \Big(3402 n_2^6 + 50424 n_2^5 + 207330 n_2^4 + 389244 n_2^3 + 375760 n_2^2 + 
 181328 n_2 + 34504   \Big)  \nonumber \\ &
      + \Big(1122 n_2^6 + 13568 n_2^5 + 50766 n_2^4 + 88248 n_2^3 + 78656 n_2^2 + 
 34504 n_2 + 577  \Big)
,}{double_star_1_numer}
and the denominator is
\all{
\beta &= 
   n_1^5 \left( 1152 n_2^5 + 7200 n_2^4 + 17472 n_2^3 + 20778 n_2^2 + 12285 n_2 + 2937  \right)
       \nonumber \\ &
   n_1^4 \left(   7200 n_2^5 + 43072 n_2^4 + 99734 n_2^3 + 112892 n_2^2 + 63433 n_2 + 14437   \right)
    \nonumber \\ &
    + n_1^3 \left(    17472 n_2^5 + 99734 n_2^4 + 218830 n_2^3 + 232766 n_2^2 + 
 121788 n_2 + 25666   \right)
     \nonumber \\ &
     + n_1^2 \left(    20778 n_2^5 + 112892 n_2^4 + 232766 n_2^3 + 228350 n_2^2 + 
 107262 n_2 + 19648     \right)
      \nonumber \\ &
      + n_1 \left(   12285 n_2^5 + 63433 n_2^4 + 121788 n_2^3 + 107262 n_2^2 + 42048 n_2 + 5472  \right) 
      \nonumber \\ &
      + 2937 n_2^5 + 14437 n_2^4 + 25666 n_2^3 + 19648 n_2^2 + 5472 n_2
.}{double_star_1_denom}

If the stars are identical, that is, if we have two stars each with $n$ leafs and then we connect their hubs through a gate node, then we have 
\all{
b^*= \fracc{(n+1) (36 n^2 + 90 n + 19)}{4n(3n^2+11n+9)} = 3 - \fracc{1}{2n} + O \left( \frac{1}{n^2} \right) 
}{identical_leafhub}
So as the size of the stars goes to infinity, the value of $b^*$ approaches 3. 

If the stars are not identical but both are large, with $n_1=n$ and $n_2= \lambda n$, then we have 

\all{
b^*= 
 \left( \fracc{5}{2}   + \fracc{1}{4\lambda}+ \fracc{\lambda}{4} \right) 
+ \fracc{(\lambda+1) (\lambda+3)(3\lambda+1)}{64 \lambda^2 n } + O\left(\frac{1}{n^2} \right)
.}{lambda_leafhub}

\clearpage

\subsection{Ring of stars:  a super-promoter of cooperation}\label{SI:star_ring}

\begin{figure}[h]
	\centering
	\includegraphics[width=.4\linewidth]{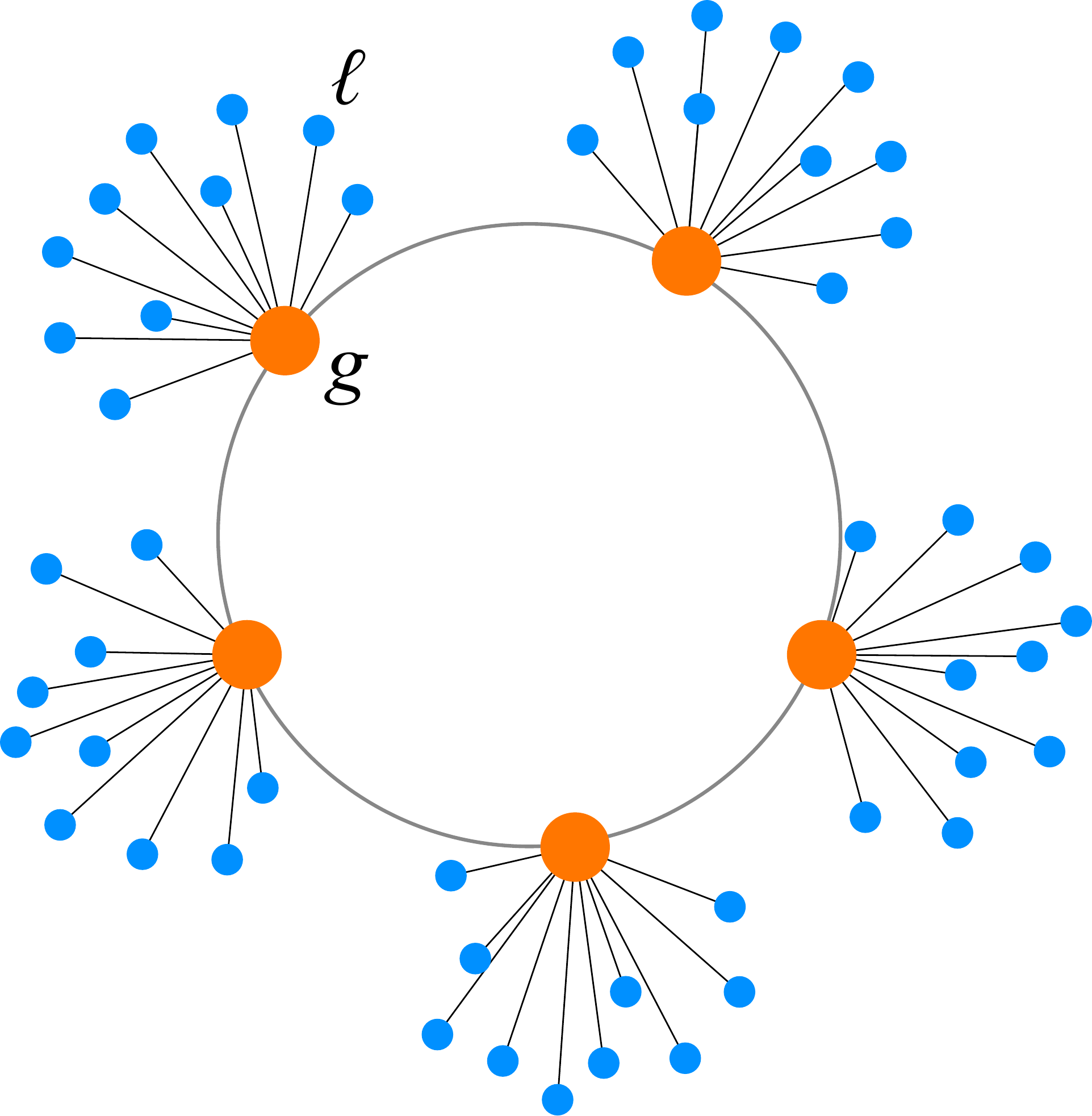}
	\caption{An example case of ring of stars  with $n=10$ and $L=5$}
	\label{fig:ring_star}
\end{figure}

Suppose there are $L$ stars, each with $n$ leafs. We situate these on a ring by connecting the hubs. 
An example with $n=10$ and $L=5$ is illustrated in Figure~\ref{fig:ring_star}. 
We define the following remeeting times: 
\begin{itemize}
\item $\tau_{h h}(x)$  is   between the hubs of two stars which are $x$ apart on the ring.  For example, two adjacent hubs have remeeting times $\tau_{hh}(1)$. 
\item $\tau_{h \ell}(x)$ is between a hub and a leaf of a star that is $x$ apart. For example, for the hub and leaf of the same star, we have $\tau_{h \ell}(0)$, and for the leaf of a star and the hub of the neighboring star, we have $\tau_{h \ell}(1)$, and so on. 
\item $\tau_{\ell \ell}(x)$ is between two distinct leafs, belonging to two stars $x$ apart. So $\tau_{\ell \ell}(0)$ is between two leafs of the same star, $\tau_{\ell \ell}(1)$ is between a leaf in one star and a leaf in a neighboring star. 
\end{itemize}
Note that $x$ varies between 0 and $L$.

The recurrence relations that we need to solve have the form of three-dimensional difference equation: 
\all{
\begin{cases}
\text{for }x=2\ldots L-2:~
&\begin{cases}
\tau_{h h} (x) = 1 +  \frac{1}{n+2} \bigg[\tau_{hh}(x-1) + \tau_{hh} (x+1) + n \tau_{h \ell}(x)   \bigg]    \\ \\
\tau_{h \ell}(x) = 1 + \fracc{1}{2(n+2)} \bigg[ n \tau_{\ell \ell} (x) + \tau_{h \ell}(x+1) + \tau_{ h \ell} (x-1) \bigg] + \frac{1}{2} \tau_{ hh} (x)    \\  \\
\tau_{\ell \ell}(x) = 1 + \tau_{h \ell(x) }  
\end{cases}
\\ \\
\text{for }x<2:~
&\begin{cases}
\tau_{h h} (1) = 1 +  \frac{1}{n+2} \bigg[ \tau_{hh} (2)  + n \tau_{h \ell}(1)  \bigg]    \\ \\
\tau_{h \ell}(1) = 1 + \fracc{1}{2(n+2)} \bigg[ n \tau_{\ell \ell} (1) + \tau_{h \ell}(2) + \tau_{ h \ell} (0) \bigg] + \frac{1}{2} \tau_{ hh} (1)    \\  \\
\tau_{h \ell}(0) = 1 + \fracc{1}{2(n+2)} \bigg[ (n-1) \tau_{\ell \ell} (0) +2  \tau_{h \ell}(1)  \bigg]     \\  \\
\tau_{\ell \ell}(0) = 1 + \tau_{h \ell }  (0)
\end{cases}
\end{cases}
.}{recur_ringstar}

This is a system of linear difference equations with constant coefficients. 
So the standard way is inserting the ansatz in the form of $r^x$ and solving the resulting characteristic equation for $r$. If there is no degenerate root, the solution will take the form ${\sum_i a_i r_i^x}$, and we will have to add this to a particular solution for the nonhomogenous system. We use the following ansatz: 
\all{
\begin{cases}
\tau_{hh}(x)=  \xi_{hh} r^x 
\\
\tau_{h\ell}(x)= \xi_{h\ell} r^x 
\\
\tau_{\ell \ell}(x)= \xi_{\ell \ell} r^x 
\end{cases}
~~ 1<x<L
}{ansatz1}
 This leads to the following characteristic equation: 
 \all{
 (r-1)^2 \bigg[ r^2 - 2 (n+2) r + 1 \bigg] = 0 
 \Longrightarrow 
 \begin{cases}
 r_1=0 \\
 r_2=0 \\
 r_3=n+2+\sqrt{(n+1)(n+3)} \\
 r_4=\fracc{1}{r_3}
 \end{cases}
 .}{roots}
 This has root zero with degeneracy two. This means that the particular solution to the nonhomogenous equation will be quadratic in $x$. 
 For brevity of notation, let us define: 
 \all{
 \begin{cases}
 \lambda \stackrel{\text{def}}{=} \sqrt{(n+1)(n+3)}, \\
 R \stackrel{\text{def}} = n+2+\sqrt{(n+1)(n+3)}.
 \end{cases}
 }{defs}
 
 Thus we plug in the following form for the solutions: 
 \all{
\begin{cases}
\tau_{hh}(x)= A_{hh} x^2 + B_{hh} x + C_{hh} + D_{hh} \big( R^x + R^{L-x} \big) 
\\
\tau_{h\ell}(x)=  A_{h\ell} x^2 + B_{h\ell} x + C_{h\ell} + D_{h\ell} \big( R^x + R^{L-x} \big) 
\\
\tau_{\ell \ell}(x)= A_{h\ell} x^2 + B_{h\ell} x + C_{h\ell} + D_{h\ell} \big( R^x + R^{L-x} \big) 
\end{cases}
}{forms1}

Plugging these into Equations~\eqref{recur_ringstar}, and setting the coefficients of different powers of $x$ and also those of $R^x$ and $R^{-x}$  identical to zero (because the equations hold for every $x$ and the functions are linearly independent),  and also using the requirement that $\tau_{hh}(x)$ must be equal to ${\tau_{hh}(L-x)}$ (and same for $\tau_{h\ell}$ and $\tau_{\ell\ell}$), we obtain: 

\all{
&A_{hh}=A_{\ell \ell} = A_{h \ell} = -(n+1), \nonumber \\ \nonumber \\
&B_{hh}=B_{\ell \ell} = B_{h \ell} = (n+1) L ,\nonumber \\ \nonumber\\
&C_{hh}= \fracc{n (L n+L-1) \left( R ^L+1\right)}{(n+1)  R ^L+n \lambda  R ^L+2 \lambda  R ^L+n-n \lambda-2 \lambda+1}, \nonumber\\  \nonumber\\
&\resizebox{\linewidth}{!}{$
C_{h \ell}= 
 \frac{2 n^4 \left((L+1)  R ^L+L-1\right)+2 n^3 \left(\left(\lambda+8\right) \left( R ^L-1\right)+L \left(\lambda+5\right) \left( R ^L+1\right)\right)+n^2 \left(3 L \left(2 \lambda+5\right) \left( R ^L+1\right)+4 \left(3 \lambda  R ^L+12  R ^L-3 \lambda-11\right)\right)+18 \lambda  R ^L+31  R ^L+n \left(25 \lambda  R ^L+64  R ^L+L \left(4 \lambda+7\right) \left( R ^L+1\right)-21 \lambda-48\right)-10 \lambda-17}{ 2 n^3 \lambda  R ^L+14 n^2 \lambda  R ^L+(n+1) (2 n (n (n+8)+20)+31)  R ^L+29 n \lambda  R ^L+18 \lambda  R ^L-2 n^4-14 n^3-2 n^3 \lambda-36 n^2-10 n^2 \lambda-41 n-17 n \lambda-10 \lambda-17}
$}
\nonumber\\ \nonumber \\
&D_{hh}= 
-\frac{n (L n+L-1)}{(n+1)  R ^L+n \lambda  R ^L+2 \lambda  R ^L+n-n \lambda-2 \lambda+1}
\nonumber\\ \nonumber\\
&D_{h \ell} = 
\frac{(n+2) (L n+L-1)}{(n+1)  R ^L+n \lambda  R ^L+2 \lambda  R ^L+n-n \lambda-2 \lambda+1} 
\nonumber\\ \nonumber\\
&\resizebox{\linewidth}{!}{$
\tau_{\ell \ell } (0) = \frac{2 \left(L \left(4 \lambda+2 n \left(n+\lambda+4\right)+7\right) (n+1)^2 \left( R ^L+1\right)+(n+2) \left(7 \lambda+2 n \left(4 \lambda+n \left(n+\lambda+6\right)+11\right)+12\right) \left( R ^L-1\right)\right)}{ 2 n^3 \lambda  R ^L+14 n^2 \lambda  R ^L+(n+1) (2 n (n (n+8)+20)+31)  R ^L+29 n \lambda  R ^L+18 \lambda  R ^L-2 n^4-14 n^3-2 n^3 \lambda-36 n^2-10 n^2 \lambda-41 n-17 n \lambda-10 \lambda-17}
$}
\nonumber\\ \nonumber\\
&\resizebox{\linewidth}{!}{$
\tau_{h \ell}(0)= 
\frac{-18 \lambda +(10 \lambda +n (17 \lambda +2 n (5 \lambda +n (R+5)+18)+41)) R^L+17 R^L+2 L (n+1)^2 (4 \lambda +2 n (\lambda +n+4)+7) \left(R^L+1\right)-n (29 \lambda +2 n (7 (\lambda +4)+n (\lambda +n+9))+71)-31}{-10 \lambda +(18 \lambda +n (29 \lambda +2 n (7 (\lambda +4)+n (\lambda +n+9))+71)) R^L+31 R^L-n (17 \lambda +2 n (5 \lambda +n (\lambda +n+7)+18)+41)-17}
$}
.
}{sols_ringStar}

Using these values, we arrive at the 
\al{
b^*
=\fracc{\alpha}{\beta}
,}
where  the numerator is
\all{
&
\alpha = 
(n+2)^2 
\Bigg[  L (n+1) \bigg(-2 (n (2 n+11)+13) R^L+(n (\lambda +2 n+9)+11)  R^{2 L}+2 n^2-\lambda  n+9 n+11\bigg)
\nonumber \\ &
+2 R^L( n^2 (n+9)+28n+26) - R^{2 L} \Big(n^2 (n+9)+ n  (\lambda+24) + 22\Big) -n^3-9 n^2+\lambda  n-24 n-22
\Bigg] 
,}{numerstarring}
and the denominator is: 
\all{
&
\beta= 
2 \Bigg[L (n+1) \bigg(-2 (n^4+7 n^3+19 n^2+24 n+13) (\lambda +n+2)^L
\nonumber \\ &
+\big(n^4+7 n^3+17 n^2+(\lambda +20) n+11\big) (\lambda +n+2)^{2 L}+n^4+7 n^3 +17 n^2-\lambda  n+20 n+11\bigg)
\nonumber \\ &
+2 (n+2) (n^4+9 n^3+30 n^2+44 n+26) (\lambda +n+2)^L
\nonumber \\ &
-\Big( n^5+11 n^4+46 n^3+94 n^2+(\lambda +98) n+44   \Big)  (\lambda +n+2)^{2 L}
\nonumber \\ &
-n^5-11 n^4-46 n^3-94 n^2+\lambda  n-98 n-44 \Bigg]
.}{gamma_ringstar}

 For large $n$, we can expand the result in powers of $1/n$. We have: 
 \all{
 b^*= \fracc{3L-1}{2L-2} + \left[  \fracc{2L^2+L+3}{2(L-1)^2} \right] \fracc{1}{n} 
 -  \left[  \fracc{3 L^3 - 6 L^2 - 15 L - 18 }{4(L-1)^3} \right] \fracc{1}{n^2} 
 + O \left(\fracc{1}{n^2}\right)
 .}{gamma_ringstar_expansion}

So in the limit as ${n \rightarrow \infty}$, the critical benefit-to-cost ratio approaches ${\frac{3L-1}{2L-2}}$. 
If $L$ is also large, this approaches $\frac{3}{2}$.
Note that the average degree is: 
\al{
\overline{k} = \fracc{L (n+2) + L n } {L+ L n}=2
.}
This is particularly interesting because very rarely graphs have critical benefit to cost ratio less than their average degree. 
A ring of stars does have this property. 
We call these structure `super-promoters of cooperation'.


\clearpage

\section{Cliques}

\subsection{Single clique (complete graph)}

In a complete graph of size $N$, the system of equations~\eqref{sys} reduces to a single equation: 
\all{
\tau_{xy}= 1+\fracc{1}{2(N-1)} (N-2) \tau_{xy} + \fracc{1}{2(N-1)} (N-2) \tau_{xy}
.}{complete_1}
This yields $\tau_{xy}=N-1$. 
Thus from~\eqref{taux} we get ${\tau_x=N}$. Plugging this into~\eqref{Gamma}, we get ${ N(N-1)(N-2)}$ in the numerator and ${-N(N-2)}$ in the denominator.   Thus we arrive at:
\all{
b^*_{ave} =
-(N-1)<0
.}{complete_2}
So regardless of $b$ and $c$, natural selection does not favor cooperation on a clique.

\vspace{3cm}

\subsection{Two cliques conjoined directly}
Suppose we have a clique of size $n_1$ and another clique of size $n_2$, and we connect one node from the first clique to a node in the second one. 
We denote these two nodes by $g_1$ and $g_2$. 
We denote the non-gate nodes in the first clique by $c_1$ and those in the second clique by $c_2$. 
The remeeting times we need to find are $\tau_{c_1,c_1'}$ (between two commoners in the first clique) and $\tau_{c_2, c_2'}$, 
$\tau_{g_1,g_2}$ (between the gates), 
$\tau_{c_1,g_1}$, $\tau_{c_2, g_2}$, 
$\tau_{c_1,g_2}$, $\tau_{g_1,c_2}$, and $\tau_{c_1,c_2}$. 
 The recurrence relations are given by: 
 \all{
 \begin{cases}
 \tau_{c_1,c_1'}=1+\fracc{\tau_{c_1,c_1'} (n_1-3)+\tau_{c1,g1}}{(n_1-1)}  ,
 \\ \\
   \tau_{c1,g1}=1+\fracc{\tau_{c_1,c_1'} (n_1-2)+\tau_{c_1,g_2}}{2 n_1}+\fracc{\tau_{c1,g1} (n_1-2)}{2(n_1-1)} ,
   \\ \\
     \tau_{c_1,c_2}=1+\fracc{\tau_{c_1,c_2} (n_2-2)+\tau_{c_1,g_2}}{2 (n_2-1)}+\fracc{\tau_{c_1,c_2} (n_1-2)+\tau_{g_1,c_2}}{2(n_1-1)} , 
     \\  \\
      \tau_{c_1,g_2}=1+\fracc{\tau_{c_1,c_2} (n_2-1)+\tau_{c1,g1}}{2 n_2}+\fracc{\tau_{c_1,g_2} (n_1-2)+\tau_{g_1,g_2}}{2(n_1-1)} ,
\\      \\
        \tau_{g_1,c_2}=1+\fracc{\tau_{c_1,c_2} (n_1-1)+\tau_{c_2,g_2}}{2 n_1}+\fracc{\tau_{g_1,c_2} (n_2-2)+\tau_{g_1,g_2}}{2 (n_2-1)} , 
\\        \\
         \tau_{g_1,g_2}=1+\fracc{\tau_{c_1,g_2} (n_1-1)}{2 n_1}+\frac{\tau_{g_1,c_2} (n_2-1)}{2 n_2} , 
\\         \\
          \tau_{c_2,c_2'}=1+\fracc{\tau_{c_2,c_2'} (n_2-3)+\tau_{c_2,g_2}}{(n_2-1)}  ,  
 \\         \\
          \tau_{c_2,g_2}=1+\fracc{\tau_{c_2,c_2'} (n_2-2)+\tau_{g_1,c_2}}{2 n_2}+\fracc{\tau_{c_2,g_2} (n_2-2)}{2 (n_2-1)} 
          \end{cases}
 }{sys_twocliques}

The closed-form of the solution is too long to present. For the case of ${n_1=n_2=n}$, we have
\all{
b^*= n^2-\frac{2n^3 (n-2) }{n (n+1) \left(n^3+2 n-1\right)-2}
.}{gamma_samecliques}

This can be expanded as
\all{
b^* = n^2 - \fracc{2}{n} + \fracc{6}{n^2} + O\left(\fracc{1}{n^3}\right) \\
}{gamma_Expand_samecliques}

This means that $b^* \approx n^2$ is a good approximation even for moderate values of $n$.

\subsection{Two cliques, conjoined  via one broker node}
If instead of being connected directly, the gate nodes were connected via one intermediary node, then we would have
\all{
b^*= \fracc{8 n^9-10 n^8+37 n^7-32 n^6+34 n^5-76 n^4+81 n^3-26 n^2}{20 n^7-42 n^6+104 n^5-100 n^4-100 n^3+222 n^2-116 n+16}
.}{one_broker}
This can be expanded for large $n$ as follows
\all{
b^*= \fracc{2}{5} n^2 + \fracc{17}{50} n + \fracc{121}{250} + O \left(\fracc{1}{n} \right) 
.}{gammapproxonebroker}


\subsection{Two cliques conjoined  via two intermediary broker nodes} 
For two identical cliques connected via a chain of two intermediary broker nodes, we can take similar steps. 
Let 
$\tau_{cc}$ be the remeeting times between two commoners of the same clique, 
$\tau_{c g}$ between a commoner and the gate node of the same clique, 
$\tau_{c b}$ between a commoner and the broker node which is closer (that is, the broker which is connected to the gate node that belongs to the same clique as the commoner), 
$\tau_{c b'}$ between a commoner and the other broker node, 
$\tau_{c g'}$ between a commoner and the gate node of the other clique, 
$\tau_{c c'}$ between a commoner from one clique and a commoner from the other clique, 
$\tau_{g b}$ between a gate node and the adjacent broker node, 
$\tau_{g b'}$ between a gate node and the  non-adjacent broker node, 
$\tau_{g g'}$ between the two gate nodes, 
$\tau_{b b'}$ between the two broker nodes. 
We also have $k_b=2$, $k_g=n$, and $k_c=n-1$. 
We have:
\all{
\begin{cases}
 \tau_{b b'} =\frac{ \tau_{g b'} }{k_b}+1 \\
  \tau_{cc}  =\frac{(n-3)  \tau_{cc} + \tau_{cg} }{k_c} +1  \\
  \tau_{cc'}  =\frac{(n-2)  \tau_{cc'} + \tau_{cg'} }{k_c}+1  \\
  \tau_{gg'}    =\frac{(n-1)  \tau_{cg'} + \tau_{g b'} }{k_g}+1  \\
  \tau_{cg}  =\frac{(n-2)  \tau_{cg} }{2 k_c}+\frac{(n-2)  \tau_{cc} + \tau_{cb} }{2 k_g}  +1  \\
     \tau_{gb}   =\frac{ \tau_{g b'} }{2 k_b}+\frac{(n-1)  \tau_{cb} }{2 k_g}   +1  \\
   \tau_{cb}    =\frac{ \tau_{cb'} + \tau_{cg} }{2 k_b}+\frac{(n-2)  \tau_{cb} +  \tau_{gb} }{2 k_c} +1  \\
   \tau_{cb'}    =\frac{ \tau_{cb} + \tau_{cg'} }{2 k_b}+\frac{(n-2)  \tau_{cb'} + \tau_{g b'} }{2 k_c}+1  \\
    \tau_{cg'}   =\frac{(n-2)  \tau_{cg'} + \tau_{gg'} }{2 k_c}+\frac{(n-1)  \tau_{cc'} + \tau_{cb'} }{2 k_g} +1  \\
   \tau_{g b'}      =\frac{  \tau_{gb} + \tau_{gg'} }{2 k_b}+\frac{(n-1)  \tau_{cb'} + \tau_{b b'} }{2 k_g}+1  .
          \end{cases}
}{sys_cliques_twoBrokers}

The solution can be expanded for large $n$ as follows: 
\all{
b^*= 4 n-\frac{224}{5}+\frac{42304}{75 n}+ O\left( \frac{1}{n^2} \right)
}{b*twobrokersCliques}

\subsection{Two cliques with longer chains}
\begin{figure}[h]
	\centering
	\includegraphics[width=.9\linewidth]{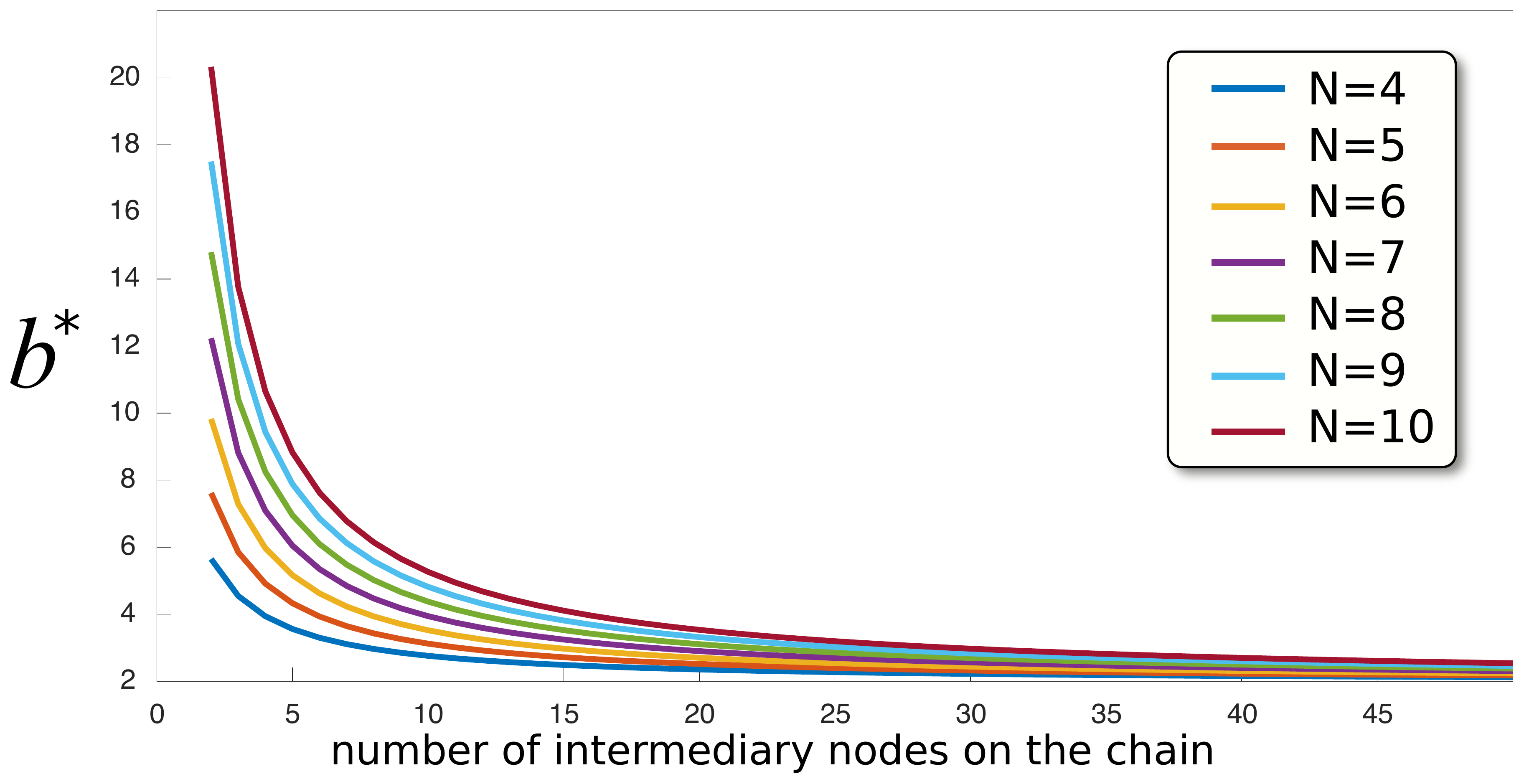}
	\caption{Conjoining two cliques with long chains. }
	\label{longChains}
\end{figure}

 When the number of intermediary nodes on the connecting chain is comparable to the number of nodes in the cliques, the analytical solutions to the system of linear equations for remeeting times become  unwieldy. But numerical solution is straightforward. 
Figure~\ref{longChains} displays the $b^*$ values for chains with 2 to 50 intermediary nodes, for  different cliques sizes. 
For large $L$, the values of $b^*$ approach 2, and the convergence rate depends on the cliques size. 
 This limiting value of 2 corresponds to an infinite chain (where the effect of the two cliques is also negligible). 

\subsection{Conjoining  large cliques}
We showed that conjoining  two cliques directly or via one broker nodes leads to a $b^*$ that grows with $n$ quadratically. 
For a single clique, natural selection does not favor cooperation over defection regardless of the benefit-to-cost ratio. 
Having two intermediary broker nodes does provide a marked reduction in $b^*$, lowering the leading behavior of $b^*$ to linear in $n$, but we note that for large network sizes, this still might not be feasible. 
That is, though cooperation is in principle possible, for large cliques, the benefit-to-cot ratio proportional to $n$ might still not be feasible to provide, though it is considerably  better than a single clique.

\clearpage

\subsection{Star of cliques}

\begin{figure}[h]
	\centering
	\includegraphics[width=.5\linewidth]{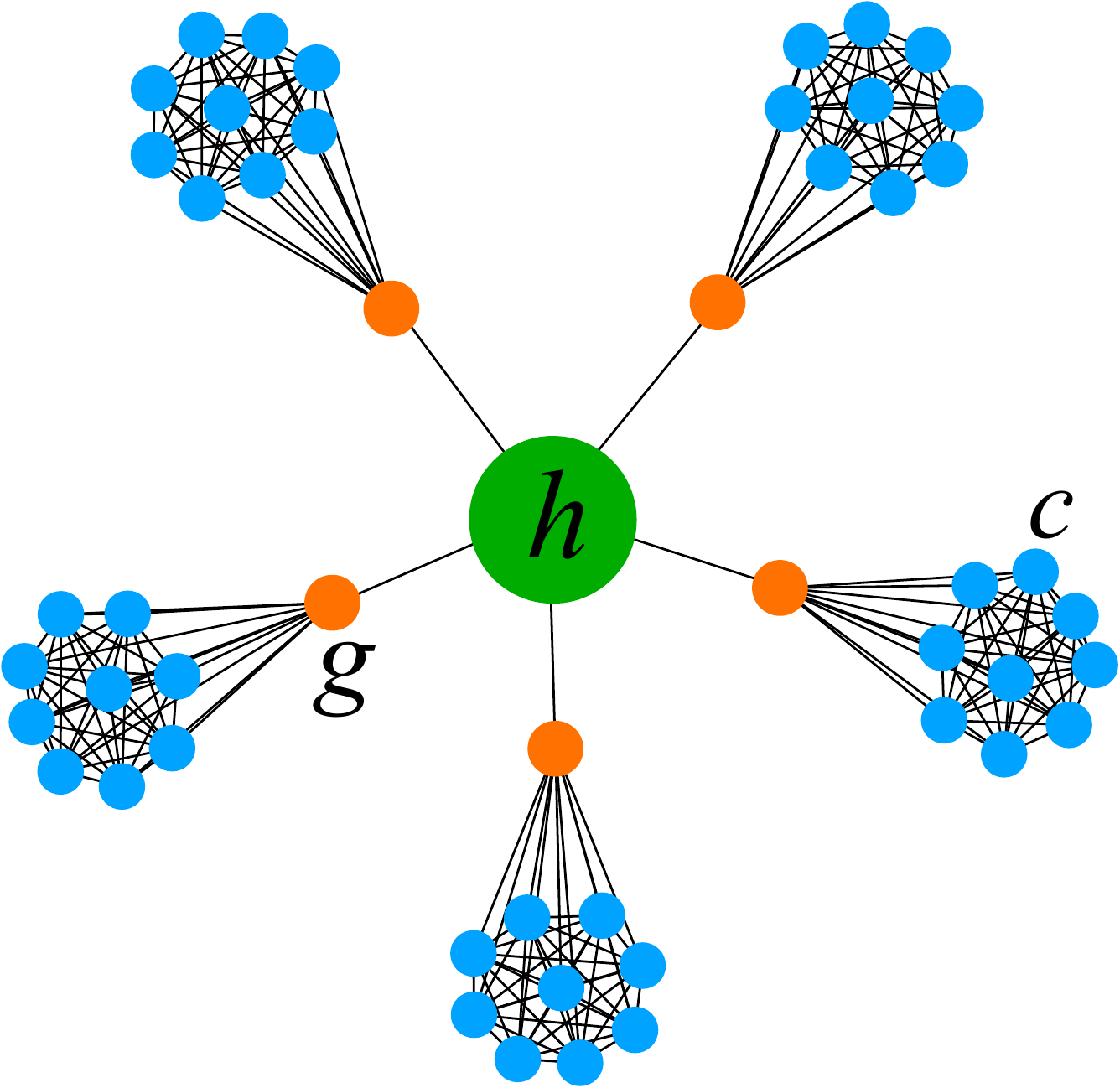}
	\caption{An example of star of cliques with $n=10$ and $m=5$. }
	\label{fig:clique_star}
\end{figure}

Suppose there are $m$ identical cliques each with $m$ nodes, and there is a hub node that is connected to one gate node in each community. 
So there are $m$ gate nodes overall, one for each community. 
An example case with $m=5$ and $n=10$ is depicted in Figure~\ref{fig:clique_star}.

\all{
\begin{cases}
\tau_{gh}= 1+ \frac{(m-1)  \tau_{gg'}}{2 m}+\frac{(n-1) \tau_{hc}}{2 n}
 \\
 \tau_{hc}=1+ \frac{(m-1)  \tau_{gc'}+ \tau_{gc}}{2 m}+\frac{(n-2) \tau_{hc}+\tau_{gh}}{2 (n-1)} 
\\
  \tau_{cc}=1+ \frac{(n-3)  \tau_{cc}+ \tau_{gc}}{n-1} 
\\
 \tau_{gc}= 1+ \frac{(n-2)  \tau_{gc}}{2 (n-1)}+\frac{(n-2)  \tau_{cc}+\tau_{hc}}{2 n} 
\\
  \tau_{gg'}= 1+ \frac{(n-1)  \tau_{gc'}+\tau_{gh}}{n}
\\
 \tau_{cc'}= 1+ \frac{(n-2)  \tau_{cc'}+ \tau_{gc'}}{n-1} 
\\ 
\tau_{gc'}= 1+ \frac{(n-2)  \tau_{gc'}+ \tau_{gg'}}{2 (n-1)}+\frac{(n-1)  \tau_{cc'}+\tau_{hc}}{2 n}
.
\end{cases}
}{starclique_eqs}

\clearpage

The solution is
\all{
\begin{cases}
\tau_{gh} =  \frac{m^2 (2 n-1) ((n-1) n+1) ((n-1) n+3) (n (n+3)-2)-m (2 n^7+n^6-15 n^5+40 n^4-52 n^3+44 n^2-24 n+6) +n ((n-3) n+1) (n-1)^3}{m^2 (2 n-1) (n (n+3)-2)+m (n (n (2 n^3+3 n-7)+6)-2)+n (n-1)^3},
\\ \\
\tau_{hc} = 
 \frac{m^2 (n (n+3)-2)(2 n^5-4 n^4+10 n^3-8 n^2+5 n-2) +m(-2 n^7+n^6+7 n^5-21 n^4+24 n^3-23 n^2+16 n-4)-(n-1)^3 n^2 (n+2)}{m^2 (2 n-1) (n (n+3)-2)+m (n (n (2 n^3+3 n-7)+6)-2)+n (n-1)^3},
 \\ \\
\tau_{cc} = 
  \frac{m (n-1) (m (2 n^5 - 4 n^4 + 12 n^3 - 5 n^2 + 3 n - 2)+(n-1) n (5 n^2+3)+2)}{m^2 (2 n-1) (n (n+3)-2)+m (n (n (2 n^3+3 n-7)+6)-2)+n (n-1)^3},
  \\ \\
\tau_{gc}= 
   \frac{(n-1) (m^2 (4 n^5-8 n^4+22 n^3-15 n^2+13 n-6) +m (-2 n^5+10 n^4-13 n^3+13 n^2-12 n+6) -(n-1)^3 n)}{m^2 (2 n-1) (n (n+3)-2)+m (n (n (2 n^3+3 n-7)+6)-2)+n (n-1)^3},
   \\ \\
\tau_{gg'} = 
    \frac{m (m (n (n+3)-2)(2 n^6-4 n^5+10 n^4-8 n^3+3 n^2+3 n-2) -(n-2) (n-1)^3 n^2 (3 n+1))}{n (m^2 (2 n-1) (n (n+3)-2)+m (n (n (2 n^3+3 n-7)+6)-2)+n (n-1)^3)},
\\ \\
\tau_{cc'}= 
    \frac{m (n (m (n (n+3)-2)  (2 n^4-4 n^3+11 n^2-7 n+2) +     n^5+10 n^4-22 n^3+14 n^2-7 n+8)-4)}{m^2 (2 n-1) (n (n+3)-2)+m (n (n (2 n^3+3 n-7)+6)-2)+n (n-1)^3},
   \\ \\
 \tau_{gc'}= 
   \frac{m^2 (n (n+3)-2) (2 n^5-4 n^4+11 n^3-9 n^2+5 n-1) -m  (n^5-11 n^4+14 n^3-10 n^2+10 n-6) (n-1)-(n-1)^4 n}{m^2 (2 n-1) (n (n+3)-2)+m (n (n (2 n^3+3 n-7)+6)-2)+n (n-1)^3},
\end{cases}
}{taus_cliquestar}

which gives

\all{
b^*(m,n)= \fracc{\sum_{i=0,j=0}^{i=2,j=9} \theta_{ij} m^i n^j }{\sum_{i=0,j=0}^{i=2,j=8} \lambda_{ij} m^i n^j }
,}{clans}
where the  polynomial coefficients are given by matrices $\theta$ and $\lambda$: 
\all{
\theta&= \left[\begin{array}{rrrrrrrrrr}
  0 & 0 & 2 & -5 & 4 & -2 & 2 & -1 & 0 & 0  \\
  0 & 0 & 2 & -9 & 22 & -22 & 1 & 3 & -5 & 0 \\
  0 & 0 & -8 & 26 & -31 & 20 & -9 & 8 & 0 &2 
  \end{array}\right], \nonumber \\ 
  \lambda&= 
  \left[\begin{array}{rrrrrrrrr}
 0 & -4 & 18 & -36 & 46 & -42 & 24 & -6 & 0  \\
   -8 & 32 & -74 & 112 & -99 & 55 & -25 & 9 & -4 \\
   8 & -44 & 88 & -72 &  13 & 9 & -4 & 2 & 2
  \end{array}\right].
  }{coefs1}

  For large $n$, we can expand the result as follows: 
  \all{
  b^*= 
  \left( \fracc{m}{m-2} \right)  n - \fracc{(m+8)(m-1)}{(m-2)^2} + \left[  \fracc{14 m^4 + 11 m^3 - 14m^2 - 50 m + 44}{2m(m-2)^3 } \right]  \fracc{1}{n}
  + O\left(\fracc{1}{n^2}\right)
  .}{expand_cliquestar}

\clearpage
\subsection{Connecting two cliques via a star}\label{sec:cliques_via_star}

\begin{figure}[h]
	\centering
	\includegraphics[width=.4\linewidth]{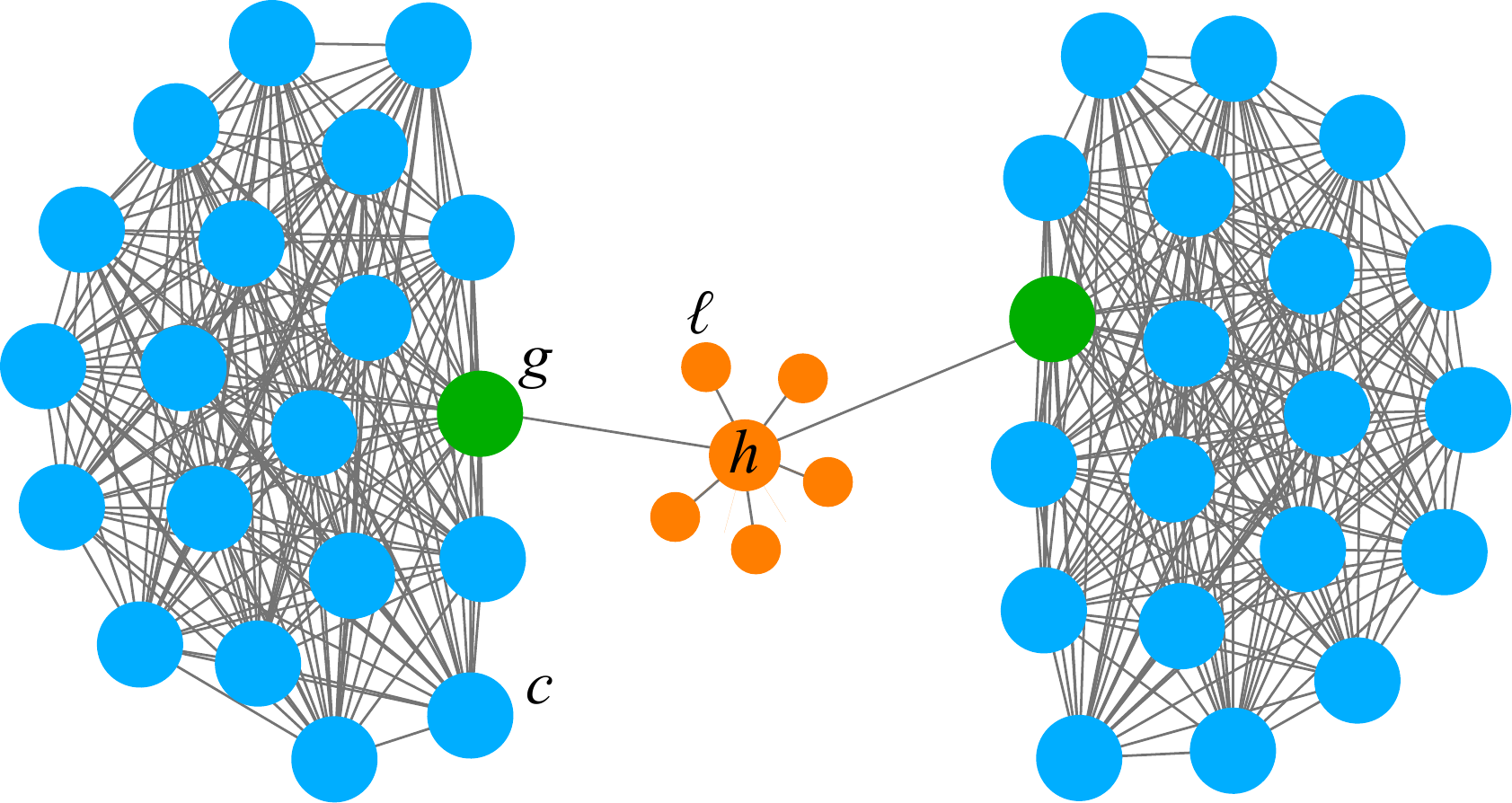}
	\caption{An example case  of connecting cliques via a star with $n=20$ and $L=5$}
	\label{cliquesviastar}
\end{figure}

If we have two cliques of size $n$ and a star of size $m$ (that is, $m-1$ leafs and one hub), 
we can connect the hub of the star to one gate node from each clique. 
An example case with $n=20,m=5$ is depicted in Figure~\ref{cliquesviastar}.
We denote the hub of the star with $h$, its leafs with $\ell$, the gate nodes with $g$, and the non-gate nodes within the cliques with $c$. 
The remeeting times satisfy the following equations: 

\begin{equation}
\begin{cases}
\tau_{cc}=1+\frac{1}{n-1} \big[ (n-3)\tau_{cc}+\tau_{cg} \big]\\ 
\tau_{cc'}=1+\frac{1}{n-1} \big[ (n-2)\tau_{cc'}+\tau_{cg'} \big] \\
\tau_{cg}=1+\frac{1}{2(n-1)} \big[ (n-2)\tau_{cg} \big]+\frac{1}{2 n} \big[ (n-2)\tau_{cc}+\tau_{ch} \big]\\
\tau_{cg'}=1+\frac{1}{2(n-1)} \big[ (n-2)\tau_{cg'}+\tau_{gg'} \big]+\frac{1}{2 n} \big[ (n-1)\tau_{cc'}+\tau_{ch} \big]\\
\tau_{ch}=1+\frac{1}{2(n-1)} \big[ (n-2)\tau_{ch}+\tau_{gh} \big]+\frac{1}{2(m+1)} \big[ (n-1)\tau_{c\ell}+\tau_{cg}+\tau_{cg'} \big]\\
\tau_{c\ell}=1+\frac{1}{2(n-1)} \big[ (n-2)\tau_{c\ell}+\tau_{g\ell} \big]+\frac{1}{2}\tau_{ch}\\
\tau_{gg'}=1+\frac{1}{ n} \big[ (n-1)\tau_{cg'}+\tau_{gh} \big]\\
\tau_{gh}=1+\frac{1}{2 n} \big[ (n-1)\tau_{ch} \big]+\frac{1}{2 (m+1)} \big[ (m-1)\tau_{g\ell}+\tau_{gg'} \big]\\
\tau_{g\ell}=1+\frac{1}{2 n} \big[ (n-1)\tau_{c\ell}+\tau_{h\ell} \big]+\frac{1}{2}\tau_{gh}\\
\tau_{h\ell}=1+\frac{1}{2(m+1)} \big[ (m-2)\tau_{\ell\ell}+2\tau_{g\ell} \big]\\
\tau_{\ell\ell}=1+\tau_{h\ell}\\
\end{cases}
\end{equation}

The solution that we obtain is

\begin{equation}
b^*=\frac{\alpha}{\beta}
\end{equation}

\all{
& \alpha=
\resizebox{.95\linewidth}{!}{$
(m+1) n \bigg[ 2 m^4 n (6 n^2-4 n+1) (n (n+3)-2)+m^3  (18 n^8+36 n^7+4 n^6-87 n^5+144 n^4-131 n^3+100 n^2-48 n+8) 
$}
 \nonumber \\ &
+m^2 n (6 n^9-n^8+116 n^7-78 n^6+43 n^5-270 n^4+473 n^3-363 n^2+160 n-34) 
\nonumber \\ &
+m (34 n^{10}-53 n^9+202 n^8-395 n^7+448 n^6-491 n^5+527 n^4-395 n^3+163 n^2-12 n-8) 
\nonumber \\ &
+2 (n-1)^2 (20 n^8-11 n^7+30 n^6-25 n^5+47 n^4-42 n^3+39 n^2-16 n)  \bigg]
}{alpha_cliquesandstar}

\al{
&\beta= -2 (80 m^4+436 m^3+1069 m^2+1201 m+428)  n^3+2(m-1) (m+2) (3 m+5) n^{10}
\nonumber \\  &
+(14 m^3+68 m^2+154 m+174) n^9+  (30 m^4+18 m^3-252 m^2-488 m-428)  n^8  
\nonumber \\ &
+ (40 m^4+318 m^3+904 m^2+1000 m+606)  n^7- 4 (41 m^4+188 m^3+330 m^2+278 m+132) n^6
\nonumber \\ &
+2 (39 m^4+143 m^3+45 m^2-55 m+48) n^5+ 2(59 m^4+310 m^3+859 m^2+931 m+261)  n^4 
\nonumber \\ &
+4  (20 m^4+132 m^3+329 m^2+395 m+165)  n^2- 8  (2 m^4+20 m^3+53 m^2+67 m+31)  n
\nonumber \\ &
+16 (m+1) (m (m+2)+2)
}

Note that with $m=1$, which means that the star is only the hub with no leafs, so that the cliques are being connected via a single node, we  recover Equation~\eqref{gammapproxonebroker}. 
For $1<m \ll n$, we can use the expansion: 
\al{
b^*= \left( \fracc{(m+4)(m+1)}{m^2+m-2} \right) n + \fracc{(m+1)(15 m^4+178 m^3+579 m^2+776 m+452)}{2(3m+5)(m^2+m-2)^2}
+ 
  O\left(\fracc{1}{n} \right) .
}
Note that in the main text,  \emph{the number of leafs} is denoted by $m$, whereas here  the size of the star is denoted by $m$. Thus, to recover the result of the main text one must simply replace $m$ with $m+1$ in the above equation.

For the simple case of $m=2$, which means that the star is simply a dyad, and the cliques are being connected via a bridging node with a leaf attached to it, we get: 
\all{
b^* \bigg|_{m=2} = 
\frac{9 n}{2}-51+\frac{27585}{44 n} + O\left(\fracc{1}{n^2} \right) .
.}{twocliques_starinmiddle_m2}

In the special case of $m=n$, we have: 

\begin{equation}
b^*=\frac{\alpha}{\beta}
\end{equation}

where the numerator is: 
\al{
\alpha=
\resizebox{.9\linewidth}{!}{$
n^2 (n+1)  (6 n^{11}+51 n^{10}+139 n^9+38 n^8-267 n^7+84 n^6+127 n^5-62 n^4+57 n^3-135 n^2+130 n-40) ,
$}
}
and the denominator is:
\al{
 \beta&=6 n^{13}+60 n^{12}+124 n^{11}+36 n^{10}-94 n^9-344 n^8+44 n^7
 \nonumber \\ &
+288 n^6+332 n^5-724 n^4+316 n^3+172 n^2-184 n+32
. }
For large $n$, we can use the following expansion: 
\al{
b^*= 
n-\frac{1}{2} +\frac{16}{n} + O\left(\fracc{1}{n^2} \right) .
}

\clearpage
\subsection{Hierarchy of cliques}\label{sec:clique_hier}

\begin{figure}[h]
	\centering
	\includegraphics[width=.9\linewidth]{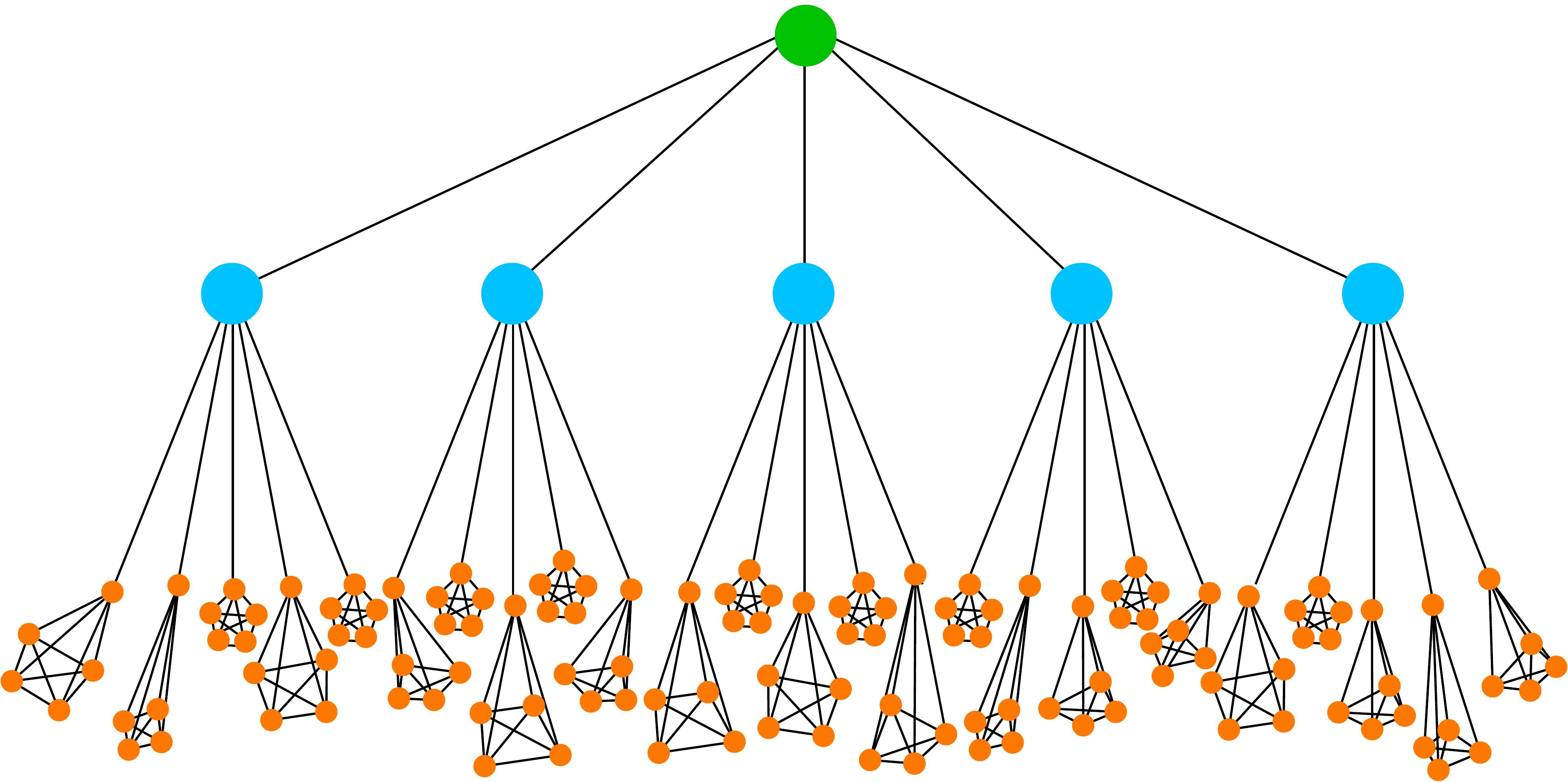}
	\caption{An example case of  hierarchical connection of cliques with $q=3$ and $n=5$.  }
	\label{fig:hier}
\end{figure}

We can also connect communities in a hierarchical structure. 
We construct the hierarchical network by connecting a base node (denoted by $b$) to $q$ middle nodes (denoted by $m$), 
and connecting each middle node to a gate node (denoted by $g$) of a clique of size $n$. 
So there are $q^2$ cliques. We denote the non-gate nodes within cliques by $c$ (for `commoner'). 
An example case with $q=3$ and $n=5$ is illustrated in Figure~\ref{fig:hier}.

The remeeting times of interest are 
$\tau_{b,m}$ (between the base node and a middle node), 
$\tau_{b,g}$ (between the base node and a gate node), 
$\tau_{b,c}$ (between the base node and a commoner), 
$\tau_{m,g}$ (between a middle node and a gate node adjacent to it), 
$\tau_{ m ,g'  }$ (between a middle node and a gate node adjacent to another middle node), 
$\tau_{ m , c }$ (between a middle node and a commoner of an adjacent clique), 
$\tau_{m  , c' }$ (between a middle node and a commoner of a clique adjacent to another middle node), 
$\tau_{m  , m'  }$ (between two middle nodes), 
$\tau_{ c , g }$ (between a commoner and a gate node in the same community), 
$\tau_{ c , g' }$ (between a commoner and the gate node of another community, the two communities being adjacent to the same middle node), 
$\tau_{ c , g'' }$ (between a commoner of a community and the gate node of another community, the two communities being adjacent to two distinct middle nodes), 
$\tau_{ c , c' }$ (between two commoners within the same community), 
$\tau_{c  , c'' }$ (between a commoner in one community and a commoner in another community, the two communities being adjacent to the same middle node), 
$\tau_{ c , c''' }$ (between a commoner in one community and a commoner in another community, the two communities being adjacent to the distinct middle nodes), 
$\tau_{ g , g' }$ (between two gate nodes adjacent to the same middle node), 
$\tau_{ g , g''  }$ (between two gate nodes adjacent to two distinct middle nodes).

\all{
\begin{cases}
\tau_{b,m}=\frac{(q-1) \tau_{m,m'}}{2q}+\frac{q \tau_{b,g}}{2(q+1)}+1
\\
 \tau_{b,g}=\frac{(q-1) \tau_{m,g'}+\tau_{m,g}}{2q}+\frac{(n-1) \tau_{b,c}+\tau_{b,m}}{2n}+1
\\ 
\tau_{b,c}=\frac{(q-1) \tau_{m,c'}+\tau_{m,c}}{2q}+\frac{(n-2) \tau_{b,c}+\tau_{b,g}}{2(n-1)}+1
\\ 
\tau_{m,g}=\frac{(n-1) \tau_{m,c}}{2n}+\frac{(q-1)  \tau_{g,g'}+\tau_{b,g}}{2(q+1)}+1
\\ 
\tau_{m,g'}=\frac{(n-1) \tau_{m,c'}+\tau_{m,m'}}{2n}+\frac{q  \tau_{g,g''}+\tau_{b,g}}{2(q+1)}+1
\\ 
\tau_{m,c}=\frac{(n-2) \tau_{m,c}+\tau_{m,g}}{2(n-1)}+\frac{(q-1) \tau_{c,g'}+\tau_{b,c}+\tau_{c,g}}{2(q+1)}+1
\\ 
\tau_{m,c'}=\frac{(n-2) \tau_{m,c'}+\tau_{m,g'}}{2(n-1)}+\frac{q \tau_{c,g''}+\tau_{b,c}}{2(q+1)}+1
\\ 
\tau_{m,m'}=\frac{q \tau_{m,g'}+\tau_{b,m}}{\text{km}}+1 
\\ 
\tau_{c,g}=\frac{(n-2) \tau_{c,g}}{2(n-1)}+\frac{(n-2) \tau_{c,c'}+\tau_{m,c}}{2n}+1
\\ 
\tau_{c,g'}=\frac{(n-2) \tau_{c,g'}+ \tau_{g,g'}}{2(n-1)}+\frac{(n-1) \tau_{c,c''}+\tau_{m,c}}{2n}+1
\\ 
\tau_{c,g''}=\frac{(n-2) \tau_{c,g''}+ \tau_{g,g''}}{2(n-1)}+\frac{(n-1) \tau_{c,c'''}+\tau_{m,c'}}{2n}+1
\\ 
 \tau_{g,g'}=\frac{(n-1) \tau_{c,g'}+\tau_{m,g}}{k_g}+1 
\\  
  \tau_{g,g''}=\frac{(n-1) \tau_{c,g''}+\tau_{m,g'}}{k_g}+1
\\ 
 \tau_{c,c'}=\frac{(n-3) \tau_{c,c'}+\tau_{c,g}}{k_c}+1
\\ 
 \tau_{c,c''}=\frac{(n-2) \tau_{c,c''}+\tau_{c,g'}}{k_c}+1
\\ 
 \tau_{c,c'''}=\frac{(n-2) \tau_{c,c'''}+\tau_{c,g''}}{k_c}+1
.
\end{cases}
}{sys_hier}

Using these results we arrive at $b^*$. 
For the numerator, the 
For the numerator, the $i-j$ element of the following matrix yields the coefficient of $n^{i-1}  q^{j-1}$ in the numerator:  
\all{
\resizebox{.8\linewidth}{!}{$
\left[
\begin{array}{cccccccccc}
 0 & 0 & 0 & 0 & 0 & 0 & 0 & 0 & 0 & 0 \\
 0 & 0 & 0 & -32 & -80 & 48 & 192 & 32 & -112 & -48 \\
 0 & 0 & 16 & 224 & 412 & -532 & -1268 & 116 & 1064 & 416 \\
 0 & 32 & 12 & -610 & -930 & 2026 & 3558 & -1488 & -4272 & -1592 \\
 0 & -120 & -168 & 871 & 982 & -4400 & -5652 & 5289 & 10158 & 3680 \\
 4 & 110 & 154 & -540 & 410 & 6546 & 5245 & -10821 & -16333 & -5815 \\
 -20 & 93 & 288 & -86 & -2494 & -7219 & -2896 & 13858 & 18330 & 6562 \\
 40 & -160 & -572 & 294 & 3689 & 6888 & 995 & -12290 & -14872 & -5452 \\
 -40 & -57 & 430 & 188 & -3357 & -5751 & 332 & 8622 & 9035 & 3398 \\
 20 & 170 & -407 & -911 & 1702 & 3253 & -996 & -4385 & -3719 & -1527 \\
 -4 & -69 & 328 & 539 & -695 & -379 & 2282 & 2535 & 1281 & 566 \\
 0 & -4 & -197 & -468 & -713 & -1400 & -1666 & -678 & -144 & -170 \\
 0 & -3 & 26 & 160 & 602 & 1235 & 1204 & 518 & 176 & 98 \\
 0 & 0 & -6 & -61 & -230 & -349 & -184 & -2 & -28 & -36 \\
 0 & 0 & 0 & 0 & 6 & 26 & 54 & 70 & 52 & 16 \\
\end{array}
\right],
$}
}{numer13446756756}
and for the denominator we have: 
\all{
\resizebox{.8\linewidth}{!}{$
\left[
\begin{array}{cccccccccc}
 0 & 0 & 0 & 0 & 0 & 32 & 80 & 0 & -80 & -32 \\
 0 & 0 & 0 & 0 & -48 & -248 & -280 & 400 & 768 & 272 \\
 0 & 0 & 0 & -16 & 184 & 624 & -200 & -2612 & -3124 & -1032 \\
 0 & 0 & 32 & 68 & -360 & -586 & 2572 & 7482 & 7348 & 2364 \\
 0 & 0 & -120 & -62 & 456 & -462 & -5714 & -11956 & -11046 & -3648 \\
 0 & 12 & 72 & -234 & -84 & 1501 & 4335 & 9459 & 10319 & 3892 \\
 0 & -50 & 266 & 762 & -684 & -265 & 4176 & 872 & -4844 & -2833 \\
 0 & 62 & -460 & -665 & 1151 & -2571 & -12452 & -10140 & -981 & 1264 \\
 0 & 20 & 184 & -421 & -1095 & 4334 & 13630 & 11894 & 3627 & -117 \\
 0 & -120 & 96 & 1260 & 914 & -3815 & -9143 & -7795 & -2917 & -216 \\
 0 & 118 & -92 & -1190 & -1310 & 1128 & 3660 & 3406 & 1464 & 152 \\
 0 & -50 & -10 & 311 & 328 & -207 & -564 & -562 & -268 & -2 \\
 0 & 8 & 22 & 97 & 272 & 293 & 136 & 80 & 24 & -20 \\
 0 & 0 & -24 & -116 & -190 & -98 & 50 & 86 & 52 & 16 \\
\end{array}
\right]
$}
}{denom5756556}

For large $n$, we can use the following expansion: 
\al{
&b^*= 
\left[ \frac{  q^2 \left(2 q^2+q+1\right)}{  q^2 \left(2 q^2+q+3\right)-6q-4} \right] n  
\nonumber \\ &
+\frac{- 16 q^{11}-60 q^{10}-152 q^9-307 q^8-689 q^7-1105 q^6-523 q^5+827 q^4+1226 q^3+619 q^2+136 q+12}{(q+1)^2 (4 q+3) \Big[q \left(q \left(2 q^2+q+3\right)-6\right)-4\Big]^2}
\nonumber \\ &
+ O\left(\fracc{1}{n}\right).
}

For example, for $q=2$, we have
\al{
b^*\bigg|_{q=2} = \fracc{11}{9} n \bigg[ 1+ O\left(\fracc{1}{n}\right)\bigg] 
.}
For $q=3$, we have
\al{
b^*\bigg|_{q=3} = \fracc{99}{97} n \bigg[ 1+ O\left(\fracc{1}{n}\right)\bigg] 
.}

The prefactor in the asymptotic expression becomes ${148/149}$ for $q=4$, and ${175/177}$ for $q=5$. 
As $q$ grows further, the prefactor approaches  unity from below.

\clearpage

\subsection{Ring of cliques}

\begin{figure}[h]
	\centering
	\includegraphics[width=.4\linewidth]{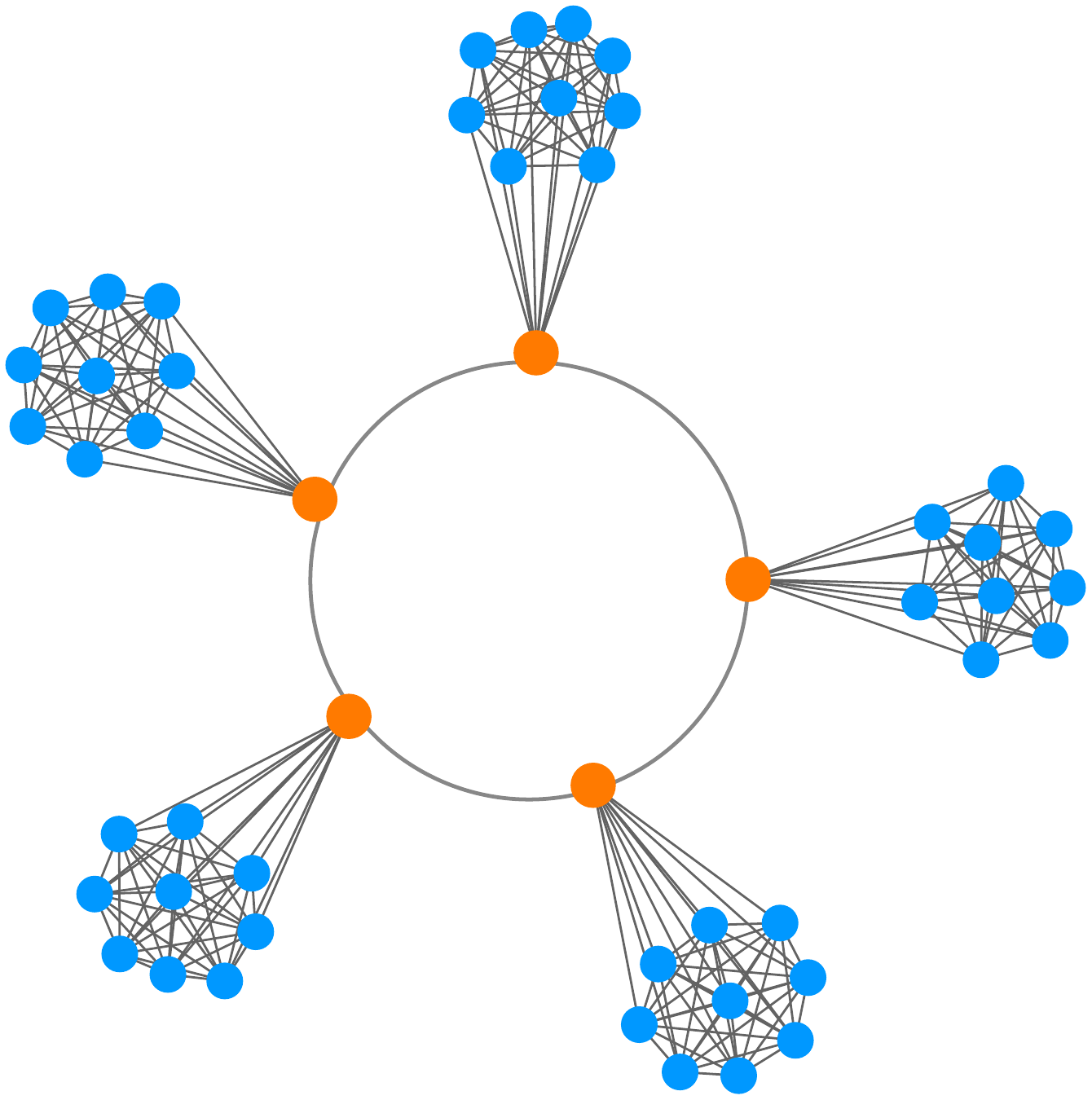}
	\caption{An example case of ring of cliques   with $n=10$ and $L=5$}
	\label{figcliquering}
\end{figure}

We assume $L$ cliques, each with $n$ nodes,  situated on a ring via `gate' nodes. 
Figure~\ref{figcliquering} shows an example of clique of rings with $L=5$ and $n=10$. 
We denote the gate nodes by $g$, and other nodes by $c$ ($c$ stands for \emph{commoner}). 
The remeeting times that we need to obtain are $\tau_{gg}(x)$ (between two gate nodes separated by a distance $x$ on the ring), 
$\tau_{gc }(x)$ (between a gate node and a commoner in another community $x$ apart), 
and $\tau_{cc} (x)$ (between a commoner from one community and a commoner from another community $x$ apart).

The recurrence relations are: 
\all{
\begin{cases}
\text{for }x=2\ldots L-2:~
&\begin{cases}
\tau_{gg} (x) = 1 +  \frac{1}{(n+1)} \bigg[\tau_{gg}(x-1) + \tau_{gg} (x+1) + (n-1) \tau_{gc}(x)   \bigg]    \\ \\
\resizebox{.7\linewidth}{!}{$
\tau_{c g}(x) = 1 + \fracc{1}{2(n+1)} \bigg[ (n-1) \tau_{cc} (x) + \tau_{cg}(x+1) + \tau_{ cg} (x-1) \bigg] + \fracc{1}{2(n-1)} (n-2)\tau_{ cc} (x) 
$}
   \\  \\
\tau_{cc}(x) = 1 +\fracc{1}{n-1} \bigg[  \tau_{ cg} (x) + (n-2) \tau_{cc}(x)  \bigg]    
\end{cases}
\\ \\
\text{for }x<2:~
&\begin{cases}
\tau_{gg} (1) = 1 +  \fracc{1}{(n+1)} \bigg[ \tau_{gg} (2)  + (n-1) \tau_{cg}(1)  \bigg]    \\ \\
\resizebox{.7\linewidth}{!}{$
\tau_{cg}(1) = 1 + \fracc{1}{2(n+1)} \bigg[ (n-1)  \tau_{cc} (1) + \tau_{cg}(2) + \tau_{ cg} (0) \bigg] + \fracc{1}{2(n-1)} \bigg[ (n-2) \tau_{cg}(1) + \tau_{ gg} (1) \bigg]  
$}
  \\  \\
\resizebox{.7\linewidth}{!}{$
\tau_{cg}(0) = 1 + \fracc{1}{2(n+1)} \bigg[ (n-2) \tau_{c c} (0) +2  \tau_{cg}(1)  \bigg]  + \fracc{1}{2(n-1)} (n-2) \tau_{cg} (0)  
$}
 \\  \\
\tau_{cc}(0) = 1 + \fracc{1}{n-1} \bigg[ \tau_{cg}(0)+ (n-3) \tau_{cc} (0) \bigg]  \\ \\
\tau_{cc} (1) = 1 + \fracc{1}{n-1} \bigg[ \tau_{cg}(1)+ (n-2) \tau_{cc} (1) \bigg] .
\end{cases}
\end{cases}
}{recur_ringstar}

 This is a system of linear difference equations with constant coefficients. Plugging in the ansatz $r^{x} + r^{L-x}$,
  we find that the characteristic equation for $r$ is given by 
\all{
(r-1)^2 \bigg[ (n-1) r^2 - n (n+1) r + (n-1) \bigg] 
=0
.}{charac_cliqueRing}
There are four roots: $r=0$ has degeneracy 2,   the third root is 
\all{
R= 
\fracc{n^2+n+\displaystyle \sqrt{ [ n(n-1)  +2]~  [n (n+3)-2] }}{ 2(n-1) } 
, }{R_cliqueRing}
and the fourth root is $1/R$. 

The double degeneracy of root zero means that the particular solution to the nonhomogenous equation is quadratic in $x$. Thus we plug in the following form into the system of equations: 

\all{
\begin{cases}
\tau_{gg}(x)= A_{gg} x^2 + B_{gg} x + C_{gg} + D_{gg} \big( R^x + R^{L-x} \big) 
\\
\tau_{cg}(x)=  A_{cg} x^2 + B_{cg} x + C_{cg} + D_{cg} \big( R^x + R^{L-x} \big) 
\\
\tau_{cc}(x)= A_{cc} x^2 + B_{cc} x + C_{cc} + D_{cc}  \big( R^x + R^{L-x} \big) 
\end{cases}
}{forms1}

For brevity of notation, we define: 
\all{
\xi \stackrel{\text{def}}{=}  \displaystyle \sqrt{ [ n(n-1)  +2]~  [n (n+3)-2] ~ } 
.}{defs_2}

Using this along with the value of $R$ obtained above, we arrive at

\all{
&A_{gg}=A_{c c} = A_{g c} =  \frac{1}{2} \left(-n^2+n-2\right) , \nonumber \\ \nonumber \\
&B_{gg}=B_{c c} = B_{g c} = \frac{1}{2} L ((n-1) n+2) ,\nonumber \\ \nonumber\\
&D_{gg}= 
\resizebox{.9\linewidth}{!}{$
-\fracc{2^{L+1} (n-1)^3 \left((L-1) n^2-L n+2 L+n\right)}{-2^{L+1} n \xi +2 n \xi  (2 R)^L+(n-1) n ((n-1) n+2) (2 R)^L+2^L (n-1) n ((n-1) n+2)-2^{L+1} \xi +2 \xi  (2 R)^L}
$}
\nonumber\\ \nonumber\\
&D_{g c} = 
\resizebox{.9\linewidth}{!}{$
\frac{2^{L+1} \left(n^2-1\right) \left((L-1) n^2-L n+2 L+n\right)}{-2^{L+1} n \xi +2 n \xi  (2 R)^L+(n-1) n ((n-1) n+2) (2 R)^L+2^L (n-1) n ((n-1) n+2)-2^{L+1} \xi +2 \xi  (2 R)^L}
$} \nonumber \\ &
D_{cc}= (n-1)+D_{gc}.
}{sol_ringclique}
The expressions for  other variables are omitted for undue length. Using the solutions, we get the critical benefit to cost ratio: 

\al{b^*= 
\fracc{\alpha}{\beta}
,}
where the numerator is given by: 
\all{
&
\resizebox{\linewidth}{!}{$
\alpha= 
 2^{2L+1} L R^L  \left(n^3+n+2\right)^2 (n^5 - 2 n^4 - 3 n^3 - 16 n^2 - 4 n + 8)     (n^8+4 n^7+2 n^6+4 n^5+11 n^4-8 n^3+8 n^2-8 n+2)
$}
\nonumber \\ &
 +
 R^L 4^L \xi    L 
 \Big( 2 n^{17}    + 2  n^{16}    - 12 n^{15}    
 - 36  n^{14}    - 140 n^{13}    - 220  n^{12}    - 488  n^{11}    - 776  n^{10}     - 678 n^9
 \nonumber \\ &
  - 1046 n^8  -  892 n^7  - 100 n^6 - 256 n^5  + 64  n^4  + 544  n^3  + 64  n^2  - 128  n  \Big) 
  \nonumber \\ &
\resizebox{\linewidth}{!}{$
  + R^{2 L} L  \left(n^3+n+2\right)^2
\Big(n^{13} + 6 n^{12} + 19 n^{11} + 48 n^{10} + 69 n^9 + 74 n^8 + 43 n^7 
 + 
 116 n^6 - 46 n^5 - 84 n^4 + 34 n^3 - 80 n^2 + 72 n - 16 \Big) 
 $}
   \nonumber \\ &
   \resizebox{\linewidth}{!}{$
+
\xi R^{2L} L 
\Big(  n^{17} +5 n^{16}  +18 n^{15}  +50 n^{14}  +98 n^{13}  +178 n^{12}  +268 n^{11}  +428 n^{10}  +317 n^9  +457 n^8  +426 n^7  -94 n^6  +152 n^5  -24 n^4  -304 n^3  +24 n^2  +48 n  \Big) 
$}
  \nonumber \\ &
     \resizebox{\linewidth}{!}{$
  +L 4^L \left(n^3+n+2\right)^2 (n^{13}-2 n^{12}-21 n^{11}+24 n^{10}+93 n^9-14 n^8+195 n^7+108 n^6-70 n^5+12 n^4-78 n^3-48 n^2+72 n-16) 
  $}
  \nonumber \\ &
       \resizebox{\linewidth}{!}{$
  +\xi L 4^L  \Big( n^{17}  -3\   n^{16}  - 28  n^{15}  + 26 n^{14}  + 18  n^{13}  + 18  n^{12}  +300  n^{11}  + 236 n^{10}  +349  n^9  +673  n^8  +362 n^7  + 234 n^6 \xi+ 168  n^5  - 72 n^4  - 240 n^3  - 88  n^2  + 80 n  \Big)    
  $}  
   \nonumber \\ &
     \resizebox{.9\linewidth}{!}{$
 -R^L 4^{L+1} (n+1)^2 (n^8-4 n^7+10 n^6-24 n^5-15 n^4-28 n^3-12 n^2-8 n+16)  ( n^8 + 4 n^7 + 2 n^6 + 4 n^5 + 11 n^4 - 8 n^3 + 8 n^2 - 8 n + 2)   
 $}
 \nonumber \\ &
 + R^L 4^{L+1} \xi 
 \Big( 
 - 2n^{16} -2^{L+1} n^{15}    +2^{L+2} n^{14}    -2^{L+2} n^{13}    +39\ 2^{L+2} n^{12}    +123\ 2^{L+2} n^{11}    +99\ 2^{L+3} n^{10}    +143\ 2^{L+3} n^9  
  \nonumber \\ &  
   +611\ 2^{L+1} n^8    +379\ 2^{L+1} n^7    +53\ 2^{L+2} n^6    -41\ 2^{L+2} n^5    -23\ 2^{L+4} n^4    -15\ 2^{L+4} n^3    +2^{L+5} n^2    +2^{L+6} n    
 \Big) 
  \nonumber \\ &
 -2 (n+1)^2 R^{2 L}  4^L \Big(n^{16} + 2 n^{15} + 8 n^{14} + 44 n^{13} + 44 n^{12} + 244 n^{11} + 256 n^{10} + 
 24 n^9 
 \nonumber \\ &
 + 969 n^8 - 782 n^7 + 760 n^6 - 468 n^5 - 254 n^4 + 280 n^3 - 
 216 n^2 + 144 n - 32\Big) 
 \nonumber \\ &
  + 2 \xi R^{2L} 4^L  
     \resizebox{.95\linewidth}{!}{$
  \Big(
 -n^{16}    -3 n^{15}    -12 n^{14}    -50 n^{13}    -106 n^{12}    -306 n^{11}    -508 n^{10}    -500 n^9    -605 n^8  g   -343 n^7    +92 n^6    +78 n^5    +176 n^4    +116 n^3    -60 n^2    -16 n    
 \Big) 
 $} 
 \nonumber \\ &
 -2^{2L+1}  (n+1)^2 
\Big( n^{16} - 2 n^{15} - 16 n^{14} + 12 n^{13} + 20 n^{12} + 100 n^{11} + 248 n^{10} + 
 144 n^9 
 \nonumber \\ &
 + 609 n^8 - 154 n^7 + 568 n^6 - 556 n^5 + 226 n^4 - 
 200 n^3 - 88 n^2 + 144 n - 32 \Big)
 \nonumber \\ & 
 + 2^{2L+1}  \xi 
 \Big( 
 - n^{16}  +  n^{15}  + 16  n^{14}  + 14  n^{13}  - 26 n^{12}  - 162  n^{11}  - 364 n^{10}  - 532  n^9 
 \nonumber \\ &
  -605  n^8  -499  n^7  - 200 ^6  + 46 n^5  + 128 n^4  +156 n^3  + 28  n^2  - 48  n   \Big)
,}{numer_cliquering}

and the denominator is given by: 
\all{
&\beta= 
     \resizebox{.9\linewidth}{!}{$
L2^{2L+1}R^L  [(n-1) n+2]^2 (n^6 + n^5 - 7 n^4 - 33 n^3 - 46 n^2 - 4 n + 24) (n^8 + 4 n^7 + 2 n^6 + 4 n^5 + 11 n^4 - 8 n^3 + 8 n^2 - 8 n + 2)   
$}
 \nonumber \\ &
 +L4^L R^L\xi  \Big( 
      \resizebox{.85\linewidth}{!}{$
 2  n^{16}   +4 n^{15}   - 12  n^{14}   - 56  n^{13}   - 148 n^{12}   - 248  n^{11}   
  - 608 n^{10}   - 832  n^9   -  638  n^8 -1228 n^7  - 868 n^6
  $}
  \nonumber \\ &  
        \resizebox{.99\linewidth}{!}{$
            + 376  n^5      - 608 n^4   + 320  n^3   + 832  n^2   - 384  n   
\bigg)
   +
   L \Big( ((n-1) n+2)^2 n^{14}+9 n^{13}+39 n^{12}+109 n^{11}   +207 n^{10} +269 n^9+131 n^8+105 n^7
   $} 
    \nonumber \\ &
+368 n^6-482 n^5+18 n^4+46 n^3-204 n^2+200 n\Big)   +L \xi 4^L R^{2L} 
      \Big( 
 -48  +n^{16}    +6 n^{15}      +22 n^{14}    +56 n^{13} 
      \nonumber \\ &
            \resizebox{.99\linewidth}{!}{$
               +114 n^{12}    
  +184 n^{11}    +204 n^{10}     +548 n^9       +29 n^8    +242 n^7    +982 n^6    -1516 n^5    +1448 n^4    -896 n^3    +16 n^2    
       +96 n    
       \bigg)   
       $}
      \nonumber \\ &
                 \resizebox{.99\linewidth}{!}{$
                     +L4^L ((n-1) n+2)^2 
( n^{14} + n^{13} - 25 n^{12} - 43 n^{11} + 127 n^{10} + 309 n^9 + 291 n^8 +673 n^7 + 480 n^6 - 194 n^5
$}
 \nonumber \\ & 
      \resizebox{.98\linewidth}{!}{$
       - 46 n^4 - 434 n^3 - 12 n^2 + 200 n - 48 ) 
    +    L 4^L 
 \Big( 
 n^{16}  - 2 n^{15}  - 18 n^{14}  +
  16  n^{13}   + 58 n^{12}  - 16 n^{11}  + 340  n^{10}  + 508 n^9  +69 n^8  
  $}
 \nonumber \\ &
       \resizebox{ \linewidth}{!}{$
  + 1378 n^7  - 74 n^6  + 772  n^5  - 296  n^4  + 384 n^3  - 848 n^2  + 288n  
  \bigg) 
  -2^{2L+2} R^L  (n+1)(n^9 - 3 n^8 + 2 n^7 - 22 n^6 - 3 n^5 - 31 n^4 - 160 n^3 + 24 n^2  + 16 n + 16)
  $}
 \nonumber \\ &
        \resizebox{ \linewidth}{!}{$
  ( n^8 + 4 n^7 + 2 n^6 + 4 n^5 + 11 n^4 - 8 n^3 + 8 n^2 - 8 n + 2)    +  2^{2L+1}  R^L \xi \Big(  -2 n^{16}    -2  n^{15}    +12  n^{14}    + 44  n^{13}    + 180 n^{12}  
$}
  \nonumber \\ &
       \resizebox{.9\linewidth}{!}{$
   + 276 n^{11}    
   + 752  n^{10}    + 1712  n^9    + 1342  n^8      + 1150  n^7    +  1140 n^6      - 620 n^5    -  928 n^4    -   64 n^3     +   64  n^2    +  64 n    
    \Big)   
 $} 
      \nonumber \\ &
             \resizebox{.95\linewidth}{!}{$
     +2^{2L+1} R^{2 L} (-n^{18} - 6 n^{17} - 23 n^{16} - 72 n^{15} - 152 n^{14} - 364 n^{13} - 620 n^{12}    -  1124 n^{11} - 1717 n^{10} - 610 n^9 - 1947 n^8
    $}
               \nonumber \\ &
                            \resizebox{.95\linewidth}{!}{$
             - 116 n^7 + 1166 n^6 +  276 n^5 + 702 n^4 - 480 n^3 - 64 n + 32) + \xi 2^{2L+1}  R^{2L }  
\Big( -n^{16}  -5 n^{15}  -20 n^{14}  -56 n^{13}
$}
                   \nonumber \\ &
    \resizebox{.95\linewidth}{!}{$
  -116 n^{12}   -268 n^{11}  -398 n^{10}  -822 n^9  -855 n^8  -211 n^7   -662 n^6  +574 n^5  +332 n^4  +36 n^3  -72 n^2  -16 n  \Big) 
  $}
\nonumber \\ &
    \resizebox{.95\linewidth}{!}{$
         -2^{2L+1}  (n+1)^2 (n^{16} - 4 n^{15} - 14 n^{14} + 56 n^{13} - 26 n^{12} - 80 n^{11} + 726 n^{10}   - 176 n^9 + 135 n^8 + 1940 n^7  - 2072 n^6 + 2008 n^5 
         $}
            \nonumber \\ &
         - 1566 n^4 +  352 n^3 - 96 n^2 + 128 n - 32) 
  + 2^{2L+1} \xi  \Big( 
-  n^{16}  +3\   n^{15}  +16  n^{14}  - 28 n^{13}  - 40 n^{12}  + 16 n^{11} 
    \nonumber \\ &
   - 434 n^{10}  -389\ 2^{2 L+1} n^9  -475  n^8 
 -1023  n^7  - 374  n^6  + 6 n^5   + 532  n^4  + 60  n^3  + 8  n^2  - 48 n  \Big) 
.}{denom_cliquering}

For large $n$, this can be expanded to give: 
\all{
b^* = \left( \fracc{L}{L-2} \right)  n - \fracc{(L+1)^2-5}{(L-2)^2} + O \left( \fracc{1}{n} \right) 
}{bign_cliquering}


\clearpage

\section{The Rich Club}

What we call the rich club is a graph with extreme core-periphery structure. 
We consider a clique of size $N_c$ as the core graph, and $N_p$ peripheral nodes. Each peripheral nodes is connected to every core node. Peripheral nodes are not connected to one another. 
So the degree of each peripheral node is $N_c$, and the degree of each core node is ${N_p+(N_c-1)}$.

It is easy to show that  natural selection does not favor cooperation in a rich-club network regardless of $b/c$.

This network comprises two sets of nodes: $m$ hubs and $N-m$ leafs. Each leaf is connected to every hub, and to no other leaf. Each hub is connected to every node in the network. In other words, there is a complete graph of $m$ nodes as a super-hub, and all the $N-m$ leaf nodes are connected to the super-hub. Denoting the hub nodes by $h$ and the leaf  nodes  by $ell$, we have 
 \all{
 \begin{cases}
 \tau_{cp} =  1 + \fracc{1}{2N_c} (N_c-1) \tau_{cc} + \fracc{1}{2(N_p+N_c-1)} \Big[ (m-1) \tau_{cp} + (n-m-1) \tau_{pp} \Big] 
 \\
 \tau_{pp} = 1 +  \frac{1}{ N_c}  ( N_c   \tau_{cp} )
 \\
 \tau_{cc} = 1 +   \fracc{1}{ (N_p+N_c-1)} \Big[  (N_c-2)  \tau_{pp} + (N_p)  \tau_{cp} \Big] 
\end{cases}
.}{core1}

Solving this system and finding $b^*$ is straightforward. Denoting the total number of nodes by $N$, we have 

\all{
 \resizebox{\linewidth}{!}{$
  b^*  =
\fracc{N_c^4+N_c^3 (6 N_p-3)+N_c^2 \big[3 N_p (4 N_p-5)+2\big]+N_c N_p \big[N_p (8 N_p-19)+8\big]+N_p \big[(7-4 N_p) N_p-3\big]}
{(N_c-1) \big[N_c \left(N_c^2-1\right)-6  N^4+(7 N_c+13) N^3-2 (N_c (N_c+6)+4) N^2+[N_c (N_c+6)+1] N\big]}
$}
}{gammacore0}

 but can be simplified if we expand the solution for large $N$:
\all{
b^*  = - \fracc{2N}{3} \fracc{2m-1}{m-1} 
+ \frac{1}{18}  \bigg[ 8m + 39 + \fracc{20}{m-1} \bigg] + O(\fracc{1}{N})  
.}{gammaCore}

Consistent with the results discussed previously, this diverges for $N_c=1$ (ordinary star), because the fraction has a pole at $N_c=1$. For any other combination of $N_c$ and $N_p$, we get a negative value for $b^*$.

Now suppose we have two rich-club graphs, one with $N_c$  core nodes and $N_p$ peripheral nodes, 
the other with $M_c$  core nodes and $M_p$ peripheral nodes. 
Suppose we connect them by attaching a core node from the first one and a core node in the second one. 
We denote these two nodes $g$, denoting `gate'. 
 The remeeting times to obtain are 
 $\tau_{p_1,p_1'}$ (between two peripheral nodes in the firs graph), 
 $\tau_{p_1,c_1}$ (between a peripheral node in the first graph and a non-gate  core node in the first graph), 
 $\tau_{p_1,g_1}$ (between a peripheral node in the first graph and the gate node of the first graph), 
$\tau_{p_1,p_2}$ (between a peripheral node in the first graph and a peripheral node in the second graph), 
$\tau_{p_1,c_2}$ (between a peripheral node in the first graph  and a core node in the second graph), 
$\tau_{p_1,g_2}$ (between a peripheral node in the first graph and the gate node of the second graph), 
 $\tau_{c_1,c_1'}$ (between two distinct core nodes in the first graph), 
 $\tau_{c_1,g_1}$ (between a non-gate core node in the first graph and the gate node of the first graph), 
 $\tau_{c_1,p_2}$ (between a core node in the first graph and the peripheral node in the second graph), 
$\tau_{c_1,c_2}$ (between a non-gate core node in the first  graph  and a non-gate core no in the second graph), 
$\tau_{c_1,g_2}$ (between a non-gate core node in the first  graph  and the gate node of the second graph), 
$\tau_{g_1,p_2}$ (between the gate node of the first graph and a peripheral node in the second graph), 
$\tau_{g_1,c_2}$ (between the gate node of the first  graph  and a non-gate core node of the second graph), 
$\tau_{g_1,g_2}$ (between the two gate nodes), 
$\tau_{p_2,p_2'}$ (between two distinct peripheral nodes in the second  graph), 
$\tau_{p_2,c_2}$ (between a peripheral node in the second graph and a non-gate core node of the second graph), 
$\tau_{p_2,g_2}$ (between a peripheral node in the second graph and the gate node of the second graph), 
$\tau_{c_2,c_2'}$ (between two distinct core nodes in the second  graph), 
and $\tau_{c_2,g_2}$ (between a non-gate core node of the second graph and the gate node of the second graph).

\all{
\begin{cases}
\tau_{p_1,p_1'}=1+\frac{(N_c-1) \tau_{p_1,c_1}+\tau_{p_1,g_1}}{N_c} 
\\ \vspace{-2mm}
\tau_{p_1,c_1}=1+\frac{\tau_{c_1,c_1} (N_c-2)+\tau_{c_1,g_1}}{2 N_c}+\frac{(N_c-2) \tau_{p_1,c_1}+(N_p-1) \tau_{p_1,p_1'}+\tau_{p_1,g_1}}{2 (N_c+N_p-1)} 
\\ \vspace{-2mm}
  \tau_{p_1,g_1}=1+\frac{\tau_{c_1,g_1} (N_c-1)}{2 N_c}+\frac{(N_c-1) \tau_{p_1,c_1}+(N_p-1) \tau_{p_1,p_1'}+\tau_{p_1,g_2}}{2 (N_c+N_p)} 
\\ \vspace{-2mm}
  \tau_{p_1,p_2}=1+\frac{\tau_{c_1,p_2} (N_c-1)+\tau_{g_1,p_2}}{2 N_c}+\frac{(M_c-1) \tau_{p_1,c_2}+\tau_{p_1,g_2}}{2 M_c} 
\\ \vspace{-2mm}
  \tau_{p_1,c_2}=1+\frac{\tau_{c_1,c_2} (N_c-1)+\tau_{g_1,c_2}}{2 N_c}+\frac{(M_c-2) \tau_{p_1,c_2}+M_p \tau_{p_1,p_2}+\tau_{p_1,g_2}}{2 (M_c+M_p-1)} 
\\ \vspace{-2mm}
  \tau_{p_1,g_2}=1+\frac{\tau_{c_1,g_2} (N_c-1)+\tau_{g_1,g_2}}{2 N_c}+\frac{(M_c-1) \tau_{p_1,c_2}+M_p \tau_{p_1,p_2}+\tau_{p_1,g_1}}{2 (M_c+M_p)} 
\\ \vspace{-2mm}
  \tau_{c_1,c_1}=1+\frac{\tau_{c_1,c_1} (N_c-3)+\tau_{c_1,g_1}+N_p \tau_{p_1,c_1}}{(N_c+N_p-1)}  
\\ \vspace{-2mm}
  \tau_{c_1,g_1}=1+\frac{\tau_{c_1,c_1} (N_c-2)+\tau_{c_1,g_2}+N_p \tau_{p_1,c_1}}{2 (N_c+N_p)}+\frac{\tau_{c_1,g_1} (N_c-2)+N_p \tau_{p_1,g_1}}{2 (N_c+N_p-1)} 
\\ \vspace{-2mm}
 \tau_{c_1,p_2}=1+\frac{\tau_{c_1,c_2} (M_c-1)+\tau_{c_1,g_2}}{2 M_c}+\frac{\tau_{c_1,p_2} (N_c-2)+\tau_{g_1,p_2}+N_p \tau_{p_1,p_2}}{2 (N_c+N_p-1)} 
\\ \vspace{-2mm}
  \tau_{c_1,c_2}=1+\frac{\tau_{c_1,c_2} (M_c-2)+\tau_{c_1,g_2}+\tau_{c_1,p_2} M_p}{2 (M_c+M_p-1)}+\frac{\tau_{c_1,c_2} (N_c-2)+\tau_{g_1,c_2}+N_p \tau_{p_1,c_2}}{2 (N_c+N_p-1)} 
\\ \vspace{-2mm}
  \tau_{c_1,g_2}=1+\frac{\tau_{c_1,c_2} (M_c-1)+\tau_{c_1,g_1}+\tau_{c_1,p_2} M_p}{2 (M_c+M_p)}+\frac{\tau_{c_1,g_2} (N_c-2)+\tau_{g_1,g_2}+N_p \tau_{p_1,g_2}}{2 (N_c+N_p-1)} 
\\ \vspace{-2mm}
  \tau_{g_1,p_2}=1+\frac{\tau_{c_1,p_2} (N_c-1)+N_p \tau_{p_1,p_2}+\tau_{p_2,g_2}}{2 (N_c+N_p)}+\frac{\tau_{g_1,c_2} (M_c-1)+\tau_{g_1,g_2}}{2 M_c} 
\\ \vspace{-2mm}
  \tau_{g_1,c_2}=1+\frac{\tau_{c_1,c_2} (N_c-1)+\tau_{c_2,g_2}+N_p \tau_{p_1,c_2}}{2 (N_c+N_p)}+\frac{\tau_{g_1,c_2} (M_c-2)+\tau_{g_1,g_2}+\tau_{g_1,p_2} M_p}{2 (M_c+M_p-1)} 
\\ \vspace{-2mm}
  \tau_{g_1,g_2}=1+\frac{\tau_{c_1,g_2} (N_c-1)+N_p \tau_{p_1,g_2}}{2 (N_c+N_p)}+\frac{\tau_{g_1,c_2} (M_c-1)+\tau_{g_1,p_2} M_p}{2 (M_c+M_p)} 
\\ \vspace{-2mm}
  \tau_{p_2,p_2}=1+\frac{(M_c-1) \tau_{p_2,c_2}+\tau_{p_2,g_2}}{M_c}  
\\ \vspace{-2mm}
  \tau_{p_2,c_2}=1+\frac{\tau_{c_2,c_2} (M_c-2)+\tau_{c_2,g_2}}{2 M_c}+\frac{(M_c-2) \tau_{p_2,c_2}+(M_p-1) \tau_{p_2,p_2}+\tau_{p_2,g_2}}{2 (M_c+M_p-1)} 
\\ \vspace{-2mm}
  \tau_{p_2,g_2}=1+\frac{\tau_{c_2,g_2} (M_c-1)}{2 M_c}+\frac{\tau_{g_1,p_2}+(M_c-1) \tau_{p_2,c_2}+(M_p-1) \tau_{p_2,p_2}}{2 (M_c+M_p)} 
\\ \vspace{-2mm}
  \tau_{c_2,c_2}=1+\frac{\tau_{c_2,c_2} (M_c-3)+\tau_{c_2,g_2}+M_p \tau_{p_2,c_2}}{(M_c+M_p-1)}  
\\ \vspace{-2mm}
  \tau_{c_2,g_2}=1+\frac{\tau_{c_2,c_2} (M_c-2)+\tau_{g_1,c_2}+M_p \tau_{p_2,c_2}}{2 (M_c+M_p)}+\frac{\tau_{c_2,g_2} (M_c-2)+M_p \tau_{p_2,g_2}}{2 (M_c+M_p-1)} 
\end{cases}
}{sys_rich}

\clearpage

The solution is too lengthy to be presentable. 
Here we only provide the solution for the symmetric, case, where $M_c=N_c$ and $M_p=N_p$. 
We represent the result   with two matrices for the polynomial coefficients of the numerator and the denominator. 
For the numerator, the $i-j$ element of the following matrix yields the coefficient of $N_c^{i-1} N_p^{j-1}$ in the numerator:  
\all{
\resizebox{\linewidth}{!}{$
\left[
\begin{array}{cccccccccccccc}
 0 & 0 & 0 & 0 & 0 & 0 & 0 & 0 & 0 & 0 & 0 & 0 & 0 & 0 \\
 0 & 0 & 0 & 2 & -16 & 56 & -120 & 168 & -144 & 64 & -8 & -2 & 0 & 0 \\
 0 & 0 & 8 & -72 & 301 & -787 & 1381 & -1633 & 1333 & -781 & 291 & -23 & -18 & 0 \\
 0 & 10 & -128 & 678 & -2173 & 4691 & -7091 & 7915 & -6610 & 3673 & -1065 & 101 & 35 & -36 \\
 4 & -104 & 794 & -3267 & 8744 & -16460 & 23351 & -25015 & 18364 & -8491 & 2630 & -365 & -227 & 42 \\
 -32 & 478 & -2826 & 9791 & -23055 & 40724 & -54479 & 50938 & -32370 & 15367 & -4575 & -202 & 97 & 144 \\
 117 & -1328 & 6625 & -20310 & 45042 & -75220 & 88791 & -73754 & 46987 & -19833 & 2605 & -1178 & 1488 & 0 \\
 -263 & 2508 & -11097 & 32195 & -68445 & 102858 & -110034 & 89955 & -48986 & 12867 & -8214 & 7048 & 0 & 0 \\
 408 & -3451 & 14504 & -41157 & 80804 & -111968 & 116070 & -79205 & 31370 & -25611 & 20256 & 0 & 0 & 0 \\
 -467 & 3760 & -15807 & 42730 & -78636 & 104212 & -88903 & 48954 & -49318 & 39401 & 0 & 0 & 0 & 0 \\
 428 & -3523 & 14610 & -37634 & 65528 & -70937 & 52755 & -64719 & 54763 & 0 & 0 & 0 & 0 & 0 \\
 -347 & 2923 & -11746 & 28386 & -40277 & 40305 & -60306 & 55941 & 0 & 0 & 0 & 0 & 0 & 0 \\
 260 & -2158 & 8084 & -15942 & 21847 & -40481 & 42515 & 0 & 0 & 0 & 0 & 0 & 0 & 0 \\
 -177 & 1363 & -4184 & 8226 & -19491 & 24035 & 0 & 0 & 0 & 0 & 0 & 0 & 0 & 0 \\
 103 & -654 & 2046 & -6575 & 9981 & 0 & 0 & 0 & 0 & 0 & 0 & 0 & 0 & 0 \\
 -46 & 302 & -1476 & 2959 & 0 & 0 & 0 & 0 & 0 & 0 & 0 & 0 & 0 & 0 \\
 20 & -198 & 593 & 0 & 0 & 0 & 0 & 0 & 0 & 0 & 0 & 0 & 0 & 0 \\
 -12 & 72 & 0 & 0 & 0 & 0 & 0 & 0 & 0 & 0 & 0 & 0 & 0 & 0 \\
 4 & 0 & 0 & 0 & 0 & 0 & 0 & 0 & 0 & 0 & 0 & 0 & 0 & 0 \\
\end{array}
\right]
$}
}{numer1344}
and for the denominator we have: 
\all{
\resizebox{\linewidth}{!}{$
 \left[
\begin{array}{cccccccccccccc}
 0 & 0 & 0 & 2 & -14 & 42 & -78 & 86 & -42 & -2 & 6 & 0 & 0 & 0 \\
 0 & 2 & -6 & -18 & 139 & -394 & 601 & -480 & 157 & 48 & -83 & 34 & 0 & 0 \\
 4 & -24 & 37 & 113 & -679 & 1465 & -1676 & 1124 & -219 & -517 & 497 & -133 & 8 & 0 \\
 -24 & 96 & -84 & -440 & 1585 & -2444 & 2615 & -1344 & -1671 & 2824 & -1271 & 202 & -68 & 24 \\
 61 & -192 & 60 & 649 & -1279 & 2275 & -2102 & -4112 & 9284 & -5802 & 1667 & -777 & 232 & 36 \\
 -85 & 212 & -148 & 506 & -396 & -445 & -7933 & 20009 & -16177 & 6941 & -3870 & 1050 & 336 & 0 \\
 68 & -200 & 910 & -2261 & 2311 & -10934 & 29547 & -29951 & 17426 & -11333 & 3016 & 1417 & 0 & 0 \\
 -39 & 382 & -1759 & 3078 & -10174 & 30336 & -38252 & 28858 & -22014 & 6249 & 3563 & 0 & 0 & 0 \\
 52 & -600 & 1825 & -6229 & 21644 & -34298 & 33086 & -30151 & 9904 & 5933 & 0 & 0 & 0 & 0 \\
 -81 & 556 & -2428 & 10561 & -21645 & 26888 & -30036 & 12202 & 6867 & 0 & 0 & 0 & 0 & 0 \\
 72 & -554 & 3379 & -9472 & 15549 & -21997 & 11585 & 5635 & 0 & 0 & 0 & 0 & 0 & 0 \\
 -57 & 642 & -2753 & 6289 & -11758 & 8305 & 3277 & 0 & 0 & 0 & 0 & 0 & 0 & 0 \\
 55 & -480 & 1696 & -4464 & 4374 & 1323 & 0 & 0 & 0 & 0 & 0 & 0 & 0 & 0 \\
 -38 & 274 & -1138 & 1630 & 353 & 0 & 0 & 0 & 0 & 0 & 0 & 0 & 0 & 0 \\
 20 & -174 & 405 & 56 & 0 & 0 & 0 & 0 & 0 & 0 & 0 & 0 & 0 & 0 \\
 -12 & 60 & 4 & 0 & 0 & 0 & 0 & 0 & 0 & 0 & 0 & 0 & 0 & 0 \\
 4 & 0 & 0 & 0 & 0 & 0 & 0 & 0 & 0 & 0 & 0 & 0 & 0 & 0 \\
\end{array}
\right]
$}
}{denom1342}

We can simplify the results in limiting cases. 
For large $N_p$, we can use the following expansion: 
\all{
b^*
=
 \left(4 N_c-\frac{3}{2}\right)
 +
\left[ 4 N_c^2-\frac{47 N_c}{4}-\frac{1}{4 N_c}-\frac{45}{3 N_c+2}+\frac{75}{4} \right] \fracc{1}{N_p}
+ O\left(\fracc{1}{N_p^2}\right)
.}{Gamma_large_samerich}

In the limit as the number of peripheral nodes approaches infinity, the critical benefit to cost ratio only depends on the number of core nodes. 

For example,  we put $N_c=1$, we recover Equation~\eqref{double_star_1_identical}, which pertains to two stars connected via hubs. 
If we set $N_c=2$, we get
\all{
b^*= 
\fracc{13}{2} + \fracc{11}{2N_p} + O\left( \fracc{1}{N_p^2} \right)
,}{riches_Nc2}
that is, $b^*$ approaches $13/2$ in the limit as $N_p\rightarrow \infty$.


\clearpage

\section{The complete bipartite graph}

A   bipartite graph is one whose nodes can be divided into to disjoint subsets such that there is no link within either of them, and every link in the graph connects a node to one of the subsets to a node in the other. 
A complete bipartite graph is a bipartite graph in which each node in the first subset is connected to every node in the other subset. 
A star graph is an example of a complete bipartite graph with the first subset being a single node, and the second subset  comprising all the leafs. 

Suppose we have two subsets, with sizes $n_x$ and $n_y$. 
The remeeting times follow the following relations: 
\al{
\begin{cases}
\tau_{xx}=1+\tau_{xy}\\
\tau_{yy}=1+\tau_{xy}\\
\tau_{xy}=1+\frac{1}{2n_y}\big[ (n_y-1)\tau_{yy}\big]+\frac{1}{2n_x}\big[ (n_x-1)\tau_{xx}\big] 
\end{cases}
}

The solution is 
\al{
\tau_x=\tau_y= \frac{4n_xn_y}{n_x+n_y}
.}

It is straightforward to check that  if we insert these values  into~\eqref{Gamma}, the denominator becomes zero. 
An alternative approach is taken in Ref (19) of the main text.

\begin{figure}[h]
	\centering
	\includegraphics[width=.4\linewidth]{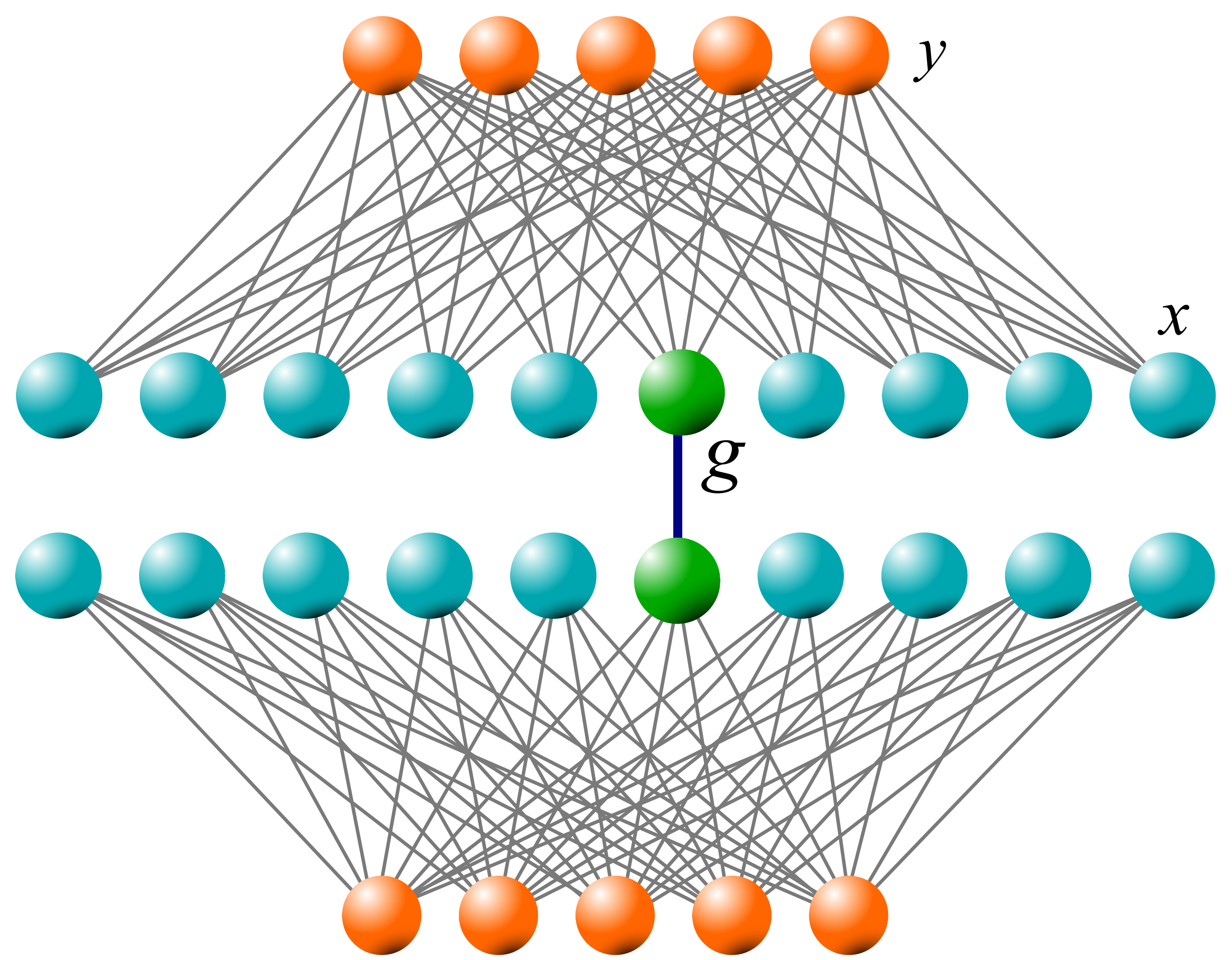}
	\caption{Connected complete bipartite graphs with $n_x=10$ and $n_y=5$ }
	\label{bipartite}
\end{figure}

Now suppose we have two identical  complete bipartite graphs, each with $n_x$ and $n_y$ nodes as above. 
We connect one $x$ node of the first graph to one $x$ node of the second graph. 
So the degree of the two gate nodes are $n_y+1$. The degree of non-gate $x$ nodes are $n_y$, and the degree of $y$ nodes are $n_x$. 
An example case with $n_x=10$ and $n_y=5$ is illustrated in Figure~\ref{bipartite}. 
The remeeting times of interest are
$\tau_{x,x'}$  (between two non-gate $x$ nodes of the same graph), 
$\tau_{x,x''}$  (between a  non-gate $x$ node of  one  graph and a non-gate $x$ node of the other graph), 
$\tau_{y,y'}$  (between two $y$ nodes of the same graph), 
$\tau_{y,y''}$  (between an  $y$ node of  one graph and a  $y$ node of  other   graph), 
$\tau_{x,y}$  (between  a non-gate $x$ node of a graph and a $y$ node in the same graph), 
$\tau_{x,y'}$  (between a non-gate $x$ node in one graph and a $y$ node in the other graph), 
$\tau_{g,x}$  (between  the gate node of a graph and a non-gate $x$ node of the same graph), 
$\tau_{g,x'}$  (between  the gate node of a graph and a non-gate $x$ node of the  other  graph), 
$\tau_{g,y}$  (between  the gate node of a graph and a non-gate $y$ node of the same graph), 
$\tau_{g,y'}$  (between  the gate node of a graph and a non-gate $x$ node of the other graph), 
 and $\tau_{g,g'}$  (between the two gate nodes),

\al{
\begin{cases}
\tau_{x,x'}=1+\tau_{x,y}\\
\tau_{x,x''}=1+\tau_{x,y'}\\
\tau_{y,y'}=1+\frac{1}{n_x} \big[ (n_x -1)\tau_{x,y} + \tau_{g,y}\big] \\
\tau_{y,y''}=1+\frac{1}{n_x} \big[(n_x-1) \tau_{x,y'}+\tau_{g,y'} \big] \\
\tau_{x,y}=1+\frac{1}{2n_y} \big[ (n_y-1)\tau_{y,y'}\big] +\frac{1}{2n_x} \big[  (n_x-2)\tau_{x,x'}+\tau_{g,x} \big]\\
\tau_{x,y'}=1+\frac{1}{2} \tau_{y,y''} +\frac{1}{2} \big[   (n_x-1)\tau_{x,x''} + \tau_{g,x'}\big]\\
\tau_{g,x}=1+\frac{1}{2} \tau_{g,y}+\frac{1}{2(n_y+1)} \big[ n_y\tau_{x,y} +\tau_{g,x'}  \big]\\
\tau_{g,x'}=1+\frac{1}{2(n_y+1)} \big[ n_y\tau_{x,y'}+\tau_{g,x}\big] +\frac{1}{2} \tau_{g,y'}\\
\tau_{g,y}=1+\frac{1}{2n_x} \big[ (n_x-1)\tau_{g,x}\big] +\frac{1}{2(n_y+1)} \big[  (n_y-1)\tau_{y,y'}+\tau_{g,y'} \big]\\
\tau_{g,y'}=1+\frac{1}{2(n_y+1)} \big[ n_y\tau_{y,y''}+\tau_{g,y}\big] +\frac{1}{2n_x} \big[  (n_x-1)\tau_{g,x'}+\tau_{g,g'} \big]\\
\tau_{g,g'}=1+\frac{1}{n_y+1} \big[ n_y\tau_{g,y'}\big] 
\end{cases}
}

The result is
\al{
b^*=\frac{\alpha}{\beta}
,}
where the numerator is given by
\al{
& \alpha=n_1 (n_2+1)^2 \bigg[ 2 n_1^4 n_2 (n_2+1) (3 n_2+2) (3 n_2+4) (8 n_2-1)
\nonumber \\ &
+n_1^3 n_2 (42 n_2^4+474 n_2^3+945 n_2^2+623 n_2+126) 
\nonumber \\ &
-n_1^2  (36 n_2^5+235 n_2^4+263 n_2^3+24 n_2^2-24 n_2+12) 
\nonumber \\ &
-n_1 (3 n_2+2) (6 n_2^3+43 n_2^2+25 n_2-6) -2  (n_2^2 + 9n_2 +6) \bigg],
}

and the denominator is given by
\al{
& \beta=4 n_1^5 n_2 (n_2+1) (3 n_2+2) (3 n_2+4) (n_2 (n_2+2)+2)
\nonumber \\ &
+2 n_1^4  (18 n_2^7+102 n_2^6+274 n_2^5+435 n_2^4+429 n_2^3+316 n_2^2+204 n_2+64) 
\nonumber \\ &
+n_1^3 (24 n2^7 + 84 n2^6 + 20 n2^5 - 155 n2^4 - 240 n2^3 - 467 n2^2 - 
 538 n2 - 192) 
\nonumber \\ &
+  
n_1^2 (8 n_2^6-89 n_2^5-531 n_2^4-848 n_2^3-412 n_2^2+56 n_2+60)
\nonumber \\ &
+n_1 (34 n_2^5+127 n_2^4+138 n_2^3+39 n_2^2+2 n_2+4)
\nonumber \\ &
+2 n_2 (n_2+1) (3 n_2 (n_2+3)+4)
}

For $n_x \ll n_y$, we can use the following expansion
\al{
b^* = 4n_x - \frac{3}{2} + \left( 14 - \frac{1}{4n_x} + n_x-4 n_x^2 - \frac{45}{3n_x+2} \right) \fracc{1}{n_y} + O \left(\fracc{1}{n_y^2}\right)
.}
Note that for $n_x=1$, we get the case of two stars connected via hubs, for which we showed that $b^* $ approaches $5/2$, as this new result reaffirms.

On the other hand, if we set $n_y=n_y=n$, we get
\al{
b^* \bigg|_{n_x=n_y=n} = 2n-1+\fracc{3}{n} + O \left(\fracc{1}{n^2}\right)
}


%
%
%
%
%
%
%
%
%
%
%
%
%
%
%
%
%
%
%
%
%
%
%
%
%
%
%

\clearpage

%
%

\section{Conjoining scale-free networks}\label{SI:SF}

\begin{figure}[t]
\centering
\includegraphics[width=.98 \columnwidth]{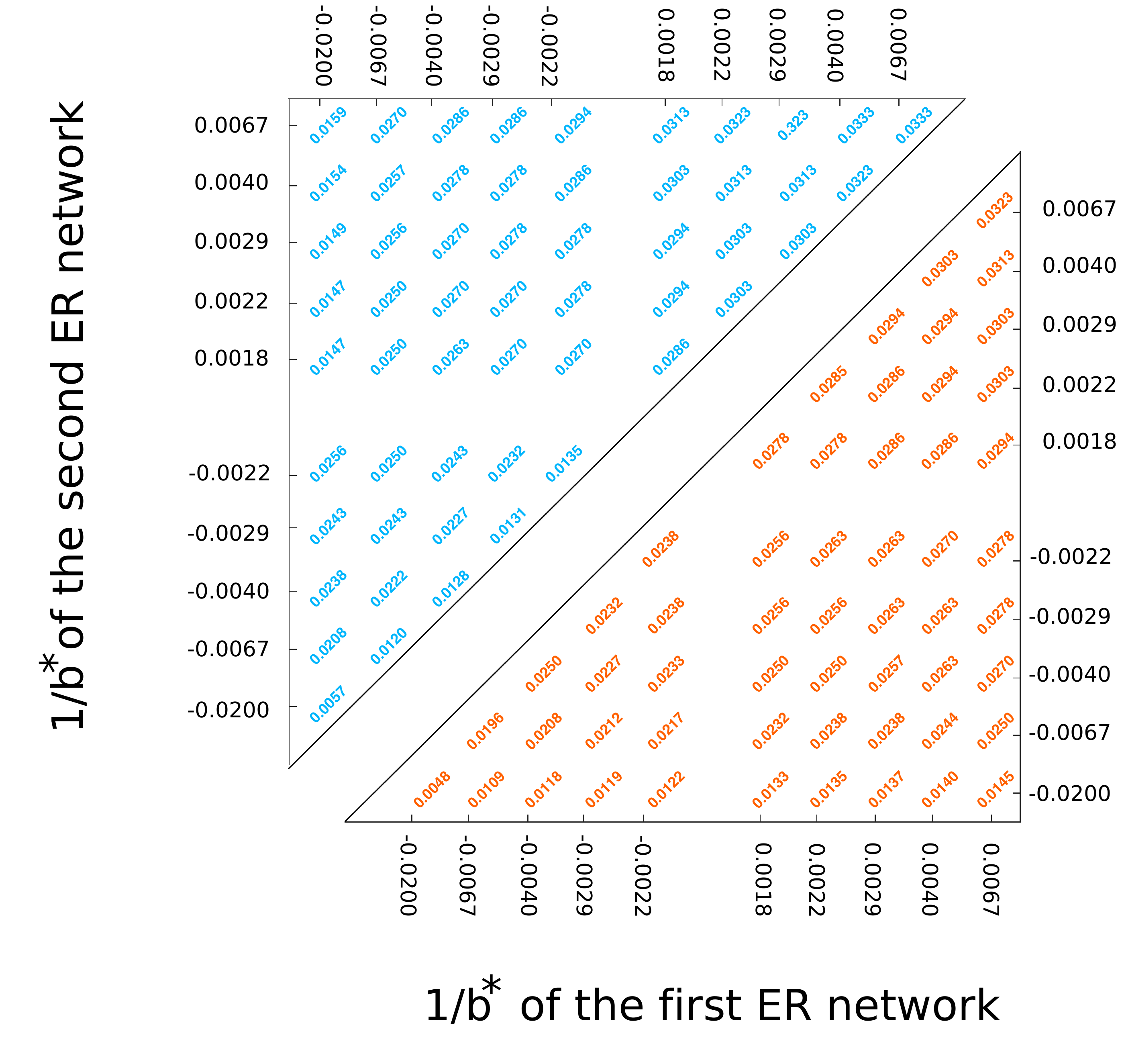}
\caption  
{
\footnotesize{
Conjoining two ER networks, the results being the same as those in Fig.~5a of the main text, but here we use $1/b^*$ to characterize the merit for cooperation instead of $b^*$. 
}
} \label{ER_inv}
\end{figure}

\begin{figure}[t]
\centering
\includegraphics[width=.98 \columnwidth]{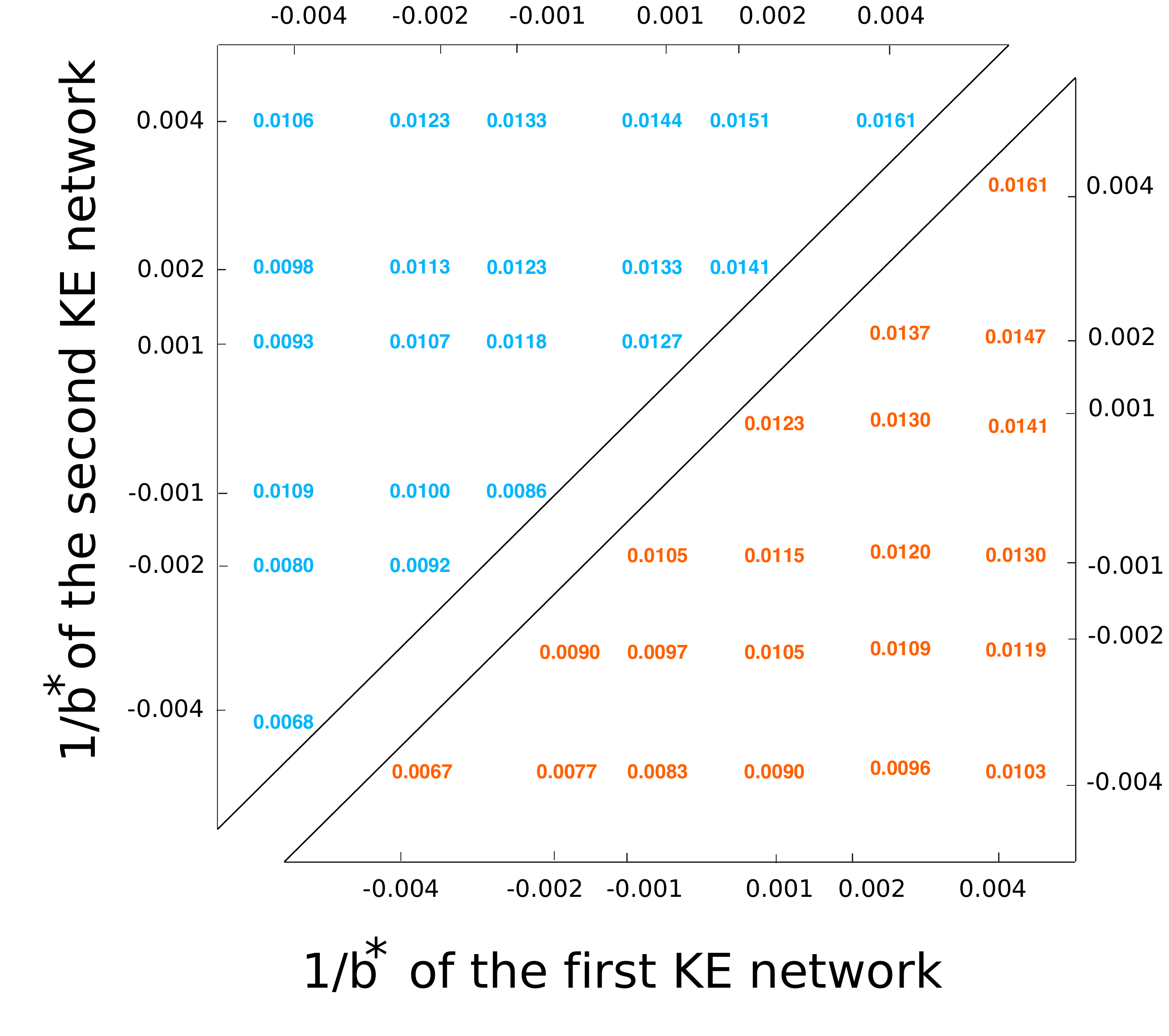}
\caption  
{
\footnotesize{
Conjoining two KE networks, the results being the same as those in Fig.~5c of the main text, but here we use $1/b^*$ to characterize the merit for cooperation instead of $b^*$. 
}
} \label{KE_inv}
\end{figure}

In the main text, Fig.~5, we presented the results for conjoining two ER networks, as well as two KE scale-free networks. 
In Figure~\ref{ER_inv}, we present the same results of Fig.~5a of the main text (which pertained to conjoining of ER networks) but with $1/b^*$ instead of $b^*$. 
In Figure~\ref{KE_inv}, we represent the results of Fig.~5c of the main text (which pertained to the conjoining of KE networks). 

In this section, we present results for the same conjoining procedure applied to three additional scale-free network models that produce networks with heavy-tailed degree distributions that exhibit structural properties that actual social networks possess. 
The first model is the model of Holme and Kim~\cite{holme2002growing} (HK), which produce networks with power-law degree distribution and high clustering. 
The second model is the Forest Fire (FF)  model of Leskovec et al.~\cite{leskovec2005graphs}, which in addition to the above properties, exhibits densification. 
The third model is Barth{\'e}lemy's  spatial scale-free model  (SSF) which combines preferential attachment with distance selection~\cite{barthelemy2003crossover}. 
 The fourth model is preferential attachment  (PA) with initial attractiveness~\cite{barabasi2016network}. 
For the first thee models, we generate five networks of size 100    whose $b^*$ values equaled -1000, -500, -250, 250, 500, 1000 (within a 5\% error margin). 
For the preferential attachment model, the generated graphs have typically better $b^*$ values and it we did not find an instance with negative $b^*$. The       $b^*$ values of the two graphs under the PA model were 50, 100, 150, 200, and 250. 

\begin{figure}[t]
\centering
\includegraphics[width=.98 \columnwidth]{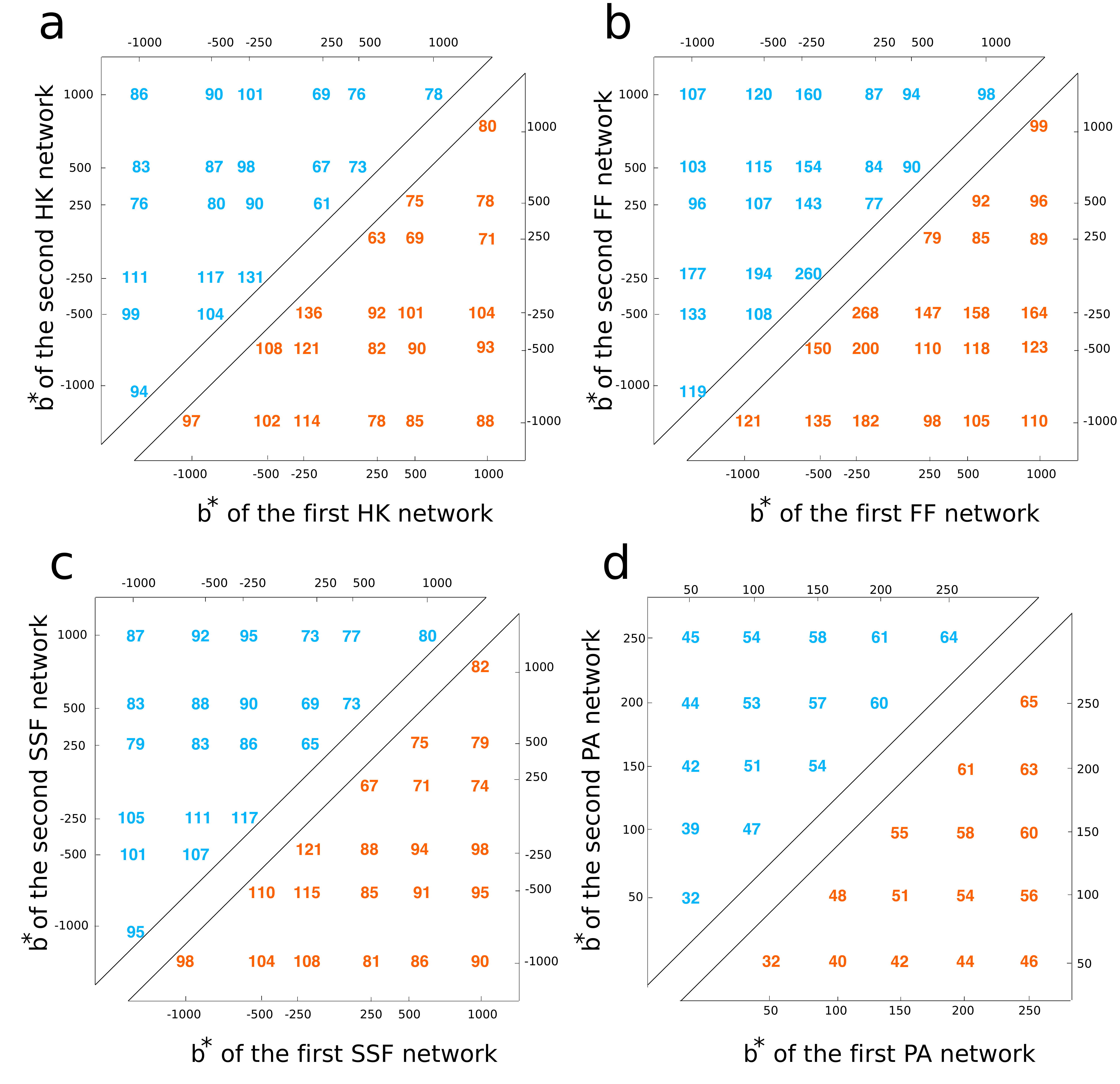}
\caption  
{
\footnotesize{
Conjoining two scale-free networks. 
(a) The model of Holme and Kim, 
(b) The Forest-Fire model of Leskovec et al.,
(c) The Spatial Sale-free model of Barthelemy. 
(d) Preferential attachment model with initial attractiveness. 
In all cases, the blue numbers (upper triangle) represent the median $b^*$ value for interconnection of the two networks via an intermediary broker nodes, 
and the orange numbers (lower triangle) pertain to direct interconnections. 
}
} \label{SI_Conjoining_SFs}
\end{figure}

\begin{figure}[t]
\centering
\includegraphics[width=.98 \columnwidth]{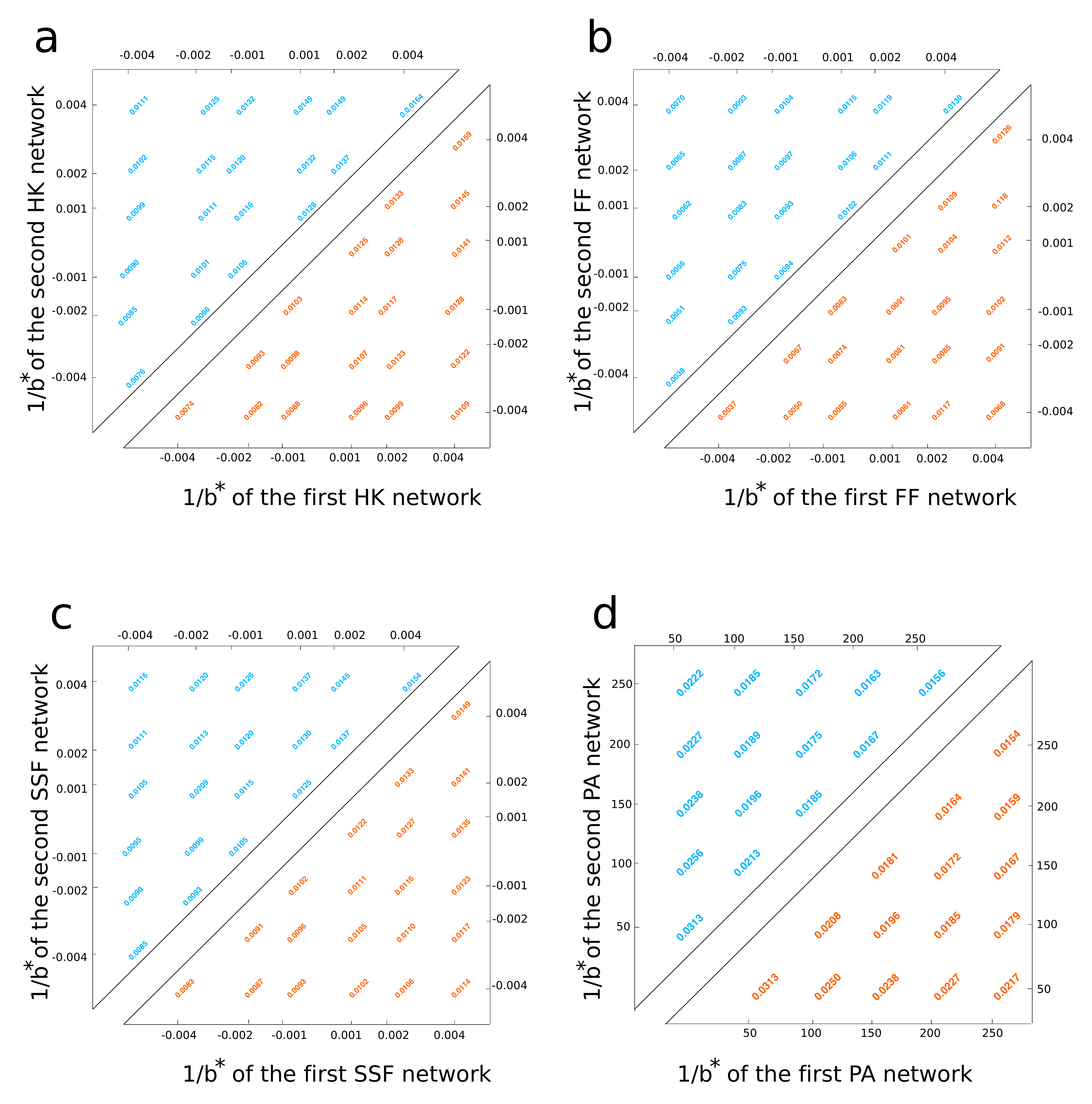}
\caption  
{
\footnotesize{
Same results as in Figure~\ref{SI_Conjoining_SFs}, but using $1/b^*$ instead of $b^*$. 
}
} \label{SI_Conjoining_SF_b_inverse}
\end{figure}

Then, for each pair of possible networks, we calculated the median $b^*$ of the composite network which results from creating a link between a node chosen from the first network and another chosen from the second network. We calculated the median value of all these $b^*$ values, and assigned it to that pair of networks. We repeated the same procedure for interconnections via one intermediary broker node as well. 
Figure~\ref{SI_Conjoining_SFs} presents the results.
Figure~\ref{SI_Conjoining_SF_b_inverse} presents the results using $1/b^*$ instead of $b^*$. 
In all cases, the $b^*$ of the composite network is better than those of the individual networks, 
and conjoining via one intermediary broker node is better than direct interconnection without an intermediary node.

%
%
%
%
%
%
%
%
%
%
%
%
%
%
%
%
%
%
%
%
%
%
%
%
%
%
%

\clearpage

\section{Community structure}\label{SI:community}

We found in this paper that dense communities promote spite, and connecting them promotes cooperation. 
We demonstrated that connections with intermediary broker nodes are better at constructing composite networks that promote cooperation as compared to direct interconnections with no intermediary nodes. 

Since sparsely interconnected cohesive groups is closely related to the  notion of community structure in the network science literature, 
here we focus on two standard frameworks for producing networks with community structure: LFR benchmarks~\cite{lancichinetti2008benchmark}, and Stochastic Block Models~\cite{decelle2011asymptotic}. 
 Employing these two frameworks, in this section we investigate the role of community structure on the evolution of cooperation.

For the LFR benchmark, we used the publicly-available code that the authors of Ref.~\cite{lancichinetti2008benchmark} have provided: 
\\
\url{https://sites.google.com/site/santofortunato/inthepress2}

We generated $10^5$ networks.  For every input parameter, we selected a value from the possible range uniformly at random, and let the algorithm decide whether a network can be constructed from the given set of parameters. The network size is $N=100$. The mixing parameter was chosen uniformly at random in $[0,1]$, the maximum  degree $k_{\text{max}}$ was chosen  uniformly from the set of integers in the $(1,N]$ interval, the average degree was  chosen uniformly from the set of integers in the $(1,k_{\text{max}}]$ interval, the degree exponent was uniformly chosen from the $[0,3]$ interval, the exponent for the community size distribution was chosen uniformly in the $[0,1]$ interval. 
We use network modularity~\cite{newman2006modularity} to quantify how strongly networks are divided into communities. 
 Stronger division into communities pertains to higher values of  modularity.

\begin{figure}[t]
\centering
\includegraphics[width=.8 \columnwidth]{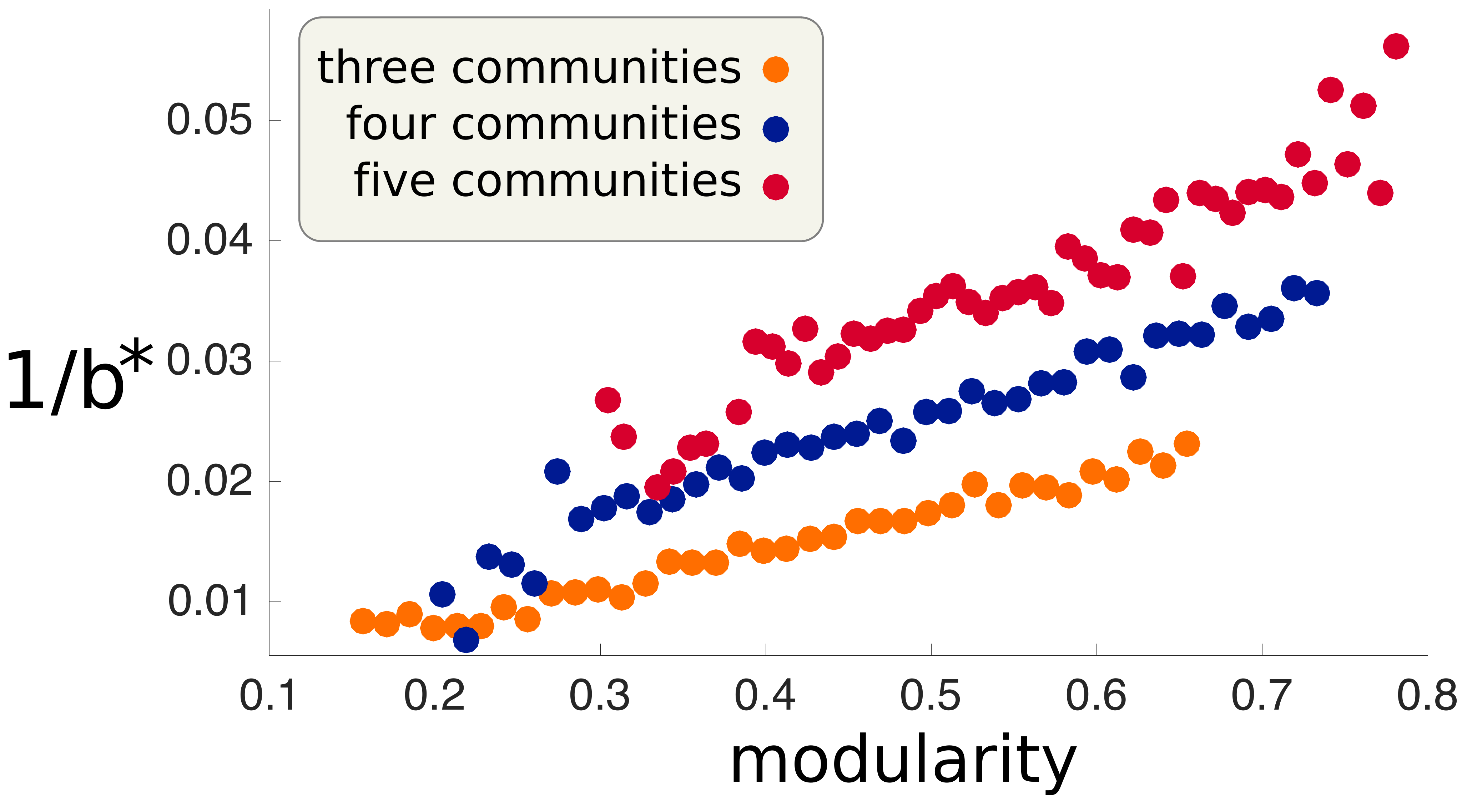}
\caption  
{
\footnotesize{
The effect of community structure on the evolution of cooperation for LFR benchmark graphs. 
Each marker represents the average value of data points falling in the corresponding bin of for modularity.
}
} \label{LFR_bc_inverse_vs_modularity}
\end{figure}

Fig.~6a shows the mean value of $1/b^*$ for different values of modularity. 
We plotted $1/b^*$ instead of $b^*$ merely because the patterns were visually better discernible.
As modularity increases, $b^*$ decreases. 
The number of communities also affects $b^*$. 
For fixed size and given modularity, greater number of communities leads to less $b^*$, hence more conduciveness to the evolution of cooperation.

\begin{figure}[t]
\centering
\includegraphics[width=.8 \columnwidth]{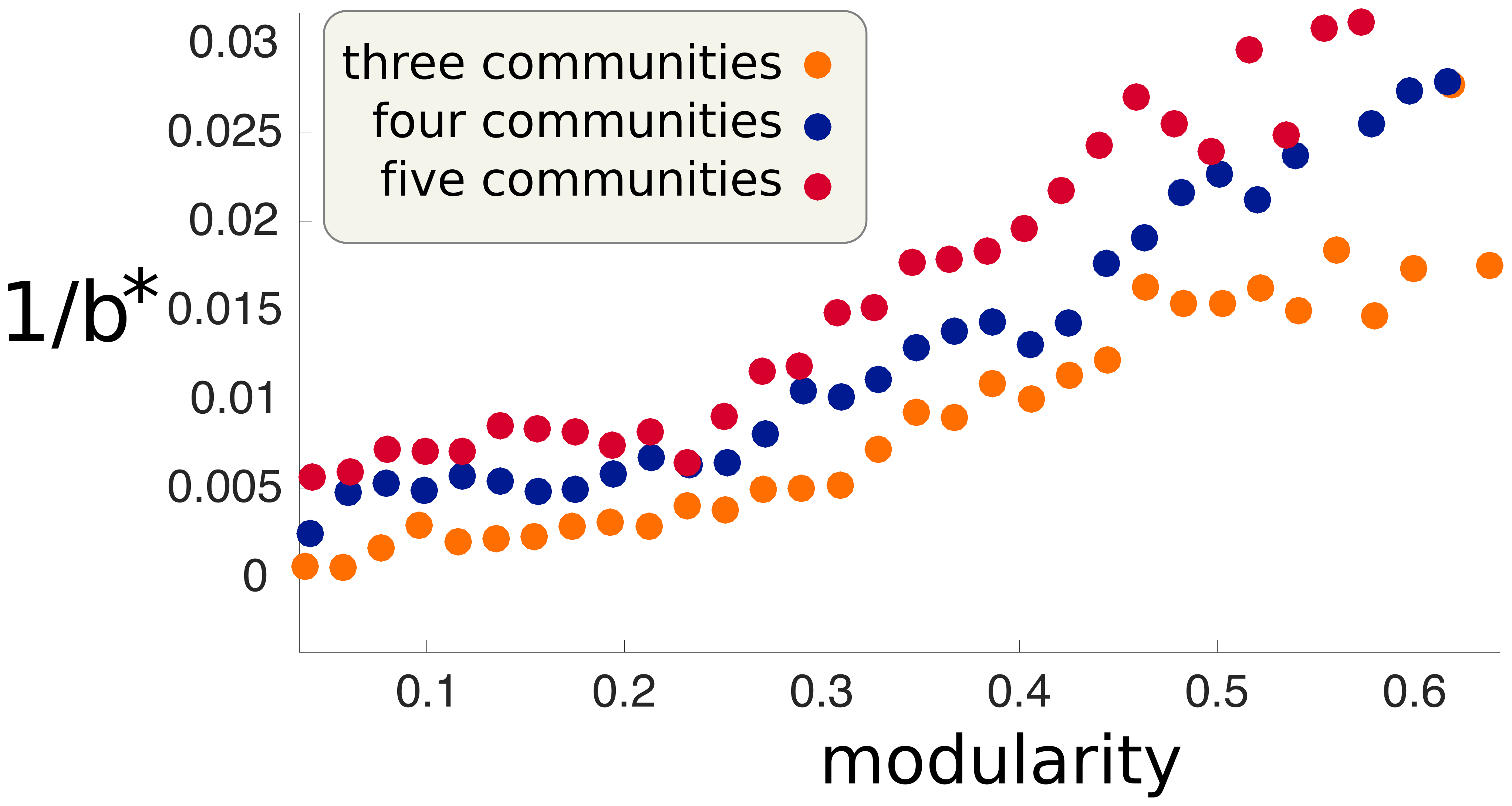}
\caption  
{
\footnotesize{
The effect of community structure on the evolution of cooperation for SBM benchmark graphs. 
Each marker represents the average value of data points falling in the corresponding bin of for modularity.
}
} \label{SBM_bc_inverse_vs_modularity}
\end{figure}

Stochastic Block Models (SBM)  is another standard modeling framework for generating networks with community structure and also for inferring community structure~\cite{decelle2011asymptotic}. 
This model simply involves two parameters: $P_{\text{within}}$ (the within-community link probability) and $P_{\text{between}}$ (the between-community link probability). 
For fixed network size $N=100$, we generated $10^6$ networks. 
Both $P_{\text{within}}$ and $P_{\text{between}}$ are selected uniformly at random from the interval $[0,1]$. 
Figure~\ref{SBM_bc_inverse_vs_modularity} depicts $1/b^*$ in terms of  modularity. 
Consistent with the above results, higher modularity leads to lower values of $b^*$. 
Similar to the case of LFR, we  plotted $1/b^*$ instead of $b^*$ simply because the patterns were visually better discernible

\begin{figure}[t]
\centering
\includegraphics[width=.7 \columnwidth]{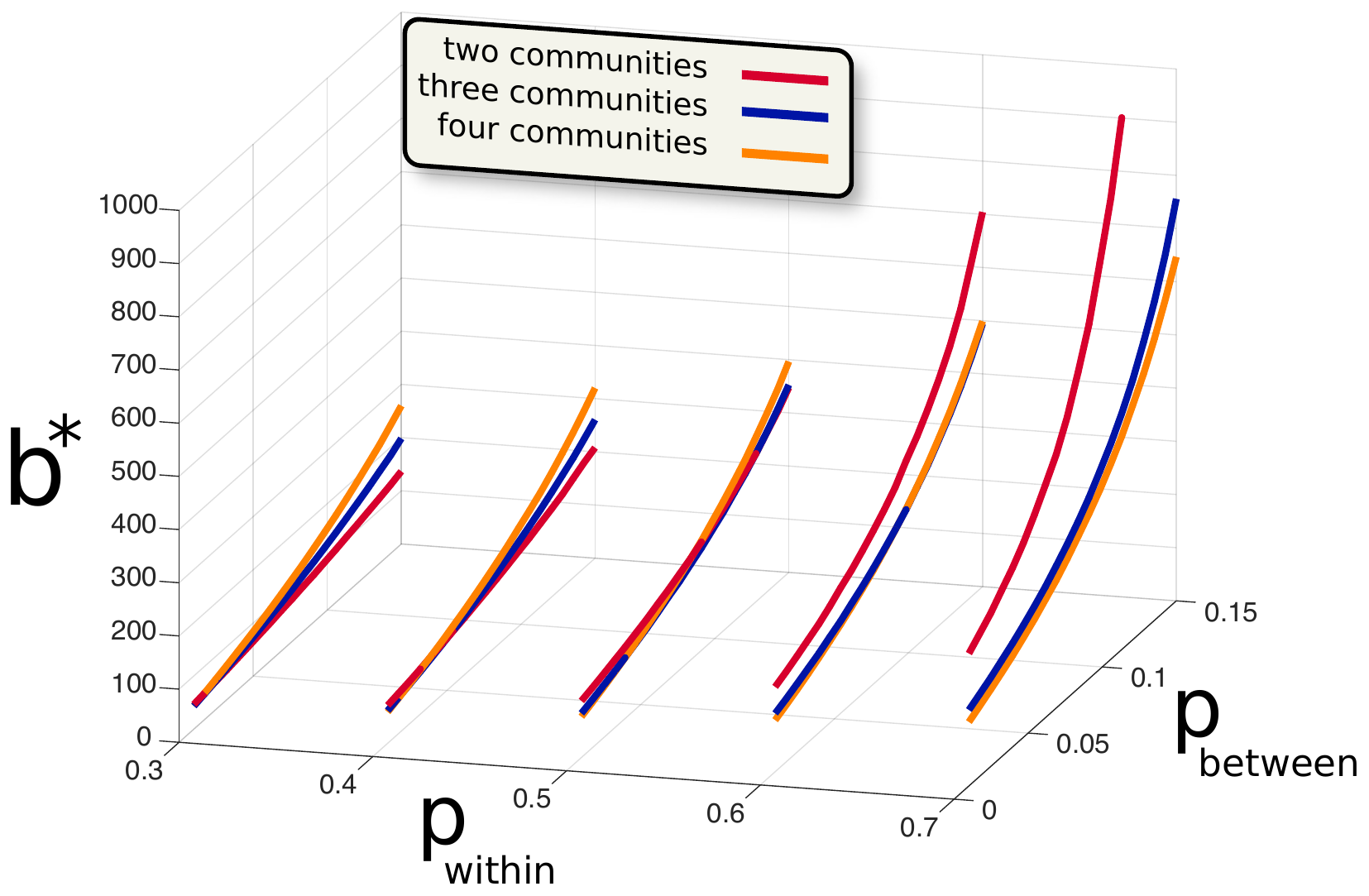}
\caption  
{
\footnotesize{
 Results for SBM networks. 
 When the communities have small within-community density, $b^*$ grows slowly by adding between-community links, as compared to dense communities. 
 Moreover, the greater the number of communities are, this increase with between-community link density becomes steeper: For $P_{\text{within}} =0.3$, the curve pertaining to four communities has the highest $b^*$, but for $P_{\text{within}} =0.7$, the trend is reversed and the curve for two communities has the highest $b^*$ and the steepest increase with $P_{\text{between}} $.
}
} \label{SBM_conjoining_PinPout}
\end{figure}

 The second experiment on SBMs that we report is to investigate how $b^*$ depends on $P_{\text{within}}$ and $P_{\text{between}}$ . 
For the range $P_{\text{within}} \in [0.3,0.7]$ and $ P_{\text{between}} \in (0,0.15]$, we plotted the expected value of $b^*$ averaged over $10^4$ realizations. 
We consider the cases of two, three, and four communities, each with size 100. 
The results are depicted in Figure~\ref{SBM_conjoining_PinPout}. 
We find that  for dense communities (which, as shown previously, support spite individually), sparse interconnections correspond to lower values of $b^*$. 
The results demonstrate that adding  too many interconnections between the communities are not beneficial.

\clearpage
 
  \section{Robustness analysis  }\label{SI:robustness}
  The above-considered topologies were ideal types amenable to exact analytical treatment. 
  More realistic scenarios of course involve more randomness. 
We test if the above-considered topologies still produce reasonably-low values of $b^*$ in presence of noise. 
That is, do  topologies close to those we considered  possess fairly similar values of $b^*$ as we calculated? or do a small amount of structural noise vary the results markedly?

\begin{figure}[h]
\centering
\includegraphics[width=.98 \columnwidth]{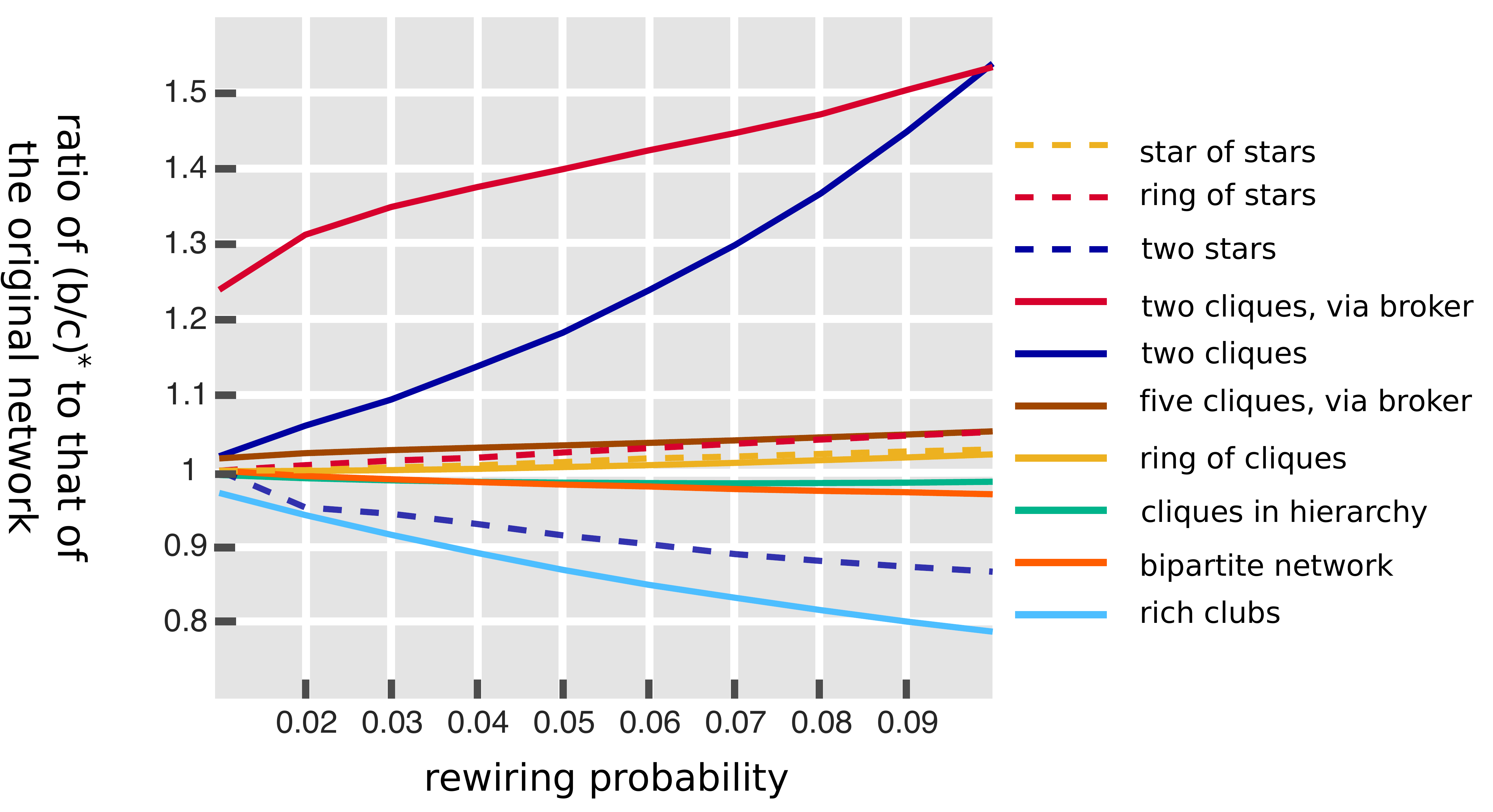}
\caption  
{
\footnotesize{
The robustness of the presented $b^*$ values for the structures discussed, as a function of the   rewiring probability 
}
} \label{rewire}
\end{figure}

For fair comparison, we retain the number of links (so that density would be intact), and rewire each link with probability $p$, and then calculate the ratio of the  $b^*$ of the rewired network to that of the original network. 
Since there are many ways to perform the rewiring, we consider the median value of the said ratio for given rewiring probability. 
Figure~\ref{rewire} presents the median  ratio of  $b^*$ to that of the original network as a function of the rewiring probability $p$ for the  topologies considered above.
In all cases, the distribution of this ratio is fairly close to unity. 
This demonstrates  that in the presence noise, the noisy networks still exhibit the cooperative merit of the original composite networks.

\clearpage

\section{Simulation results}\label{SI:simulation}
\begin{figure}[t]
\centering
\includegraphics[width=.5 \columnwidth]{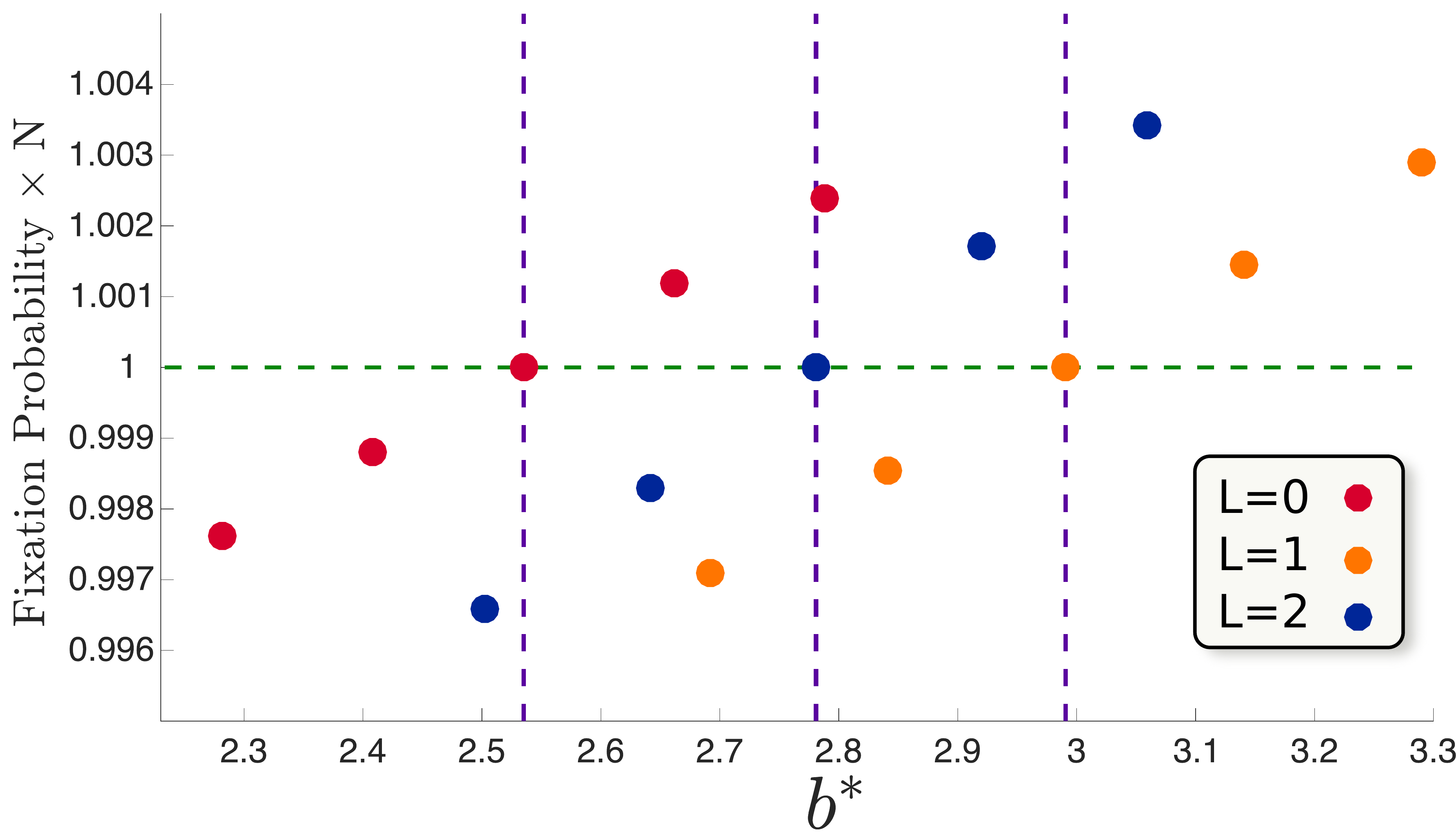}
\caption  
{
\footnotesize{
 Simulation results for conjoining two stars, each with 50 leaves. Hubs are connected via zero, one, and two intermediary nodes. 
}
} \label{simul_2stars}
\end{figure}

\begin{figure}[t]
\centering
\includegraphics[width=.5 \columnwidth]{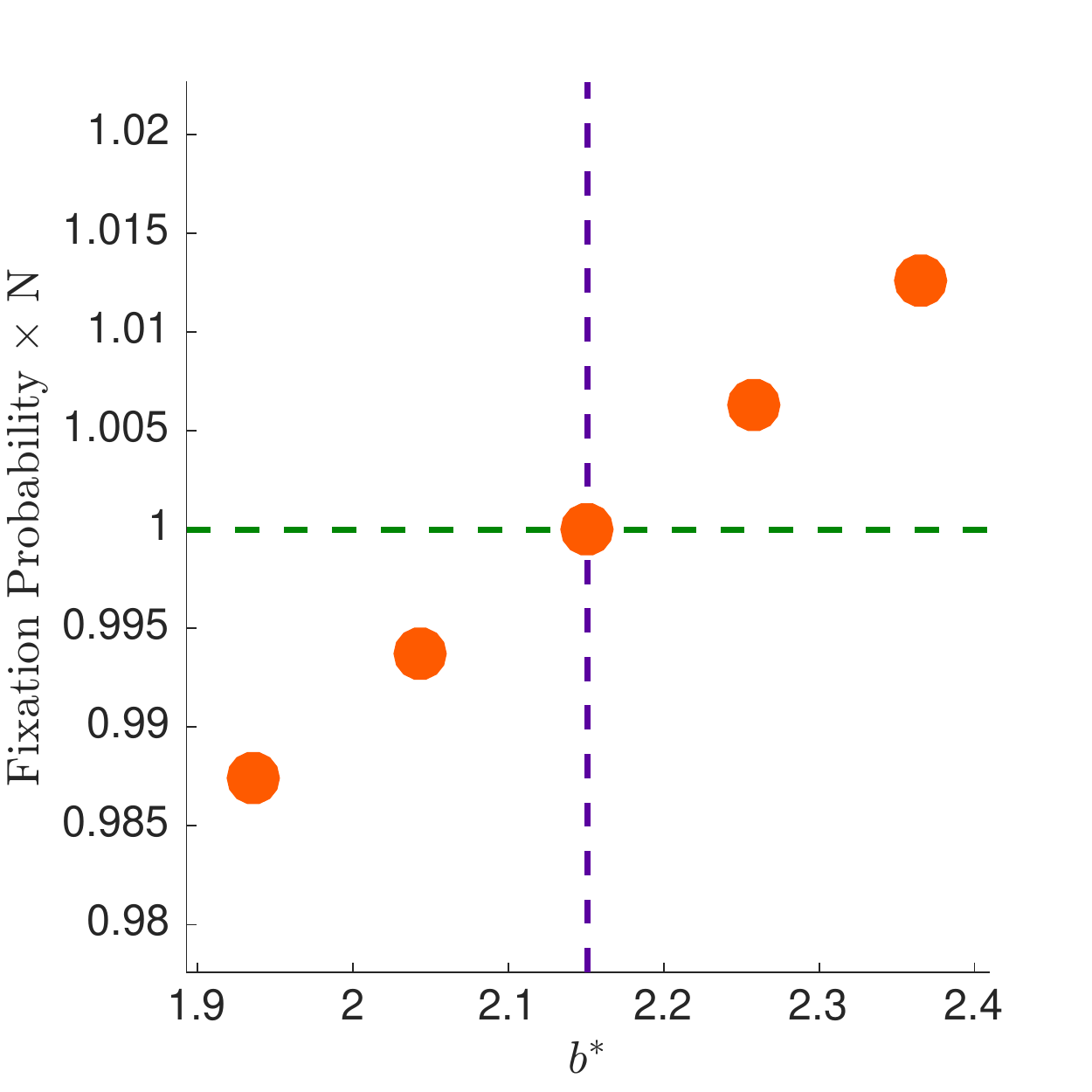}
\caption  
{
\footnotesize{
 Simulation results for the imperfect star of stars:  for  a star graph with 15 leaves, we select five leaf nodes and to each of them connect 10 new leaf nodes.
 The theoretical prediction is $b^*\approx 2.15$. 
}
} \label{simul_imperfect_starOfstars}
\end{figure}

\begin{figure}[t]
\centering
\includegraphics[width=.5 \columnwidth]{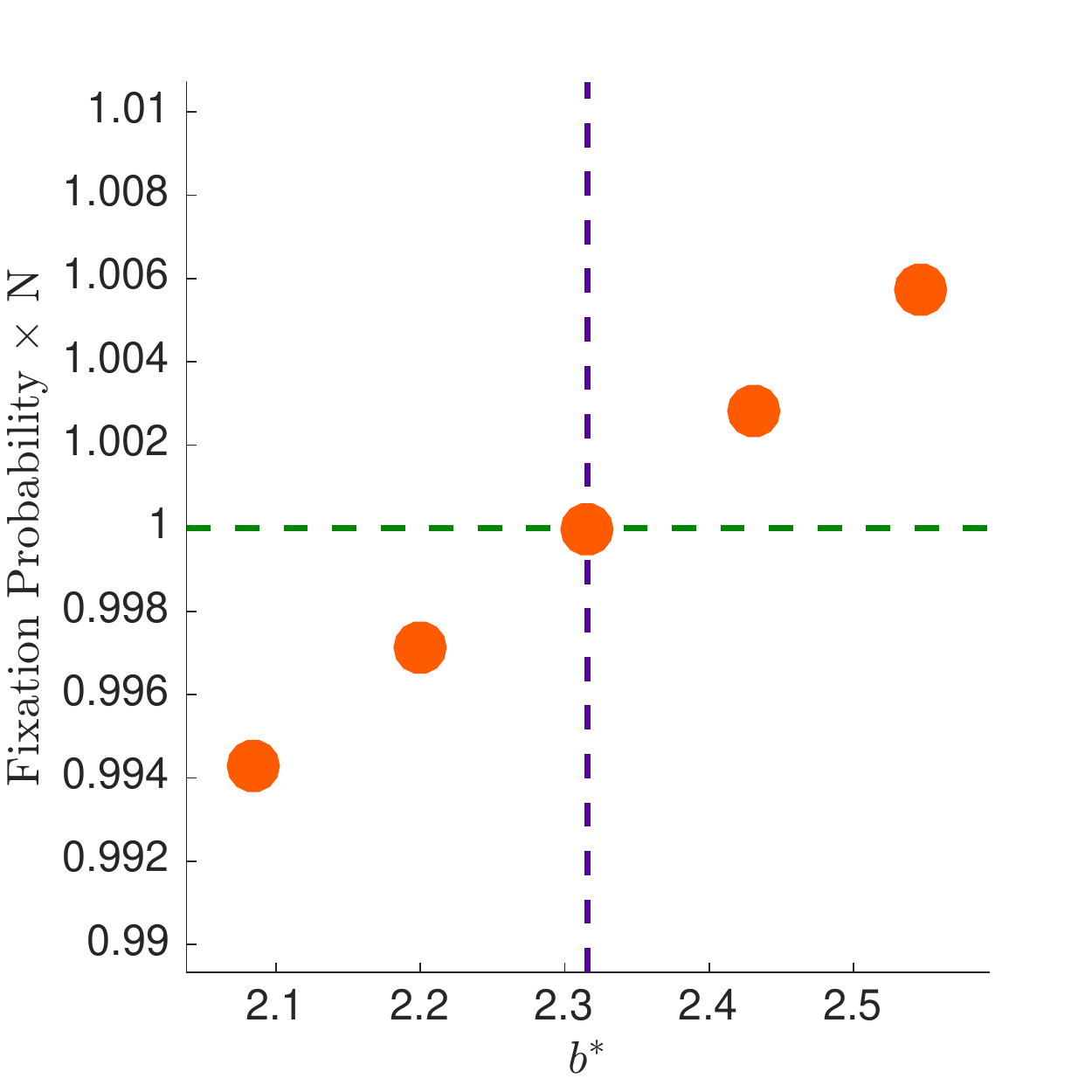}
\caption  
{
\footnotesize{
 Simulation results for star of stars: 
We start from a star graph with 5  leaves, and to each leaf node we attach 5 new nodes. 
 The theoretical prediction is $b^* \approx 2.32$. 
}
} \label{simul_star2}
\end{figure}

\begin{figure}[t]
\centering
\includegraphics[width=.5 \columnwidth]{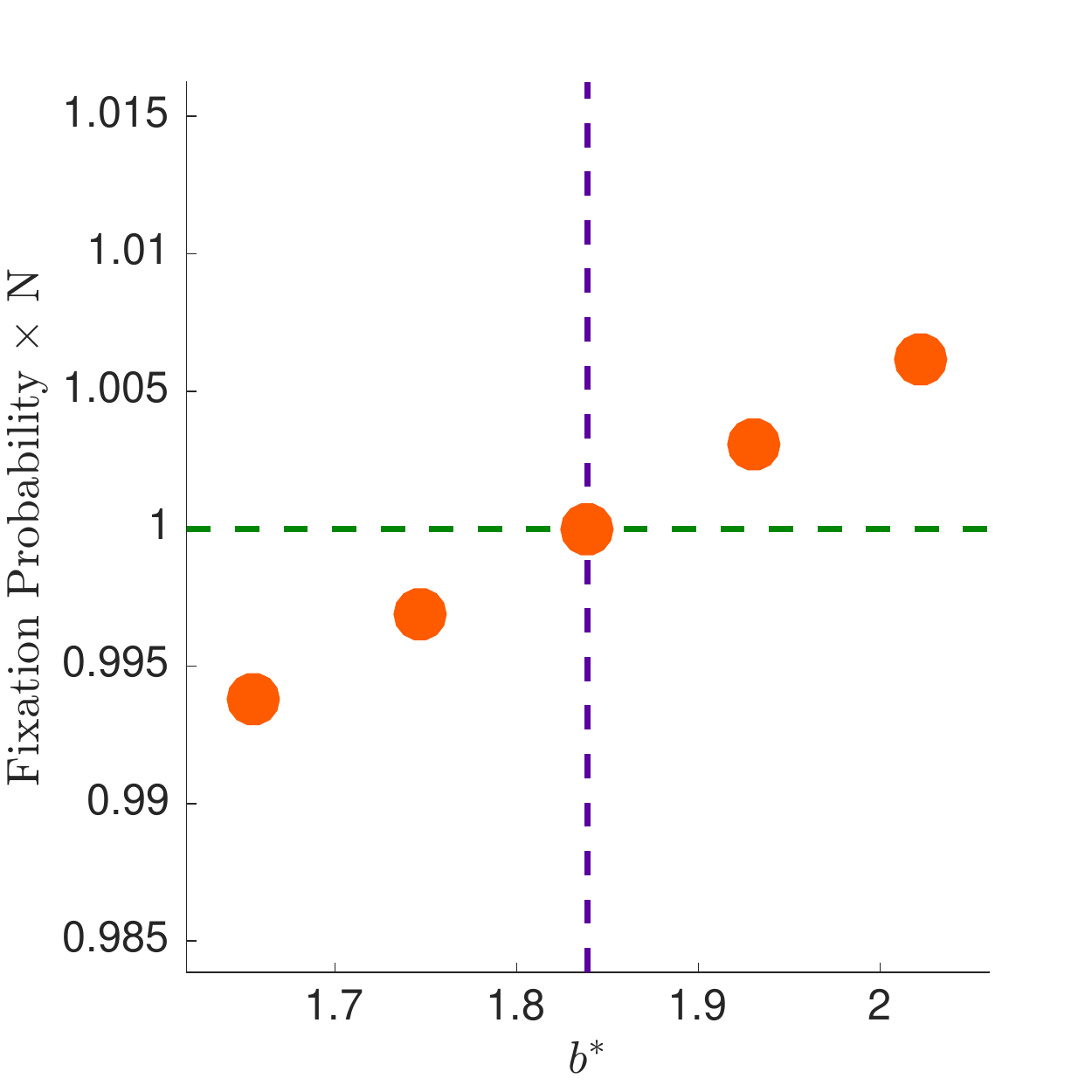}
\caption  
{
\footnotesize{
 Simulation results for the ring of five stars, each with 20 leaf nodes. 
 The theoretical prediction is $b^* \approx 1.84$. 
 It is notable that the average degree is 2, and $b^*<2$. This is one of the \emph{superpromoter} graphs discussed in the text. 
}
} \label{simul_ringOfStars}
\end{figure}

\begin{figure}[t]
\centering
\includegraphics[width=.5 \columnwidth]{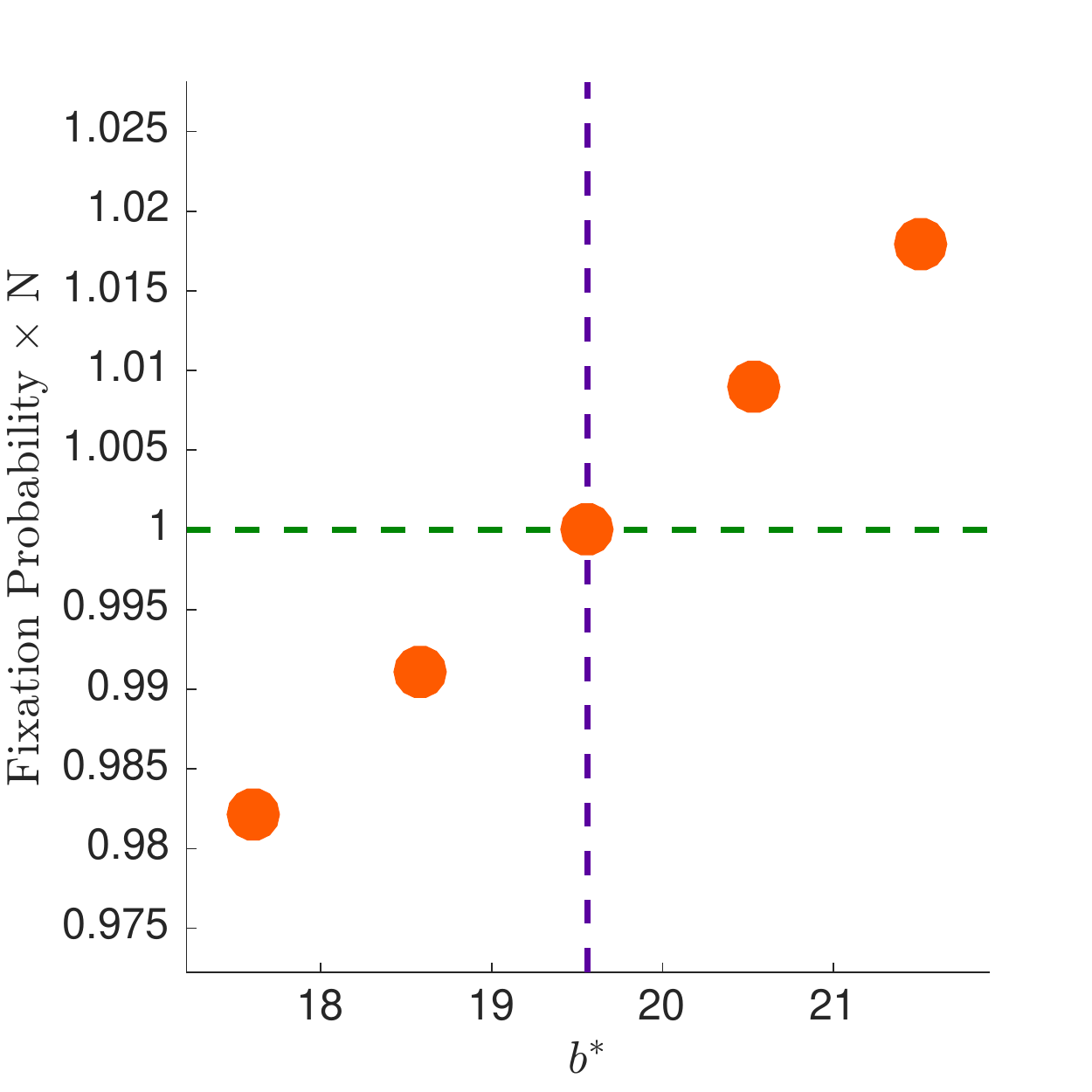}
\caption  
{
\footnotesize{
 Simulation results for the star of cliques: four cliques, each of size 10, connected to a single broker node. The theoretical prediction is $b^*\approx19.56$
}
} \label{simul_star_of_cliques}
\end{figure}

\begin{figure}[t]
\centering
\includegraphics[width=.5 \columnwidth]{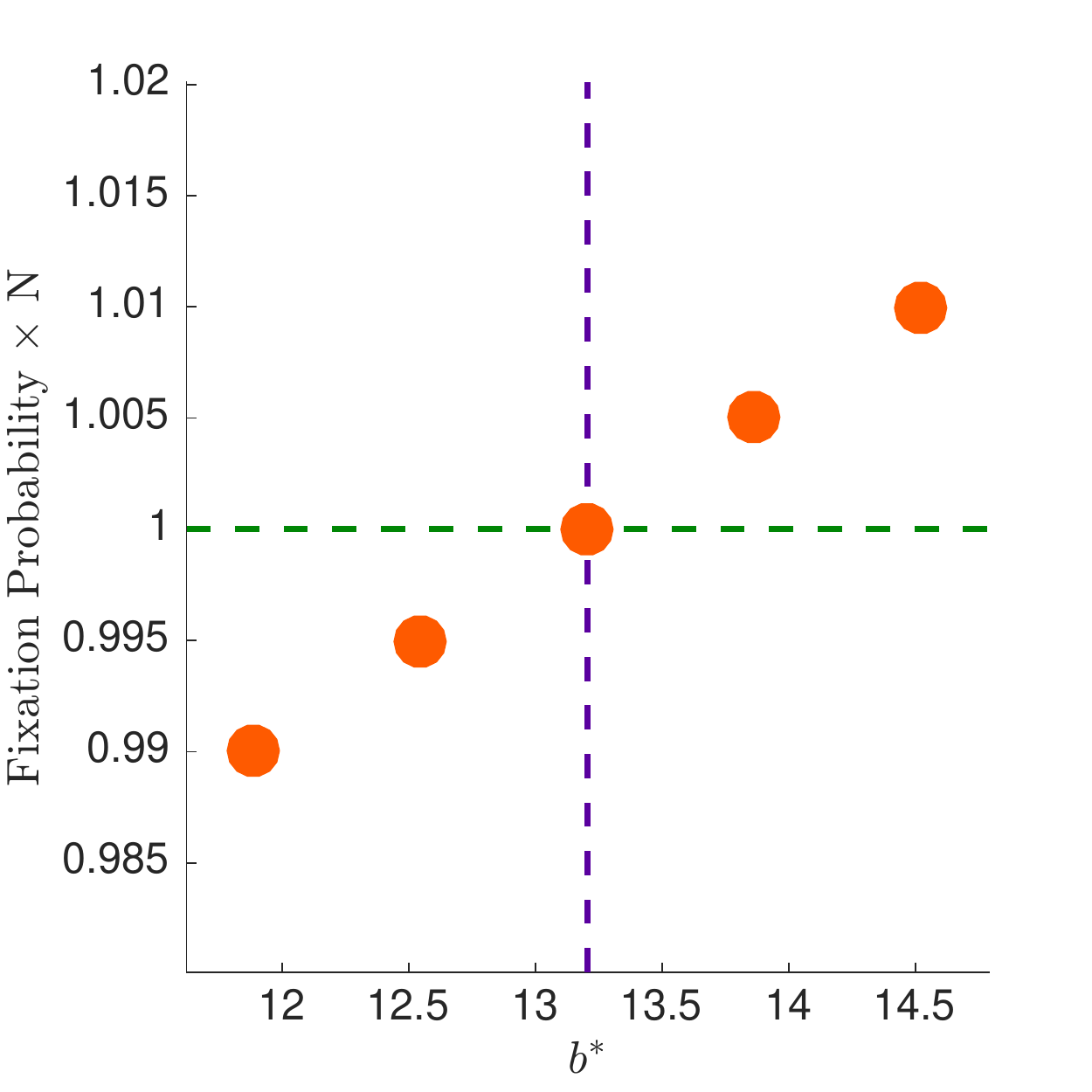}
\caption  
{
\footnotesize{
 Simulation results for the conjoining of two cliques via a star graph. Two cliques of size 10, connected via a broker node, with four leaf nodes connected to the broker node. The theoretical prediction is $b^* \approx 13.21$. 
}
} \label{simul_2cliques_starMiddle}
\end{figure}

\begin{figure}[t]
\centering
\includegraphics[width=.5 \columnwidth]{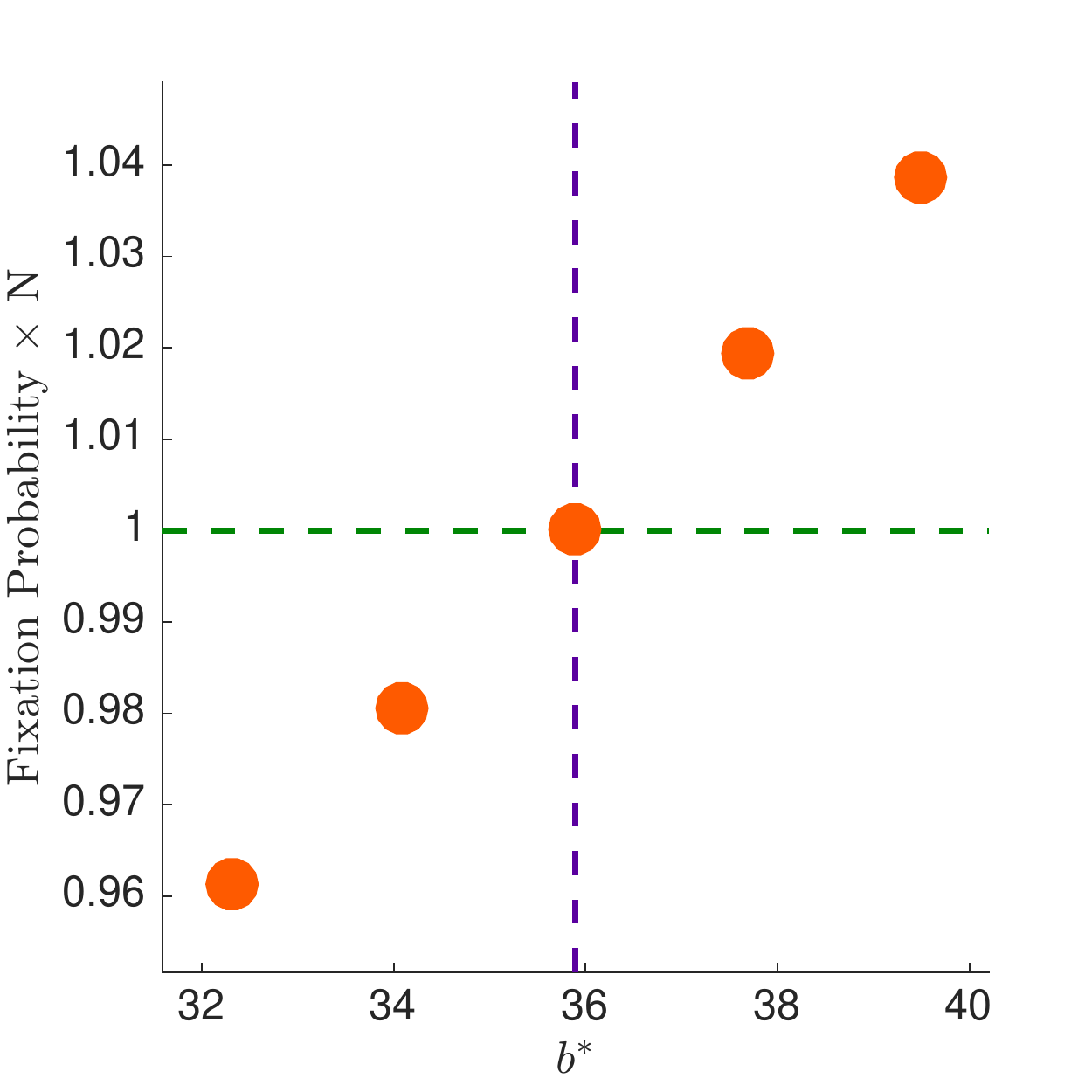}
\caption  
{
\footnotesize{
 Simulation results for the ring of cliques: four cliques, each with size 20. The theoretical prediction is $b^* \approx 35.90$. 
}
} \label{simul_cliqueRing}
\end{figure}

In addition to the robustness checks performed above, we also present simulation results for confirming the accuracy of the reported results for different topologies. 
Figure~\ref{simul_2stars} presents the results for conjoining two stars each with 50 leaves, whose hubs are connected via a chain of zero, one, or two intermediary nodes (three cases depicted together). The vertical axis is the fixation probability times network size $N$. The dashed vertical purple line marks the theoretical prediction, that is, the value of $b^*$ at which the value of fixation probability times $N$ equals one. The markers represent simulation results for different values of $b$. The simulation results closely agree with the theoretical predictions.
Figure~\ref{simul_imperfect_starOfstars} depicts the results for imperfect star of stars, 
Figure~\ref{simul_star2} for star of stars, 
and Figure~\ref{simul_ringOfStars} for ring of stars. 
Figure~\ref{simul_star_of_cliques} depicts the results for the star of cliques, 
Figure~\ref{simul_2cliques_starMiddle} for conjoining to cliques via a star as described in Section~\ref{sec:cliques_via_star}, 
Figure~\ref{simul_cliqueRing} for ring of cliques, 
and Figure~\ref{clique_hier} for hierarchy of cliques as discussed in Section~\ref{sec:clique_hier}.
Finally, Figure~\ref{simul_richClub} presents the results for conjoining two rich clubs and Figure~\ref{simul_bipartite} pertains to the conjoining of two complete bipartite graphs. 
In all cases, the selection strengths for simulations is 0.01. 
The results are calculated over $10^6$ Monte Carlo trials. 
That is, for each network, we run $10^6$ trials in which we take a network in which every node is a defector, we select one node uniformly at random, we make it a cooperator, and initiate the dynamics according to the game and update mechanisms described in the paper. 
The fraction of trials that terminate in an all-C state is the fixation probability. 
The value of $c$ is set to 1 and the test values of $b$ are $0.9,0.95,1,1.05,$ and 1.1 of the theoretically-predicted $b^*$, so that the crossover can be visually observed from the figure with convenience.

\begin{figure}[t]
\centering
\includegraphics[width=.5 \columnwidth]{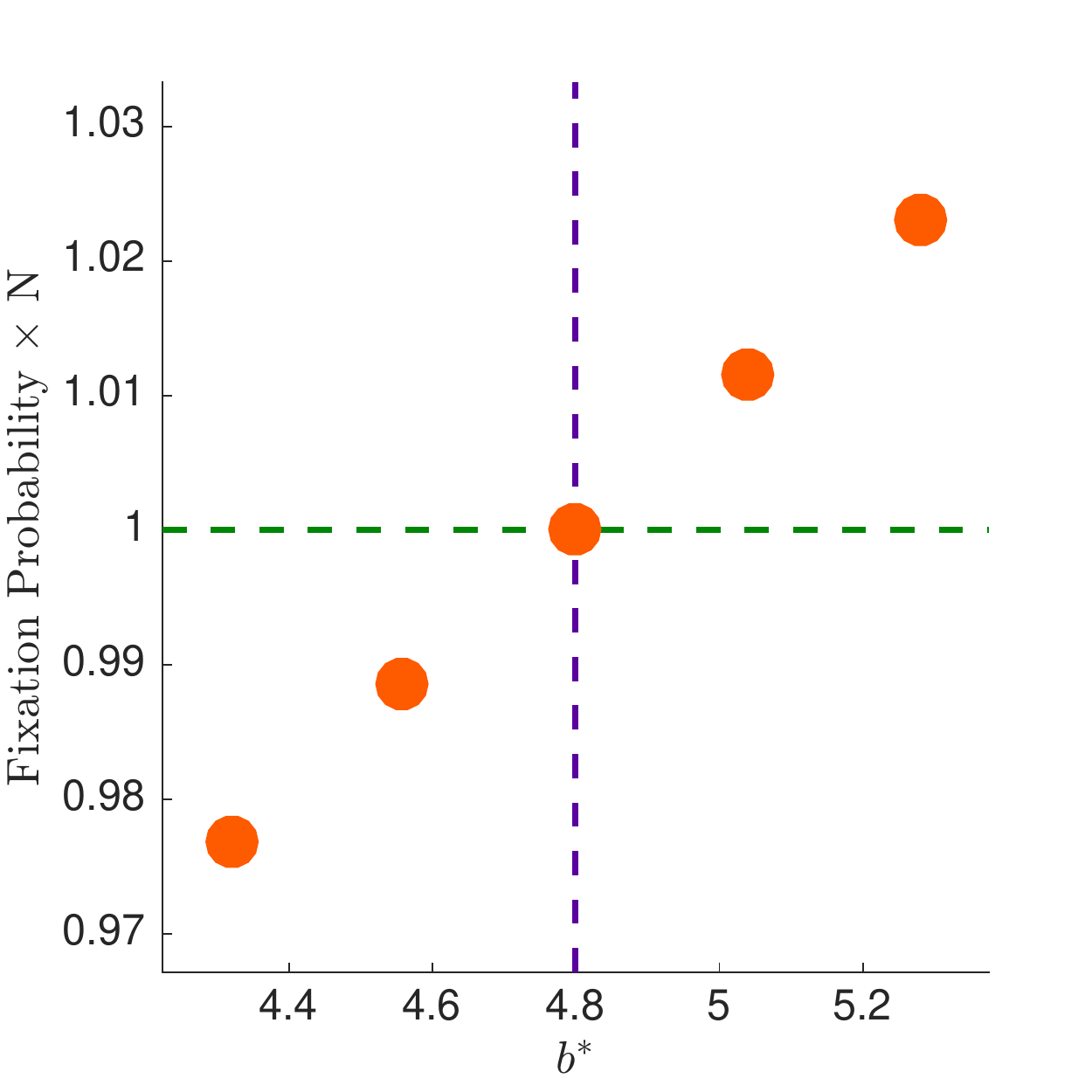}
\caption  
{
\footnotesize{
 Simulation results for hierarchy of cliques, identical to the graph illustrated in Figure~\ref{fig:hier}. The theoretical prediction is $b^*\approx 4.80$.
}
} \label{clique_hier}
\end{figure}

\begin{figure}[t]
\centering
\includegraphics[width=.5 \columnwidth]{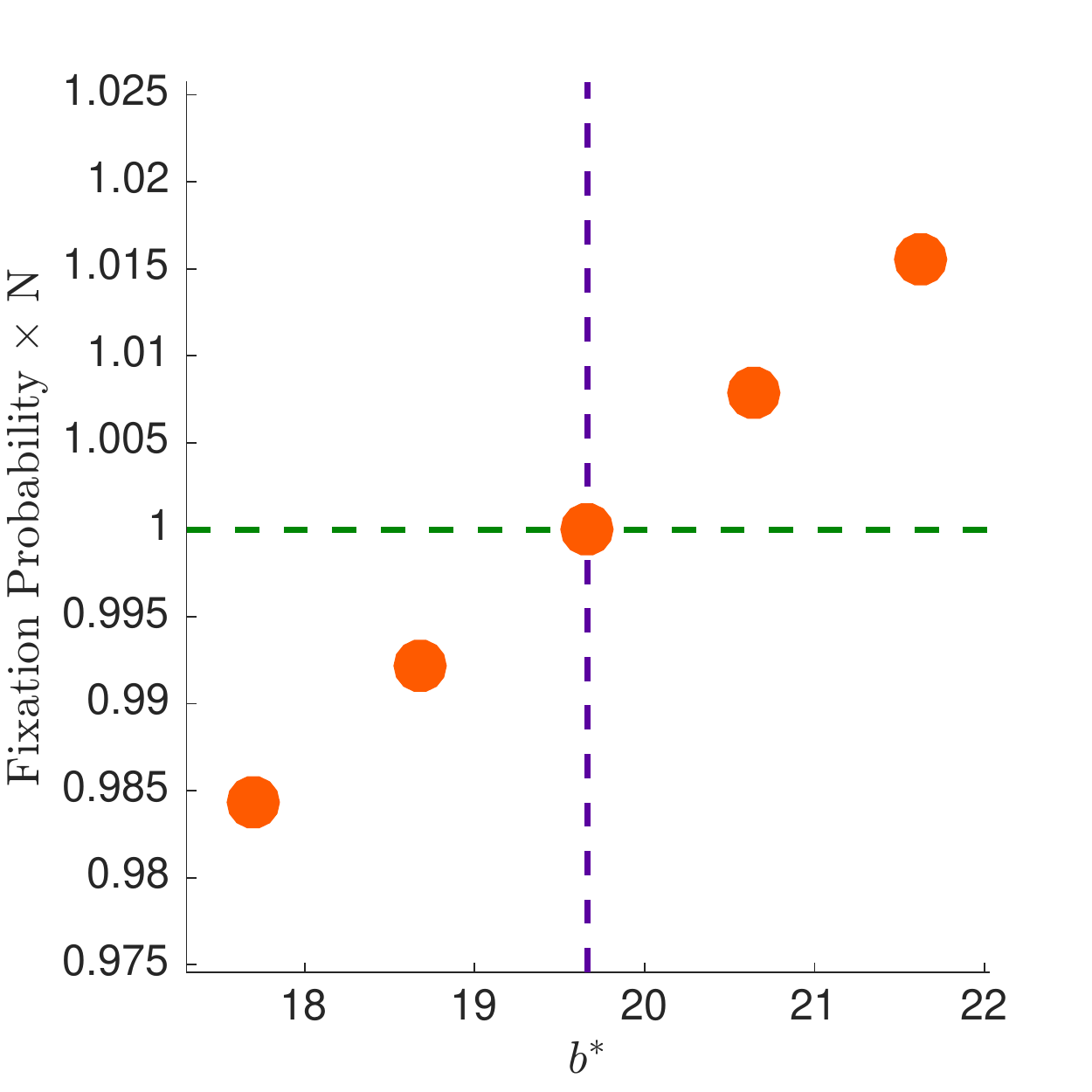}
\caption  
{
\footnotesize{
Simulation results for conjoining  two rich clubs, each with 50 peripheral nodes and 5 core nodes. 
The theoretical prediction is $b^* \approx 19.66$. 
}
} \label{simul_richClub}
\end{figure}

\begin{figure}[t]
\centering
\includegraphics[width=.5 \columnwidth]{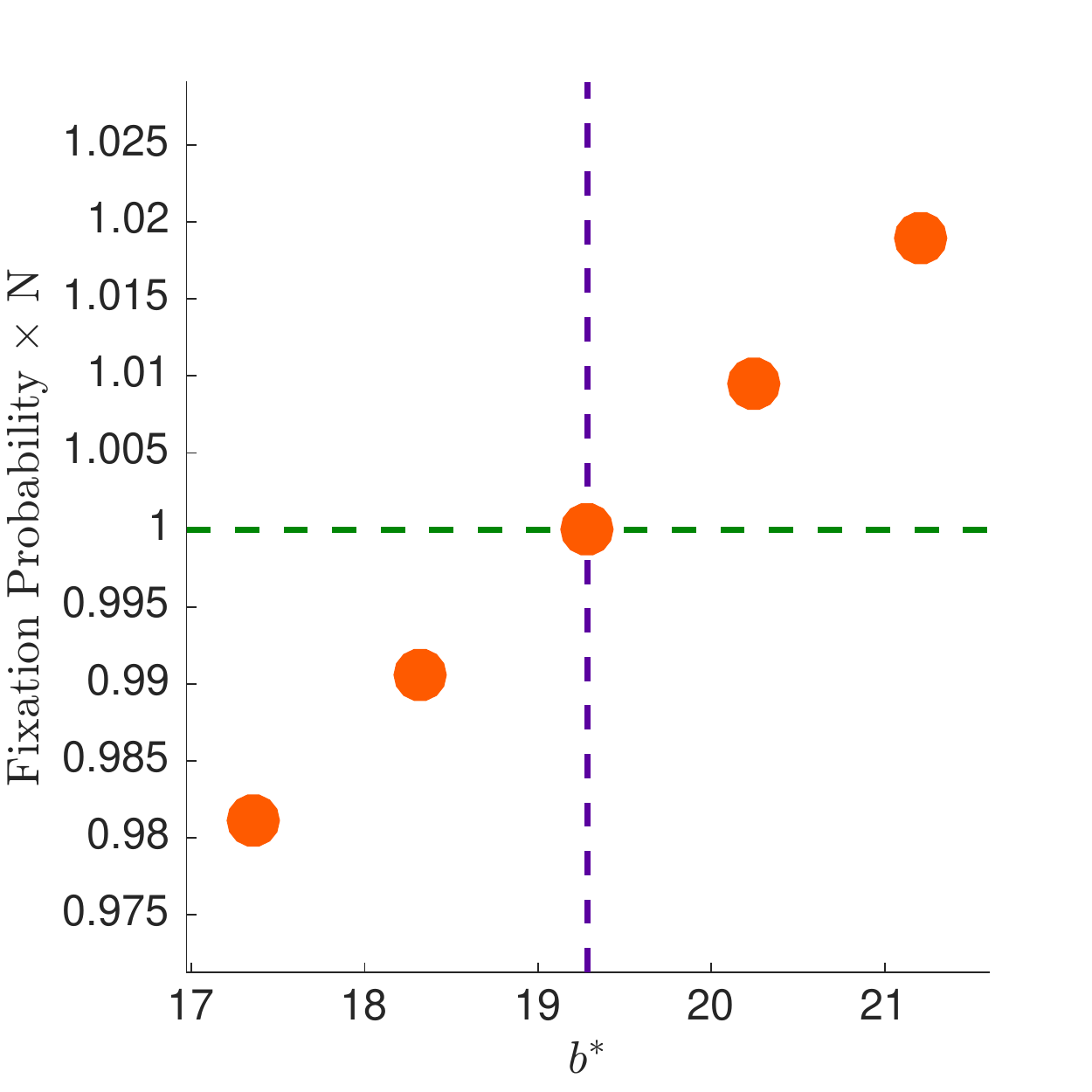}
\caption  
{
\footnotesize{
 Simulation results for conjoining two complete bipartite graphs, each consisting of two sets of 10 nodes. 
 The theoretical prediction is $b^* \approx 19.29$. 
}
} \label{simul_bipartite}
\end{figure}

\clearpage

\section{Imitation updating}\label{SI:imitation}

In addition to the DB updating considered in the main text, here we also provide the solution for imitation (IM) updating, in which each node also has the option of not updating its strategy at all  (equivalently, copies its own strategy). 

DB and IM updating can both be considered special cases of a general process, which we now describe.  The population structure is defined by two sets of probabilities: the \emph{replacement probability} $p_{xy}$ and the \emph{interaction probability} $\tilde{p}_{xy}$.  At each time-step, each individual $x$ interacts with other individuals $y$ with probability (or frequency) $\tilde{p}_{xy}$, receiving an expected payoff of $f_x$.  This payoff is rescaled to $F_x = 1+ \delta f_x$, where $\delta >0$ represents the strength of selection.  Then, an individual is chosen, uniformly at random, to update its strategy.  The probability that individual $x$ imitates individual $y$ is proportional to $p_{xy} F_y$.  

For DB updating on an unweighted, undirected graph, the replacement and interaction probabilities are the same:
\begin{equation}
\label{eq:DBps}
p_{xy} = \tilde{p}_{xy} = \begin{cases} 
\frac{1}{k_x} & \text{ $y \in \cN_x$}\\
0 & \text{otherwise}
\end{cases}
\end{equation}
For IM updating, the interaction probabilities $\tilde{p}_{xy}$ are again given by Eq.~\eqref{eq:DBps}, but the replacement probabilities are instead given by 
\begin{equation}
\label{eq:IMp}
p_{xy} = \begin{cases} 
\frac{1}{k_x+1} & \text{if $y=x$ or $y \in \cN_x$}\\
0 & \text{otherwise}
\end{cases}
\end{equation}
This reflects the fact that in IM updating, an individual can ``imitate" itself (i.e. choose not to change its strategy).

We now provide a general derivation, valid for both DB and IM updating, of the conditions for cooperation to be favored under weak selection. More explicit details can be found in \cite{allen2017evolutionary}. We represent the population state by a binary vector $\vs = (s_1, \ldots, s_N)$, where $s_x=1$ if $x$ is a (C)ooperator and $s_x=0$ if vertex $x$ is a (D)efector.  For the donation game
\begin{equation}
\label{eq:PD}
\bordermatrix{
& \C & \D\cr
\C & b-c & -c\cr
\D & b & 0 },
\end{equation}
the payoff to vertex $x$ can be expressed as
\begin{equation}
\label{eq:fPD}
f_x(\vs) = -cs_x + b \sum_y \tilde{p}_{xy} s_y.
\end{equation}

We require some notation for random walks.  Consider a random walk starting at vertex $x$ that takes $m$ steps according to the replacement probabilities $p$, followed by $n$ steps according to interaction probabilities $\tilde{p}$.  We denote by $p_{xy}^{(m,n)}$ the probability that such a walk terminates at vertex $y$. We let $\pi_x$ denote the stationary probability of vertex $x$ for random walks using the replacement probabilities $p$.  These stationary probabilities are given by
\begin{equation}
\label{eq:pidef}
\pi_x = \begin{cases} \displaystyle \frac{k_x}{\sum_y k_y} & \text{DB updating}\\[5mm]
\displaystyle \frac{k_x+1}{\sum_y (k_y+1)} & \text{IM updating} \end{cases}
\end{equation}
The stationary probability $\pi_x$ can also be understood as the \emph{reproductive value} (RV) of vertex $x$ \cite{maciejewski2014reproductive}.  Since random walks on undirected graphs are reversible (e.g.~\cite{aldous2002reversible}), we have the reversibility property $\pi_x p_{xy} = \pi_y p_{yx}$, and more generally, $\pi_x p_{xy}^{(m,0)} = \pi_y p_{yx}^{(m,0)}$, valid for all vertices $x,y$ and all $m \geq 0$.

We study selection using the RV-weighted frequency $\hat{s} = \sum_x \pi_x s_x$.  We denote expected change in $\hat{s}$ over a single time-step by $D(\vs)$, which can be calculated as
\begin{align*}
D(\vs) & = \frac{1}{N}  \sum_x s_x \left( - \pi_x + \sum_y \pi_y \frac{F_x(\vs)p_{yx}}{\sum_z F_z(\vs) p_{yz}}\right)\\
& = \frac{1}{N} \sum_x s_x \left(- \pi_x + \sum_y \pi_y \left(p_{yx} + \delta \left(f_x(\vs)p_{yx} - \sum_z f_z(\vs) p_{yz} \right) \right) \right) + \mathcal{O}(\delta^2)\\
& = \frac{\delta}{N} \sum_x \pi_x  s_x \left( f_x^{(0,0)}(\vs) - f_x^{(2,0)}(\vs) \right) + \mathcal{O}(\delta^2)\\
& =  \frac{\delta}{N} \sum_x \pi_x  s_x 
\left( -c \left(s_x^{(0,0)} - s_x^{(2,0)} \right) + b \left(s_x^{(0,1)} - s_x^{(2,1)} \right)  \right) + \mathcal{O}(\delta^2).
\end{align*}
Above, we have made use of the reversibility property $\pi_y p_{yx} = \pi_x p_{xy}$ and we have introduced the notation
\[
f_x^{(n,m)} = \sum_y p_{xy}^{(n,m)} f_y, \qquad s_x^{(n,m)} = \sum_y p_{xy}^{(n,m)} s_y.
\]

We define $D'(\vs)$ to be the $\delta$-derivative of $D(\vs)$ at $\delta=0$:
\begin{equation}
\label{eq:D'sGI}
D'(\vs) = \frac{1}{N} \sum_x \pi_x  s_x \left( -c \left(s_x^{(0,0)} - s_x^{(2,0)} \right) + b \left(s_x^{(0,1)} - s_x^{(2,1)} \right)  \right).
\end{equation}

\cite{chen2013sharp} showed that the fixation probability of cooperators under weak selection can be expressed as
\begin{equation}
\label{eq:rhoweaku1}
\rho_\C= \frac{1}{N} + \delta \langle D' \rangle + \mathcal{O}(\delta^2),
\end{equation}
where $\langle \; \rangle$ denotes an expectation over states arising under neutral drift, from an initial state with a single cooperator placed at a uniformly chosen random vertex.  It follows that cooperation is favored under weak selection if and only if $\langle D' \rangle>0$.

We can calculate $\langle D' \rangle$ using the method of coalescing random walks \cite{holley1975ergodic,cox1989coalescing,liggett2012interacting,chen2013sharp}.  Consider a process involving two random walkers, initially located at vertices $x$ and $y$.  At each time-step, one of the two walkers is chosen to take a step, using the step probabilities for replacement.  The \emph{coalescence time} $\tau_{xy}$ is defined as the expected time to until the two walkers meet.  Coalescence times satisfy the recurrence relation
\begin{equation}
\label{eq:taurecur}
\tau_{xy} = \tau_{yx} = (1-\delta_{xy}) 
\left[ 1+ \sum_{z} (p_{xz} \tau_{zy} +  p_{yz} \tau_{xz})\right] 
\end{equation}

Duality between CRW's and the voter model \cite{holley1975ergodic,liggett2012interacting} implies that $\tau_{xy}$ is proportional to the expected time since $x$ and $y$ diverged from their most recent common ancestor.  Using this duality, \cite{allen2017evolutionary} obtain the identity
\begin{equation}
\label{eq:stau}
\left \langle s_x s_y - s_z s_w \right \rangle = \frac{\tau_{zw} - \tau_{xy}}{2}.
\end{equation}

We introduce the notation
\begin{equation}
\label{eq:taumndef}
\tau^{(n,m)} = \sum_{x,y} \pi_x p_{xy}^{(n,m)} \tau_{xy}.
\end{equation}
Then substituting Eqs.~\eqref{eq:D'sGI} and \eqref{eq:stau} into Eq.~\eqref{eq:rhoweaku1} we obtain the fixation probability of cooperation:
\[
\rho_\C = \frac{1}{N} +  
\frac{\delta}{2N} \Big(  -c \tau^{(2,0)} + b \left( \tau^{(2,1)} - \tau^{(0,1)} \right) \Big)
+ \mathcal{O}(\delta^2).
\]
Cooperation is therefore favored under weak selection if and only if 
\[
  -c \tau^{(2,0)} + b \left( \tau^{(2,1)} - \tau^{(0,1)} \right) > 0.
\]
\cite{allen2017evolutionary} show that this condition implies both $\rho_\C > 1/N$ and $\rho_\D < 1/N$ under weak selection.  The critical benefit-cost ratio is given by
\begin{equation}
\label{eq:critbc}
b^* = \frac{\tau^{(2,0)}}{\tau^{(2,1)} - \tau^{(0,1)}}.
\end{equation}

For DB updating (which we considered so far), since the replacement and interaction probabilities coincide, we can write the critical benefit-cost ratio more simply as
\begin{equation}
\label{eq:DBbc1}
b^* = \frac{\tau^{(2)}}{\tau^{(3)} - \tau^{(1)}},
\end{equation}
where $\tau^{(n)} = \sum_x \pi_x p_{xy}^{(n)} \tau_{xy}$.  Eq.~\eqref{eq:taurecur} leads to a recurrence for $\tau^{(n)}$:
\begin{equation}
\label{eq:taunrecur}
\tau^{(n+1)} = \tau^{(n)} + \sum_x \pi_x p_{xx}^{(n)} \tau_x - 1,
\end{equation}
where $\tau_x = 1 + \sum_y p_{xy} \tau_{xy}$ is the remeeting time for two random walks from $x$. In particular (assuming the graph has no self-loops), we have
\begin{subequations}
\label{eq:taus}
\begin{align}
\label{eq:tau0}
\tau^{(0)} & = 0\\
\label{eq:tau1}
\tau^{(1)} & = \sum_x\pi_x \tau_x - 1\\
\label{eq:tau2}
\tau^{(2)} & = \sum_x\pi_x \tau_x - 2\\
\label{eq:tau3}
\tau^{(3)} & = \sum_x\pi_x \tau_x \left(1 +p_x \right) - 3.
\end{align}
\end{subequations}
Above, $p_x = p_{xx}^{(2)}$ is the probability that a two-step random walk from $x$ returns to $x$.  Using Eq.~\eqref{eq:taus}, we can rewrite the critical benefit-cost ratio in Eq.~\eqref{eq:DBbc1} as
\begin{align}
\label{eq:DBbc1}
b^* & = \frac{\sum_x\pi_x \tau_x - 2}{\sum_x\pi_x \tau_x p_x - 2} 
= \frac{\sum_xk_x \tau_x - 2N\bar{k}}{\sum_xk_x \tau_x p_x - 2N\bar{k}}
\end{align}

In the case of IM updating, substituting from Eqs.~\eqref{eq:IMp}, \eqref{eq:pidef}, and \eqref{eq:taumndef} into Eq.~\eqref{eq:critbc} gives the critical benefit-cost ratio for IM:
\[ 
b^* = \frac{\sum_{x,y,z}(k_x+1)p_{xy}p_{yz}\tau_{xz}}
{\sum_{x,y,z,w}(k_x+1)p_{xy}p_{yz}\tilde{p}_{zw}\tau_{xw} - \sum_{x,y}(k_x+1)p_{xy} \tau_{xy}}.
\]

\begin{figure}[t]
\centering
\includegraphics[width=.8 \columnwidth]{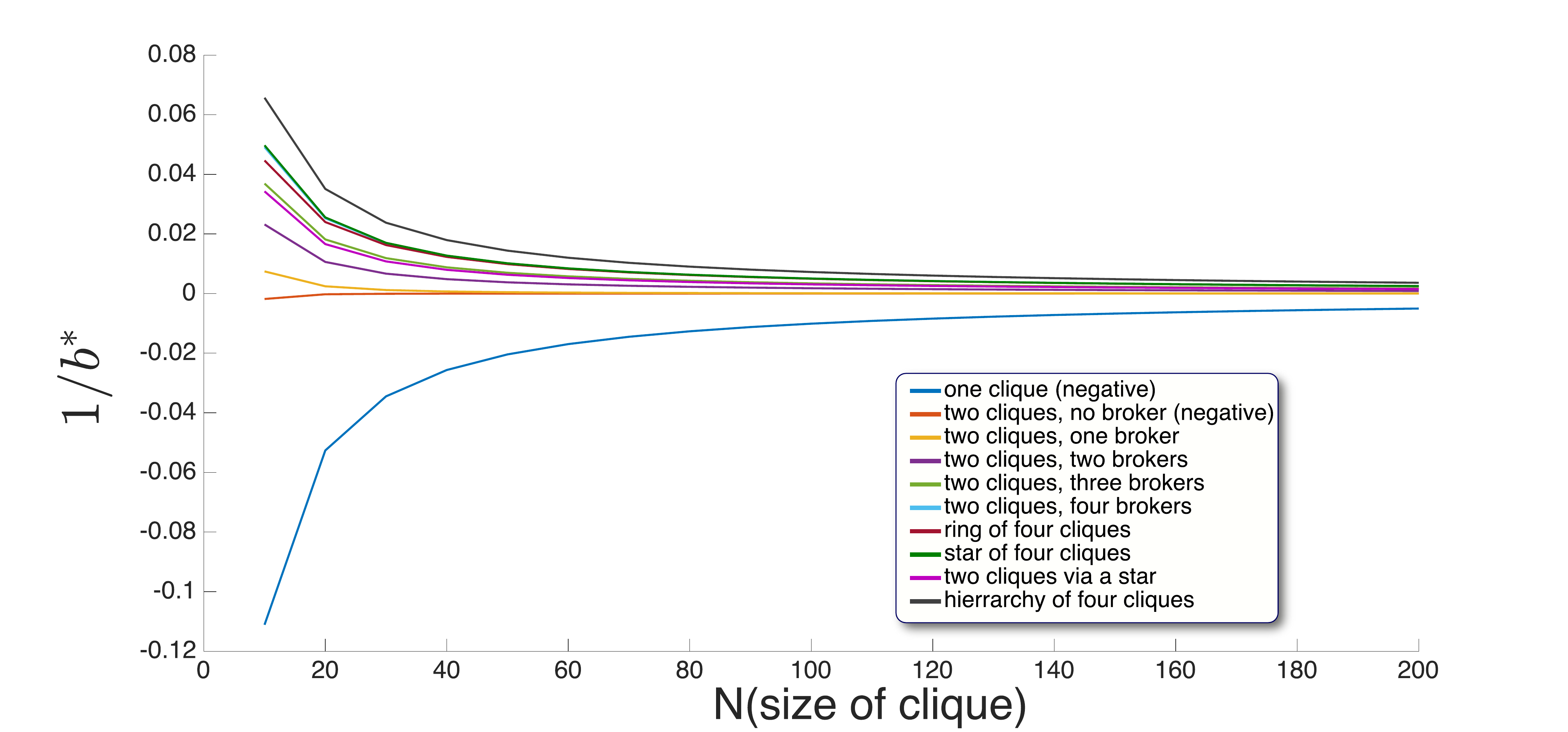}
\caption  
{
\footnotesize{
 IM updating for conjoining cliques
}
} \label{clique_hier}
\end{figure}

\begin{figure}[t]
\centering
\includegraphics[width=.8 \columnwidth]{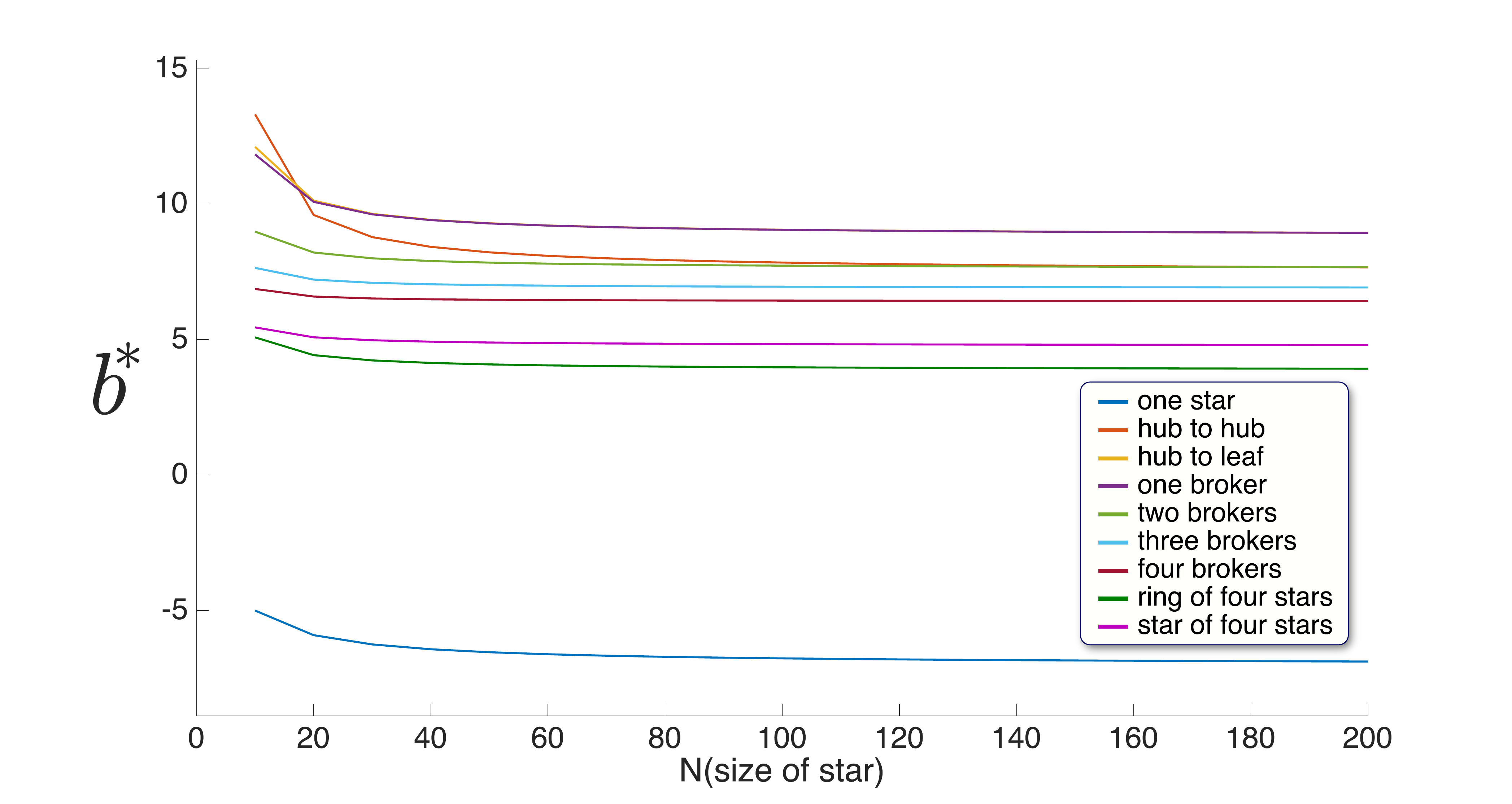}
\caption  
{
\footnotesize{
 IM updating for conjoining stars
}
} \label{clique_hier}
\end{figure}

\begin{figure}[t]
\centering
\includegraphics[width=.8 \columnwidth]{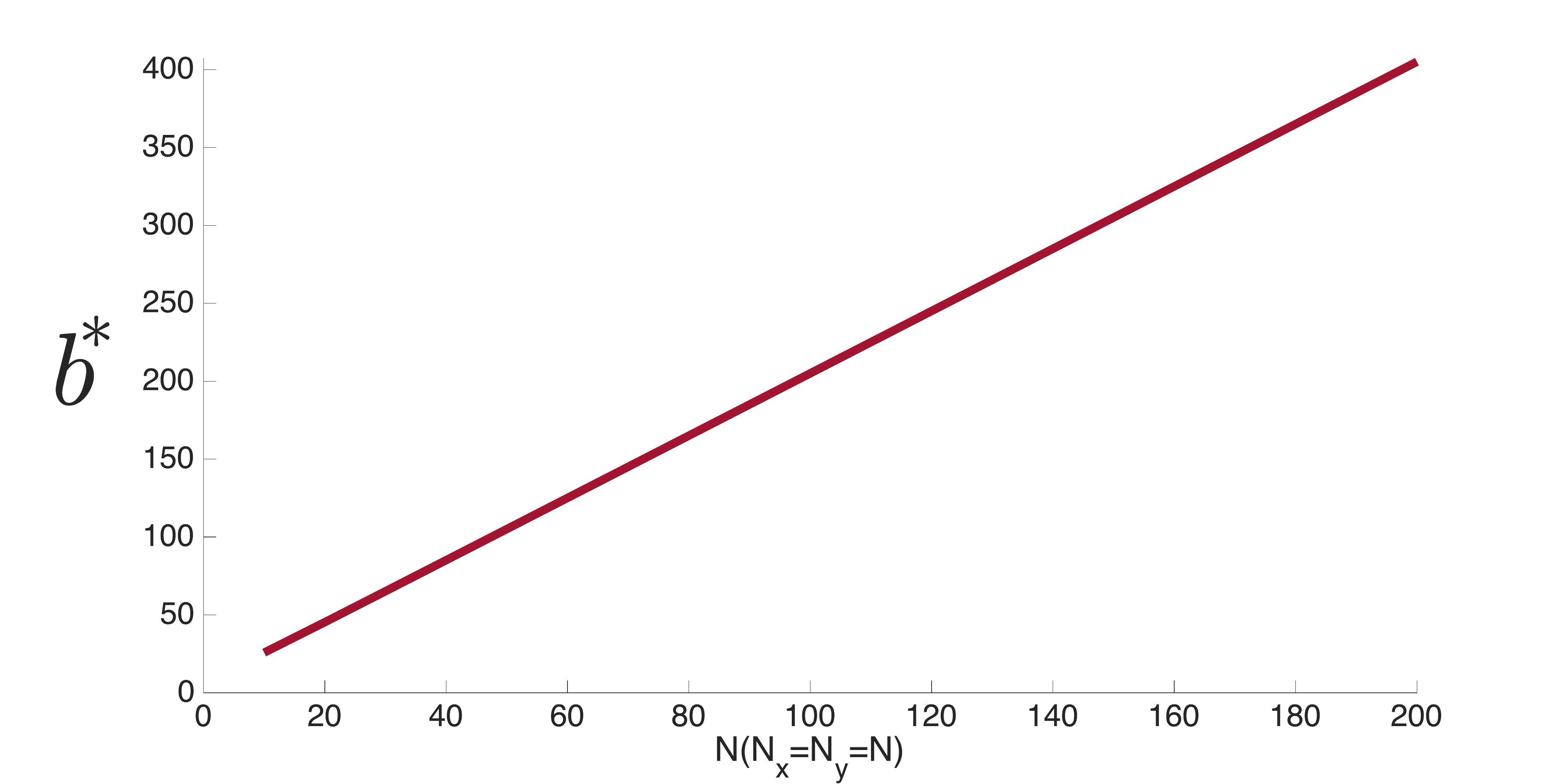}
\caption  
{
\footnotesize{
 IM updating for conjoining complete bipartite networks
}
} \label{clique_hier}
\end{figure}

\begin{figure}[t]
\centering
\includegraphics[width=.8 \columnwidth]{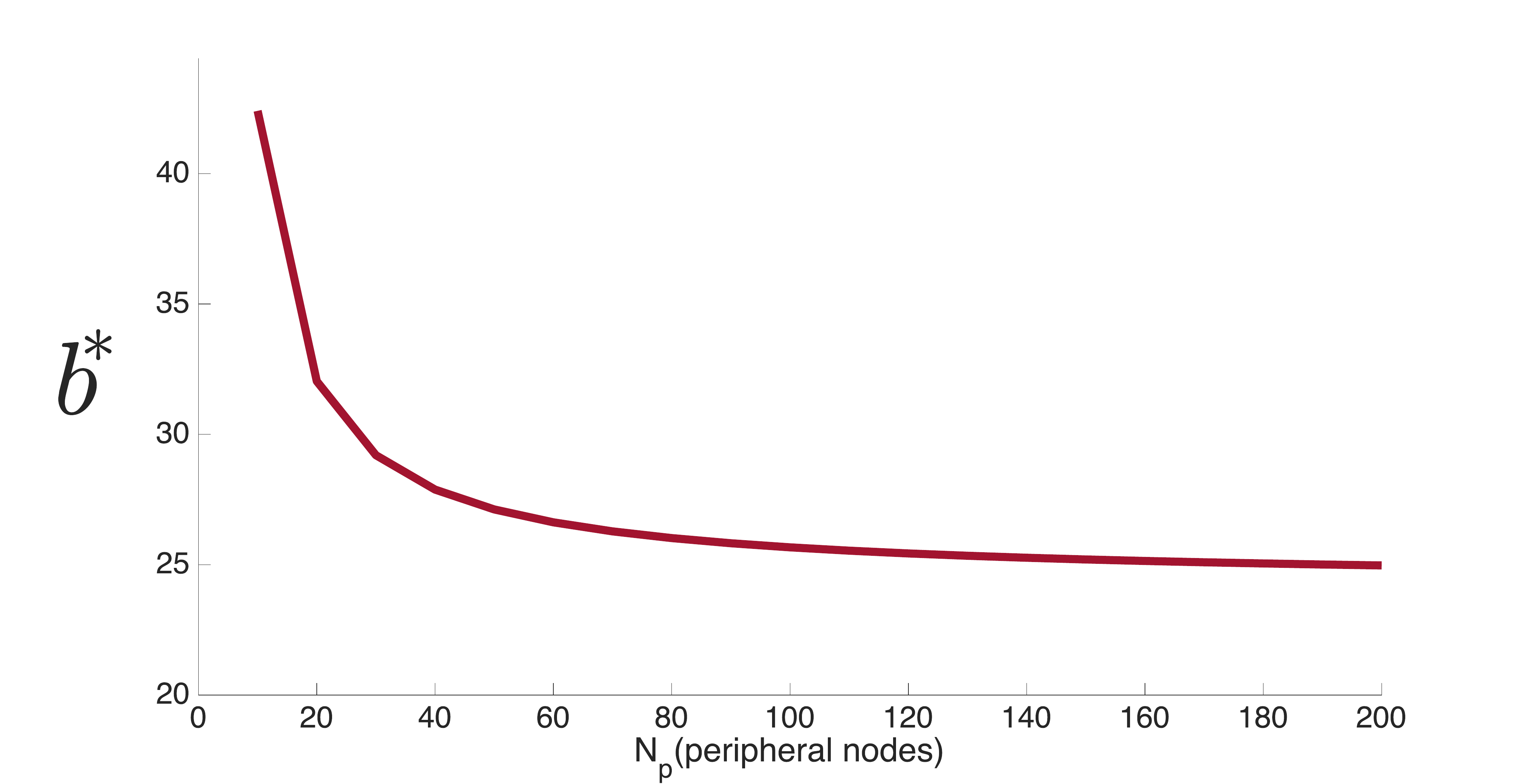}
\caption  
{
\footnotesize{
 IM updating for conjoining rich clubs
}
} \label{clique_hier}
\end{figure}

\clearpage
\section{Summary of the results}\label{SI:summary}

The summary of the results for the real social network data sets (as discussed in main text) is presented in Table~\ref{real}. 
For the high school network, division to both two and three communities returned reasonable groupings, so we reported results for both. 

\begin{table}[h]
\centering
\caption{results for real social network data sets}
\label{real}
\begin{tabular}{lllll}
Network                                           & N                                        & m                      & bc                              & BC                   \\ \hline
\multicolumn{1}{l|}{fourth grade}                 & \multicolumn{1}{l|}{24}                  & \multicolumn{1}{l|}{2} & \multicolumn{1}{l|}{-11,-39}    & 41                   \\ \hline
\multicolumn{1}{l|}{fifth grade}                  & \multicolumn{1}{l|}{22}                  & \multicolumn{1}{l|}{2} & \multicolumn{1}{l|}{-23,-47}    & 19                   \\ \hline
\multicolumn{1}{l|}{\multirow{2}{*}{high school}} & \multicolumn{1}{c|}{\multirow{2}{*}{37}} & \multicolumn{1}{l|}{2} & \multicolumn{1}{l|}{10,18}      & \multirow{2}{*}{7.9} \\ \cline{3-4}
\multicolumn{1}{l|}{}                             & \multicolumn{1}{c|}{}                    & \multicolumn{1}{l|}{3} & \multicolumn{1}{l|}{10,-9,-151} &                      \\ \hline
\multicolumn{1}{l|}{zachary club}                 & \multicolumn{1}{l|}{34}                  & \multicolumn{1}{l|}{2} & \multicolumn{1}{l|}{14.5,11.5}  & 7.5                  
\end{tabular}
\end{table}

A condensed summary for the limiting behavior of $(b/c)^*$  for more notable topologies is presented in Table~\ref{sum2}. 
A more comprehensive summary of the limiting behaviors of the conjoined graphs is presented in Table~\ref{summaryTable}. 

\begin{table}[t]
\centering
\caption{Limiting behavior of results for notable topologies}\label{sum2}
\begin{tabular}{lc}
Graph                                                      & $\left(\frac{b}{c}\right)^\ast$ \\ \hline
\multicolumn{1}{l|}{Star}                                  & Inf                             \\ \hline
\multicolumn{1}{l|}{Star of Stars}                         & $2$                             \\ \hline
\multicolumn{1}{l|}{Two Stars}                             & $\frac{5}{2}, 3$                \\ \hline
\multicolumn{1}{l|}{Ring of Stars}                         & $\frac{3}{2}$                   \\ \hline
\multicolumn{1}{l|}{Clique}                                & $-(N-1)$                        \\ \hline
\multicolumn{1}{l|}{Conjoined Cliques}                     & $N^2$                           \\ \hline
\multicolumn{1}{l|}{Star of Cliques}                       & $N$                             \\ \hline
\multicolumn{1}{l|}{Hierarchy of Cliques}                  & $N$                             \\ \hline
\multicolumn{1}{l|}{Ring of Cliques}                       & $N$                             \\ \hline
\multicolumn{1}{l|}{The Rich Club}                         & $-N$                            \\ \hline
\multicolumn{1}{l|}{\multirow{2}{*}{Conjoined Rich Clubs}} & \multirow{2}{*}{$N_c$}               \\
\multicolumn{1}{l|}{}                                      &                                 \\ \hline
\multicolumn{1}{l|}{Complete Bipartite}                    & Inf                             \\ \hline
\multicolumn{1}{l|}{Conjoined Complete Bipartites}   & $2N$                           
\end{tabular}
\end{table}

\begin{table}[t]
\centering
\caption{Summary of the limiting behavior of the $(b/c)^*$ of conjoined networks. }
\label{summaryTable}
\resizebox{\linewidth}{!}{
\begin{tabular}{llll}
Graph                                                      & Parameters                                                                                                           & Limit/Special case                            & $\left(\dfrac{b}{c}\right)^\ast$                                                         \\ \hline
\multicolumn{1}{l|}{Star}                                  & \multicolumn{1}{l|}{$n$ leaf nodes}                                                                                  & \multicolumn{1}{l|}{N/A}                      & Inf                                                                                     \\ \hline
 \multicolumn{1}{l|}{Extended Star}                         & \multicolumn{1}{l|}{two layers, each $n$ nodes}                                                                      & \multicolumn{1}{l|}{$n \to \infty$}& $\dfrac{28}{11}\approx 2.55$ \T                                                             \\ \hline
\multicolumn{1}{l|}{3-Layer Extended Star}                 & \multicolumn{1}{l|}{three layers, each $n$ nodes}                                                                    & \multicolumn{1}{l|}{$n \to \infty$}& $\dfrac{106940}{41723}\approx 2.56$                                                      \\ \hline
\multicolumn{1}{l|}{Imperfect Extended Star}               & \multicolumn{1}{l|}{Layer one $n$ nodes, Layer two $n_g$ nodes}                                                      & \multicolumn{1}{l|}{$n_g \ll n$}& $\dfrac{34}{89 n_g}n \approx \dfrac{0.38}{n_g}n$                                          \\ \hline
\multicolumn{1}{l|}{Star of Stars}                         & \multicolumn{1}{l|}{$n$ leaf nodes, each connected to $n_d$ nodes}                                                   & \multicolumn{1}{l|}{$n \ll n_d$}& $\left( 2+\dfrac{1}{n-1}\right) + \left(\dfrac{1}{2}+\dfrac{1}{1-n} \right) \dfrac{1}{n_d}$ \\ \hline
\multicolumn{1}{l|}{Imperfect Star of Stars}               & \multicolumn{1}{p{4.5cm}|}{$n$ leaf nodes, $n_g$ of them connected to $n_d$ nodes}                                          & \multicolumn{1}{l|}{$n_g \ll n\ll n_d$}& $\dfrac{3}{2}+\dfrac{1}{2(n_g-1)}$                                                        \\ \hline
\multicolumn{1}{l|}{Two Stars: Hub to Leaf}                & \multicolumn{1}{l|}{Two stars, each with $n$ leaf nodes}                                                             & \multicolumn{1}{l|}{$n \to \infty$}& $\dfrac{5}{2}$                                                                           \\ \hline
\multicolumn{1}{l|}{Two Stars: Leaf to Leaf}               & \multicolumn{1}{l|}{Two stars, each with $n$ leaf nodes}                                                             & \multicolumn{1}{l|}{$n \to \infty$}& $3$                                                                                     \\ \hline
\multicolumn{1}{l|}{Ring of Stars}                         & \multicolumn{1}{l|}{$L$ stars, each with $n$ leaf nodes}                                                             & \multicolumn{1}{l|}{$n \to \infty$}& $\dfrac{3L-1}{2L-2}$                                                                     \\ \hline
\multicolumn{1}{l|}{Clique}                                & \multicolumn{1}{l|}{$n$ nodes}                                                                                       & \multicolumn{1}{l|}{N/A}                      & $-(n-1)$                                                                                \\ \hline
\multicolumn{1}{l|}{Conjoined Cliques}                     & \multicolumn{1}{l|}{two cliques, each with $n$ nodes}                                                                & \multicolumn{1}{l|}{$n \to \infty$}& $n^2$                                                                                   \\ \hline
\multicolumn{1}{l|}{Star of Cliques}                       & \multicolumn{1}{l|}{$m$ cliques, each with $n$ nodes}                                                                & \multicolumn{1}{l|}{$n \to \infty$}& $\left(\dfrac{m}{m-2}\right)n -\dfrac{(m+8)(m-1)}{(m-2)^2}$                               \\ \hline
\multicolumn{1}{l|}{Connecting Two Cliques via a Star}     & \multicolumn{1}{p{4.8cm}|}{two cliques of size $n$ connecting via a star of size $m$}                                       & \multicolumn{1}{l|}{$m=n, n \to \infty$}& $n-\dfrac{1}{2}$                                                                         \\ \hline
\multicolumn{1}{l|}{Hierarchy of Cliques}                  & \multicolumn{1}{p{4.8cm}|}{$q^2$ cliques of size $n$, at the bottom layer, connected to the base node via $q$ middle nodes} & \multicolumn{1}{l|}{$n \to \infty$}           & $\left[ \dfrac{q^2(2q^2+q+1)}{q^2(2q^2+q+3)-6q-4}  \right] n$                            \\ \hline
\multicolumn{1}{l|}{Ring of Cliques} & \multicolumn{1}{l|}{$L$ cliques of size $n$}                                                                         & \multicolumn{1}{l|}{$n \to \infty$}& $\left(\dfrac{L}{L-2}\right)n-\dfrac{(L+1)^2-5}{(L-2)^2}$                                  \\ \hline
\multicolumn{1}{l|}{The Rich Club}& \multicolumn{1}{l|}{$N_c$ core and $N_p=N-N_c$ peripheral nodes}                                                                       & \multicolumn{1}{l|}{$N \to \infty$}& $-\dfrac{2N}{3}\dfrac{2N_c-1}{N_c-1}+\dfrac{1}{18}\big[ 8N_c+39+\dfrac{20}{N_c-1} \big]$            \\ \hline
\multicolumn{1}{l|}{Conjoined Rich Clubs} & \multicolumn{1}{p{4.8cm}|}{two rich clubs, each with $N_c$ core and $N_p$ peripheral nodes}                & \multicolumn{1}{l|}{ $N_p \to \infty$ }& $4N_c-\frac{3}{2}$  \\ \hline
\multicolumn{1}{l|}{Complete Bipartite}& \multicolumn{1}{l|}{two subsets of nodes with sizes $n_x$ and $n_y$}                                                 & \multicolumn{1}{l|}{N/A}& Inf                                                                                     \\ \hline
\multicolumn{1}{l|}{Conjoined Complete Bipartite Graphs}   & \multicolumn{1}{p{4.8cm}|}{two identical complete bipartite graphs with sizes $n_x$ and $n_y$}                              & \multicolumn{1}{l|}{$n_x=n_y=n \to \infty$} & $2n-1$                                                                                 
\end{tabular}}
\end{table}

\clearpage
%

\begin{thebibliography}{100}
\expandafter\ifx\csname url\endcsname\relax
  \def\url#1{\texttt{#1}}\fi
\expandafter\ifx\csname urlprefix\endcsname\relax\def\urlprefix{URL }\fi
\providecommand{\bibinfo}[2]{#2}
\providecommand{\eprint}[2][]{\url{#2}}


%
%

\bibitem{nowak2006five}
\bibinfo{author}{Nowak, M.~A.}
\newblock \bibinfo{title}{Five rules for the evolution of cooperation}.
\newblock \emph{\bibinfo{journal}{Science}} \textbf{\bibinfo{volume}{314}},
  \bibinfo{pages}{1560--1563} (\bibinfo{year}{2006}).
  
  \bibitem{simpson2015beyond}
\bibinfo{author}{Simpson, B.} \& \bibinfo{author}{Willer, R.}
\newblock \bibinfo{title}{Beyond altruism: Sociological foundations of
  cooperation and prosocial behavior}.
\newblock \emph{\bibinfo{journal}{Annual Review of Sociology}}
  \textbf{\bibinfo{volume}{41}}, \bibinfo{pages}{43--63}
  (\bibinfo{year}{2015}).
%


\bibitem{jordan2013contagion}
\bibinfo{author}{Jordan, J.~J.}, \bibinfo{author}{Rand, D.~G.},
  \bibinfo{author}{Arbesman, S.}, \bibinfo{author}{Fowler, J.~H.} \&
  \bibinfo{author}{Christakis, N.~A.}
\newblock \bibinfo{title}{Contagion of cooperation in static and fluid social
  networks}.
\newblock \emph{\bibinfo{journal}{PloS one}} \textbf{\bibinfo{volume}{8}},
  \bibinfo{pages}{e66199} (\bibinfo{year}{2013}).

\bibitem{rand2014static}
\bibinfo{author}{Rand, D.~G.}, \bibinfo{author}{Nowak, M.~A.},
  \bibinfo{author}{Fowler, J.~H.} \& \bibinfo{author}{Christakis, N.~A.}
\newblock \bibinfo{title}{Static network structure can stabilize human
  cooperation}.
\newblock \emph{\bibinfo{journal}{Proceedings of the National Academy of
  Sciences}} \textbf{\bibinfo{volume}{111}}, \bibinfo{pages}{17093--17098}
  (\bibinfo{year}{2014}).
%

\bibitem{hauert2004spatial}
\bibinfo{author}{Hauert, C.} \& \bibinfo{author}{Doebeli, M.}
\newblock \bibinfo{title}{Spatial structure often inhibits the evolution of
  cooperation in the snowdrift game}.
\newblock \emph{\bibinfo{journal}{Nature}} \textbf{\bibinfo{volume}{428}},
  \bibinfo{pages}{643} (\bibinfo{year}{2004}).

\bibitem{lieberman2005evolutionary}
\bibinfo{author}{Lieberman, E.}, \bibinfo{author}{Hauert, C.} \&
  \bibinfo{author}{Nowak, M.~A.}
\newblock \bibinfo{title}{Evolutionary dynamics on graphs}.
\newblock \emph{\bibinfo{journal}{Nature}} \textbf{\bibinfo{volume}{433}},
  \bibinfo{pages}{312--316} (\bibinfo{year}{2005}).

\bibitem{ohtsuki2006simple}
\bibinfo{author}{Ohtsuki, H.}, \bibinfo{author}{Hauert, C.},
  \bibinfo{author}{Lieberman, E.} \& \bibinfo{author}{Nowak, M.~A.}
\newblock \bibinfo{title}{A simple rule for the evolution of cooperation on
  graphs and social networks}.
\newblock \emph{\bibinfo{journal}{Nature}} \textbf{\bibinfo{volume}{441}},
  \bibinfo{pages}{502--505} (\bibinfo{year}{2006}).

\bibitem{szabo2007evolutionary}
\bibinfo{author}{Szab{\'o}, G.} \& \bibinfo{author}{Fath, G.}
\newblock \bibinfo{title}{Evolutionary games on graphs}.
\newblock \emph{\bibinfo{journal}{Physics Reports}}
  \textbf{\bibinfo{volume}{446}}, \bibinfo{pages}{97--216}
  (\bibinfo{year}{2007}).
%


\bibitem{debarre2014social}
\bibinfo{author}{D{\'e}barre, F.}, \bibinfo{author}{Hauert, C.} \&
  \bibinfo{author}{Doebeli, M.}
\newblock \bibinfo{title}{Social evolution in structured populations}.
\newblock \emph{\bibinfo{journal}{Nature Communications}}
  \textbf{\bibinfo{volume}{5}}, \bibinfo{pages}{3409} (\bibinfo{year}{2014}).

\bibitem{allen2017evolutionary}
\bibinfo{author}{Allen, B.} \emph{et~al.}
\newblock \bibinfo{title}{Evolutionary dynamics on any population structure}.
\newblock \emph{\bibinfo{journal}{Nature}} \textbf{\bibinfo{volume}{544}},
  \bibinfo{pages}{227--230} (\bibinfo{year}{2017}).
%
%
%

\bibitem{centola2010spread}
\bibinfo{author}{Centola, D.}
\newblock \bibinfo{title}{The spread of behavior in an online social network
  experiment}.
\newblock \emph{\bibinfo{journal}{Science}} \textbf{\bibinfo{volume}{329}},
  \bibinfo{pages}{1194--1197} (\bibinfo{year}{2010}).

\bibitem{centola2007complex}
\bibinfo{author}{Centola, D.} \& \bibinfo{author}{Macy, M.}
\newblock \bibinfo{title}{Complex contagions and the weakness of long ties}.
\newblock \emph{\bibinfo{journal}{American Journal of Sociology}}
  \textbf{\bibinfo{volume}{113}}, \bibinfo{pages}{702--734}
  (\bibinfo{year}{2007}).

\bibitem{nowak1992evolutionary}
\bibinfo{author}{Nowak, M.~A.} \& \bibinfo{author}{May, R.~M.}
\newblock \bibinfo{title}{Evolutionary games and spatial chaos}.
\newblock \emph{\bibinfo{journal}{Nature}} \textbf{\bibinfo{volume}{359}},
  \bibinfo{pages}{826} (\bibinfo{year}{1992}).

%

\bibitem{long2013bridges}
\bibinfo{author}{Long, J.~C.}, \bibinfo{author}{Cunningham, F.~C.} \&
  \bibinfo{author}{Braithwaite, J.}
\newblock \bibinfo{title}{Bridges, brokers and boundary spanners in
  collaborative networks: a systematic review}.
\newblock \emph{\bibinfo{journal}{BMC Health Services Research}}
  \textbf{\bibinfo{volume}{13}}, \bibinfo{pages}{158} (\bibinfo{year}{2013}).

\bibitem{fischer1975toward}
\bibinfo{author}{Fischer, C.~S.}
\newblock \bibinfo{title}{Toward a subcultural theory of urbanism}.
\newblock \emph{\bibinfo{journal}{American Journal of Sociology}}
  \textbf{\bibinfo{volume}{80}}, \bibinfo{pages}{1319--1341}
  (\bibinfo{year}{1975}).

\bibitem{wellman2001persistence}
\bibinfo{author}{Wellman, B.}
\newblock \bibinfo{title}{The persistence and transformation of community: from
  neighbourhood groups to social networks}.
\newblock \emph{\bibinfo{journal}{Report to the law commission of Canada}}
  (\bibinfo{year}{2001}).

\bibitem{burt2009structural}
\bibinfo{author}{Burt, R.~S.}
\newblock \emph{\bibinfo{title}{Structural holes: The social structure of
  competition}} (\bibinfo{publisher}{Harvard university press},
  \bibinfo{year}{2009}).

\bibitem{rosenthal1997social}
\bibinfo{author}{Rosenthal, E.}
\newblock \bibinfo{title}{Social networks and team performance}.
\newblock \emph{\bibinfo{journal}{Team Performance Management: An International
  Journal}} \textbf{\bibinfo{volume}{3}}, \bibinfo{pages}{288--294}
  (\bibinfo{year}{1997}).

\bibitem{watts1999networks}
\bibinfo{author}{Watts, D.~J.}
\newblock \bibinfo{title}{Networks, dynamics, and the small-world phenomenon}.
\newblock \emph{\bibinfo{journal}{American Journal of sociology}}
  \textbf{\bibinfo{volume}{105}}, \bibinfo{pages}{493--527}
  (\bibinfo{year}{1999}).

\bibitem{benkler2011penguin}
\bibinfo{author}{Benkler, Y.}
\newblock \emph{\bibinfo{title}{The penguin and the leviathan: How cooperation
  triumphs over self-interest}} (\bibinfo{publisher}{Crown Business},
  \bibinfo{address}{New York}, \bibinfo{year}{2011}).

\bibitem{ansell2016says}
\bibinfo{author}{Ansell, C.}, \bibinfo{author}{Bichir, R.} \&
  \bibinfo{author}{Zhou, S.}
\newblock \bibinfo{title}{Who says networks, says oligarchy? oligarchies as"
  rich club" networks.}
\newblock \emph{\bibinfo{journal}{Connections (02261766)}}
  \textbf{\bibinfo{volume}{35}} (\bibinfo{year}{2016}).

\bibitem{burt2010neighbor}
\bibinfo{author}{Burt, R.~S.}
\newblock \emph{\bibinfo{title}{Neighbor networks: Competitive advantage local
  and personal}} (\bibinfo{publisher}{Oxford University Press},
  \bibinfo{year}{2010}).

\bibitem{fracassi2016corporate}
\bibinfo{author}{Fracassi, C.}
\newblock \bibinfo{title}{Corporate finance policies and social networks}.
\newblock \emph{\bibinfo{journal}{Management Science}}  (\bibinfo{year}{2016}).

\bibitem{heemskerk2016corporate}
\bibinfo{author}{Heemskerk, E.~M.} \& \bibinfo{author}{Takes, F.~W.}
\newblock \bibinfo{title}{The corporate elite community structure of global
  capitalism}.
\newblock \emph{\bibinfo{journal}{New Political Economy}}
  \textbf{\bibinfo{volume}{21}}, \bibinfo{pages}{90--118}
  (\bibinfo{year}{2016}).

\bibitem{crane1969social}
\bibinfo{author}{Crane, D.}
\newblock \bibinfo{title}{Social structure in a group of scientists: A test of
  the" invisible college" hypothesis}.
\newblock \emph{\bibinfo{journal}{American Sociological Review}}
  \bibinfo{pages}{335--352} (\bibinfo{year}{1969}).

\bibitem{zhou2004rich}
\bibinfo{author}{Zhou, S.} \& \bibinfo{author}{Mondrag{\'o}n, R.~J.}
\newblock \bibinfo{title}{The rich-club phenomenon in the internet topology}.
\newblock \emph{\bibinfo{journal}{IEEE Communications Letters}}
  \textbf{\bibinfo{volume}{8}}, \bibinfo{pages}{180--182}
  (\bibinfo{year}{2004}).

\bibitem{davis1996significance}
\bibinfo{author}{Davis, G.~F.}
\newblock \bibinfo{title}{The significance of board interlocks for corporate
  governance}.
\newblock \emph{\bibinfo{journal}{Corporate Governance: An International
  Review}} \textbf{\bibinfo{volume}{4}}, \bibinfo{pages}{154--159}
  (\bibinfo{year}{1996}).

\bibitem{kranton2003theory}
\bibinfo{author}{Kranton, R.~E.} \& \bibinfo{author}{Minehart, D.~F.}
\newblock \bibinfo{title}{A theory of buyer-seller networks}.
\newblock In \emph{\bibinfo{booktitle}{Networks and Groups}},
  \bibinfo{pages}{347--378} (\bibinfo{publisher}{Springer},
  \bibinfo{year}{2003}).

\bibitem{rocha2010information}
\bibinfo{author}{Rocha, L.~E.}, \bibinfo{author}{Liljeros, F.} \&
  \bibinfo{author}{Holme, P.}
\newblock \bibinfo{title}{Information dynamics shape the sexual networks of
  internet-mediated prostitution}.
\newblock \emph{\bibinfo{journal}{Proceedings of the National Academy of
  Sciences}} \textbf{\bibinfo{volume}{107}}, \bibinfo{pages}{5706--5711}
  (\bibinfo{year}{2010}).

\bibitem{ER1}
\bibinfo{author}{Erd\H{o}s, P.} \& \bibinfo{author}{R\'enyi, A.}
\newblock \bibinfo{title}{On random graphs}.
\newblock \emph{\bibinfo{journal}{Publicationes Mathematicae}}
  \textbf{\bibinfo{volume}{6}}, \bibinfo{pages}{290--297}
  (\bibinfo{year}{1959}).


\bibitem{klemm2002growing}
\bibinfo{author}{Klemm, K.} \& \bibinfo{author}{Eguiluz, V.~M.}
\newblock \bibinfo{title}{Growing scale-free networks with small-world
  behavior}.
\newblock \emph{\bibinfo{journal}{Physical Review E}}
  \textbf{\bibinfo{volume}{65}}, \bibinfo{pages}{057102}
  (\bibinfo{year}{2002}).

\bibitem{lancichinetti2008benchmark}
\bibinfo{author}{Lancichinetti, A.}, \bibinfo{author}{Fortunato, S.} \&
  \bibinfo{author}{Radicchi, F.}
\newblock \bibinfo{title}{Benchmark graphs for testing community detection
  algorithms}.
\newblock \emph{\bibinfo{journal}{Physical Review E}}
  \textbf{\bibinfo{volume}{78}}, \bibinfo{pages}{046110}
  (\bibinfo{year}{2008}).

\bibitem{parker1993friendship}
\bibinfo{author}{Parker, J.~G.} \& \bibinfo{author}{Asher, S.~R.}
\newblock \bibinfo{title}{Friendship and friendship quality in middle
  childhood: Links with peer group acceptance and feelings of loneliness and
  social dissatisfaction.}
\newblock \emph{\bibinfo{journal}{Developmental Psychology}}
  \textbf{\bibinfo{volume}{29}}, \bibinfo{pages}{611} (\bibinfo{year}{1993}).

\bibitem{anderson1999p}
\bibinfo{author}{Anderson, C.~J.}, \bibinfo{author}{Wasserman, S.} \&
  \bibinfo{author}{Crouch, B.}
\newblock \bibinfo{title}{A p* primer: Logit models for social networks}.
\newblock \emph{\bibinfo{journal}{Social Networks}}
  \textbf{\bibinfo{volume}{21}}, \bibinfo{pages}{37--66}
  (\bibinfo{year}{1999}).

\bibitem{zachary1977information}
\bibinfo{author}{Zachary, W.~W.}
\newblock \bibinfo{title}{An information flow model for conflict and fission in
  small groups}.
\newblock \emph{\bibinfo{journal}{Journal of Anthropological Research}}
  \textbf{\bibinfo{volume}{33}}, \bibinfo{pages}{452--473}
  (\bibinfo{year}{1977}).

\bibitem{coleman1964introduction}
\bibinfo{author}{Coleman, J.~S.} \emph{et~al.}
\newblock \bibinfo{title}{Introduction to mathematical sociology.}
\newblock \emph{\bibinfo{journal}{Introduction to mathematical sociology}}
  (\bibinfo{year}{1964}).

\bibitem{girvan2002community}
\bibinfo{author}{Girvan, M.} \& \bibinfo{author}{Newman, M.~E.}
\newblock \bibinfo{title}{Community structure in social and biological
  networks}.
\newblock \emph{\bibinfo{journal}{Proceedings of the National Academy of
  Sciences}} \textbf{\bibinfo{volume}{99}}, \bibinfo{pages}{7821--7826}
  (\bibinfo{year}{2002}).

\bibitem{ohtsuki2007direct}
\bibinfo{author}{Ohtsuki, H.} \& \bibinfo{author}{Nowak, M.~A.}
\newblock \bibinfo{title}{Direct reciprocity on graphs}.
\newblock \emph{\bibinfo{journal}{Journal of Theoretical Biology}}
  \textbf{\bibinfo{volume}{247}}, \bibinfo{pages}{462--470}
  (\bibinfo{year}{2007}).

\bibitem{reiter2018crosstalk}
\bibinfo{author}{Reiter, J.~G.}, \bibinfo{author}{Hilbe, C.},
  \bibinfo{author}{Rand, D.~G.}, \bibinfo{author}{Chatterjee, K.} \&
  \bibinfo{author}{Nowak, M.~A.}
\newblock \bibinfo{title}{Crosstalk in concurrent repeated games impedes direct
  reciprocity and requires stronger levels of forgiveness}.
\newblock \emph{\bibinfo{journal}{Nature Communications}}
  \textbf{\bibinfo{volume}{9}}, \bibinfo{pages}{555} (\bibinfo{year}{2018}).

\bibitem{olejarz2015indirect}
\bibinfo{author}{Olejarz, J.}, \bibinfo{author}{Ghang, W.} \&
  \bibinfo{author}{Nowak, M.~A.}
\newblock \bibinfo{title}{Indirect reciprocity with optional interactions and
  private information}.
\newblock \emph{\bibinfo{journal}{Games}} \textbf{\bibinfo{volume}{6}},
  \bibinfo{pages}{438--457} (\bibinfo{year}{2015}).

\bibitem{willer1999network}
\bibinfo{author}{Willer, D.}
\newblock \emph{\bibinfo{title}{Network exchange theory}}
  (\bibinfo{publisher}{Greenwood Publishing Group}, \bibinfo{year}{1999}).
%

\bibitem{wang2013optimal}
\bibinfo{author}{Wang, Z.}, \bibinfo{author}{Szolnoki, A.} \&
  \bibinfo{author}{Perc, M.}
\newblock \bibinfo{title}{Optimal interdependence between networks for the
  evolution of cooperation}.
\newblock \emph{\bibinfo{journal}{Scientific Reports}}
  \textbf{\bibinfo{volume}{3}}, \bibinfo{pages}{2470} (\bibinfo{year}{2013}).

\bibitem{wang2013interdependent}
\bibinfo{author}{Wang, Z.}, \bibinfo{author}{Szolnoki, A.} \&
  \bibinfo{author}{Perc, M.}
\newblock \bibinfo{title}{Interdependent network reciprocity in evolutionary
  games}.
\newblock \emph{\bibinfo{journal}{Scientific Reports}}
  \textbf{\bibinfo{volume}{3}}, \bibinfo{pages}{1183} (\bibinfo{year}{2013}).

\bibitem{jiang2013spreading}
\bibinfo{author}{Jiang, L.-L.} \& \bibinfo{author}{Perc, M.}
\newblock \bibinfo{title}{Spreading of cooperative behaviour across
  interdependent groups}.
\newblock \emph{\bibinfo{journal}{Scientific Reports}}
  \textbf{\bibinfo{volume}{3}}, \bibinfo{pages}{2483} (\bibinfo{year}{2013}).

\bibitem{wang2014rewarding}
\bibinfo{author}{Wang, Z.}, \bibinfo{author}{Szolnoki, A.} \&
  \bibinfo{author}{Perc, M.}
\newblock \bibinfo{title}{Rewarding evolutionary fitness with links between
  populations promotes cooperation}.
\newblock \emph{\bibinfo{journal}{Journal of Theoretical Biology}}
  \textbf{\bibinfo{volume}{349}}, \bibinfo{pages}{50--56}
  (\bibinfo{year}{2014}).
%

\bibitem{battiston2017determinants}
\bibinfo{author}{Battiston, F.}, \bibinfo{author}{Perc, M.} \&
  \bibinfo{author}{Latora, V.}
\newblock \bibinfo{title}{Determinants of public cooperation in multiplex
  networks}.
\newblock \emph{\bibinfo{journal}{New Journal of Physics}}
  \textbf{\bibinfo{volume}{19}}, \bibinfo{pages}{073017}
  (\bibinfo{year}{2017}).

\bibitem{tarnita2009strategy}
\bibinfo{author}{Tarnita, C.~E.}, \bibinfo{author}{Ohtsuki, H.},
  \bibinfo{author}{Antal, T.}, \bibinfo{author}{Fu, F.} \&
  \bibinfo{author}{Nowak, M.~A.}
\newblock \bibinfo{title}{Strategy selection in structured populations}.
\newblock \emph{\bibinfo{journal}{Journal of Theoretical Biology}}
  \textbf{\bibinfo{volume}{259}}, \bibinfo{pages}{570--581}
  (\bibinfo{year}{2009}).

\end{thebibliography}

\begin{thebibliography}{100}

\bibitem{allen2017evolutionary}
Benjamin Allen, Gabor Lippner, Yu-Ting Chen, Babak {Fotouhi}, Naghmeh {Momeni},
  Shing-Tung Yau, and Martin~A Nowak.
\newblock Evolutionary dynamics on any population structure.
\newblock {\em Nature}, 544(7649):227--230, 2017.

\bibitem{holme2002growing}
Petter Holme and Beom~Jun Kim.
\newblock Growing scale-free networks with tunable clustering.
\newblock {\em Physical Review E}, 65(2):026107, 2002.

\bibitem{leskovec2005graphs}
Jure Leskovec, Jon Kleinberg, and Christos Faloutsos.
\newblock Graphs over time: densification laws, shrinking diameters and
  possible explanations.
\newblock In {\em Proceedings of the eleventh ACM SIGKDD international
  conference on Knowledge discovery in data mining}, pages 177--187. ACM, 2005.

\bibitem{barthelemy2003crossover}
Marc Barth{\'e}lemy.
\newblock Crossover from scale-free to spatial networks.
\newblock {\em EPL (Europhysics Letters)}, 63(6):915, 2003.

\bibitem{barabasi2016network}
Albert-L{\'a}szl{\'o} Barab{\'a}si.
\newblock {\em Network Science}.
\newblock Cambridge university press, 2016.

\bibitem{lancichinetti2008benchmark}
Andrea Lancichinetti, Santo Fortunato, and Filippo Radicchi.
\newblock Benchmark graphs for testing community detection algorithms.
\newblock {\em Physical Review E}, 78(4):046110, 2008.

\bibitem{decelle2011asymptotic}
Aurelien Decelle, Florent Krzakala, Cristopher Moore, and Lenka Zdeborov{\'a}.
\newblock Asymptotic analysis of the stochastic block model for modular
  networks and its algorithmic applications.
\newblock {\em Physical Review E}, 84(6):066106, 2011.

\bibitem{newman2006modularity}
Mark~EJ Newman.
\newblock Modularity and community structure in networks.
\newblock {\em Proceedings of the National Academy of Sciences},
  103(23):8577--8582, 2006.

\bibitem{maciejewski2014reproductive}
Wes Maciejewski.
\newblock Reproductive value in graph-structured populations.
\newblock {\em Journal of Theoretical Biology}, 340:285--293, 2014.

\bibitem{aldous2002reversible}
David Aldous and Jim Fill.
\newblock Reversible markov chains and random walks on graphs, 2002.

\bibitem{chen2013sharp}
Yu-Ting Chen et~al.
\newblock Sharp benefit-to-cost rules for the evolution of cooperation on
  regular graphs.
\newblock {\em The Annals of Applied Probability}, 23(2):637--664, 2013.

\bibitem{holley1975ergodic}
Richard~A Holley and Thomas~M Liggett.
\newblock Ergodic theorems for weakly interacting infinite systems and the
  voter model.
\newblock {\em The Annals of Probability}, pages 643--663, 1975.

\bibitem{cox1989coalescing}
J~Theodore Cox.
\newblock Coalescing random walks and voter model consensus times on the torus
  in  $\mathbb{Z}^d$.
\newblock {\em The Annals of Probability}, pages 1333--1366, 1989.

\bibitem{liggett2012interacting}
Thomas~Milton Liggett.
\newblock {\em Interacting particle systems}, volume 276.
\newblock Springer Science \& Business Media, 2012.

\end{thebibliography}
\addcontentsline{toc}{section}{Supplementary   References}

\begingroup
\renewcommand{\addcontentsline}[3]{}
\renewcommand{\section}[2]{}

\endgroup

\end{document}